\documentclass[preprint]{aastex6}

\usepackage{hyperref}
\usepackage{breakurl}
\usepackage{amsmath}
\usepackage{graphicx}
\usepackage{verbatim}
\usepackage{booktabs}
\usepackage{textcomp}   

\begin{document}

\title[Iodine-free spectra]{Deriving Iodine-free spectra for high-resolution echelle spectrographs\footnote{Based on observations obtained with the Magellan Telescopes, operated by the Carnegie Institution, Harvard University, University of Michigan, University of Arizona, and the Massachusetts Institute of Technology.}}
\shorttitle{Iodine-free spectra for echelle spectrographs}
\shortauthors{D\'iaz et al.}

\author{Mat\'ias R. D\'iaz\altaffilmark{1,4}, Stephen A. Shectman\altaffilmark{2}, R. Paul Butler\altaffilmark{3} \& James S. Jenkins\altaffilmark{1}}

\altaffiltext{1}{Departamento de Astronom\'ia, Universidad de Chile, Camino El Observatorio 1515, Las Condes, Santiago, Chile.}
\altaffiltext{2}{The Observatories, Carnegie Institution for Science, 813 Santa Barbara Street, Pasadena, CA 91101, USA.}
\altaffiltext{3}{Department of Terrestrial Magnetism, Carnegie Institution for Science, 5241 Broad Branch Road NW, Washington, DC 20015, USA.}
\altaffiltext{4}{Carnegie-Chile Graduate Fellow 2015-2017, The Observatories, Carnegie Institution for Science.}
\email{$\dagger$ matias.diaz.m@ug.uchile.cl}

\begin{abstract}
We describe a new method to derive clean, iodine-free spectra directly from observations acquired using high resolution echelle spectrographs equipped with iodine cells. 
The main motivation to obtain iodine-free spectra is to use portions of the spectrum that are superimposed with the dense forest of iodine absorption lines, in order to retrieve lines that can be used to monitor the magnetic activity of the star, helping to validate candidate planets. 
In short, we provide a straight-forward methodology to clean the spectra by using the forward model used to derive radial velocities, the Line Spread Function information plus the stellar spectrum without iodine to reconstruct and subtract the iodine spectrum from the observations. We show our results using observations of the star $\tau$ Ceti acquired with the PFS, HIRES and UCLES spectrographs, reaching an iodine-free spectrum correction at the $\sim$1\% RMS level. We additionally discuss the limitations and further applications of the method.
\end{abstract}


\section{Introduction} \label{sec:intro}
Detection methods and instrumental techniques have dramatically improved  their precision since the detection of the first extrasolar planet, a little over 20 years ago. Doppler spectroscopy has been one of the two most successful detection techniques to search for exoplanets orbiting other stars. In order to achieve the radial velocity precision needed to detect companions orbiting the host star, precise wavelength references are needed. Over the years, two main methods have stood out as the best to derive precise radial velocities: one is using simultaneous calibration, within a highly stabilized echelle platform, by applying two fibre observations to simultaneously record the spectrum of the star and an `in-house' calibration source (e.g. Thorium-Argon). This method has been improved and new calibration sources such as laser frequency combs can deliver radial velocities with a precision of less than half a meter per second (e.g. HARPS, \citealt{Pepe2002}; ESPRESSO, \citealt{Pepe2010}). This method is often used with the cross-correlation technique and it produced the very first detection of an exoplanet orbiting a solar-type star \citep{MayorQueloz1995}. 

The second way to obtain precise radial velocity measurements is done by superimposing a gaseous absorption reference spectrum onto the stellar spectrum, providing a precise wavelength scale that is required for such work.  A general benefit of this method is that the gas cell is small and portable, and can be applied to unstabilized systems. It has been found that cells mounted at the entrance of the spectrograph dramatically improved radial velocity precision, going from hundreds of meters per second down to $\sim$10 meters per second in the late 1980s. The idea, originally discussed in \citet{Griffin1973}, was improved  by \citet{CampbellWalker1979} by using hydrogen fluoride to generate a reference using absorption lines. They achieved a precision at the level of $\sim$15 meters per second. The main problem of the method used by \citet{CampbellWalker1979} is that hydrogen fluoride gas provided few lines over a small wavelength range, besides being poisonous. Later, \cite{MarcyButler1992} improved the precision by making use of molecular iodine to generate reference absorption lines in the stellar spectra. While the initial precision was $\sim$20 meters per second, further improvements in the technique \citep{Valenti1995} allowed them to achieve a radial velocity precision of $\sim$3 meters per second nearly a decade later \citep{Butler1996}. The precision our team has achieved is now at the 1 meter per second level \citep{Vogt2015,Butler2017,Diaz2018}. 

The iodine cell method brought with it a large number of exoplanet discoveries during the next few years \cite[e.g.,][]{MarcyButler1996, ButlerMarcy1996, Marcy1999, Butler1999, Rivera2005}. Given its obvious successes in allowing the detection of exoplanets using currently functioning echelle spectrographs, the problem always remained that the spectra were essentially rendered useless for any follow-up work beyond velocity calculations.  In particular, it has subsequently been found that chromospheric lines, or indeed other photospheric absorption lines can be used in the modelling process to detect smaller signals, and hence smaller planets, or also to better validate signals as being Doppler in nature \citep{Tuomi2013b,JenkinsTuomi2014,Tuomi2014a, AngladaEscude2012,Anglada2016,Boisse2011,Boisse2012}. Planet search teams benefit from being able to calculate line bisector inverse slopes \cite[BIS,][]{Dravnis1987,TonerGray1988,Queloz2001} in the region where the radial velocities were calculated with iodine, or have access to additional activity sensitive lines to use in their analysis, e.g. Na D, Fe {\sc i}, Mg {\sc i} \citep{Dumusque2018,Wise2018}. 

The main goal of this work is to provide a detailed description for deriving iodine-free spectra for every observation that we have made with echelle spectrographs that are equipped with iodine cells that are used to obtain accurate radial velocities. We initially developed this algorithm to derive iodine-free spectra for PFS, but the application has extended to other spectrographs such as HIRES and UCLES. We also discuss the possibilities that this method enables when analyzing the clean spectrum of the star.

We note the existence of previous works in an attempt to correct or use clean spectra from observations acquired with iodine, as described in \citet{Niedzielski2009} and \citet{Nowak2010} where they cleaned the spectra using a numerical mask constructed by a sum of delta functions centered in the position of selected iodine lines. The approach we present herein differs on previous works as we use our forward model to provide a full iodine-cleaned spectrum for every observation.
 
\section{The iodine cell method for precise radial velocity measurements}

The advantage of the iodine cell is that thousands of absorption lines are superimposed onto the stellar spectrum at the time of the observation, providing a precise wavelength calibration that is required to achieve $\sim$1 m s$^{-1}$ radial velocity precision to search for small exoplanets around other stars. As the precision has been improving over the last 20 years, new challenges have arisen when trying to search for small signatures of exoplanet candidates, mainly because the precision of the measurements are now reaching below the intrinsic stellar noise, often referred to as stellar jitter, of even the most quiescent stars. Efforts have been made to not only understand how the star varies with time, i.e., monitoring its activity with spectral indices that quantify the chromospheric activity, but to also `correct' for this additional noise effect. 

Several stellar absorption lines serve as proxies to determine the activity of a star. One well known example is the Mount Wilson S-indices derived from the Ca {\sc ii} H \& K lines \citep{Duncan1991}, which have been considered as one standard proxy of stellar activity. Also, H$\alpha$ lines are used as diagnostics to analyze the stellar activity as a function of time. Radial velocity searches have also made use of these stellar activity indicators, as it is crucial to estimate rotation periods and stellar jitter \citep{Tinney2002,Wright2005,Jenkins2006,Jenkins2008,Jenkins2011,Arriagada2011} and in some cases to validate the true nature of a planetary candidate \citep[see e.g.,][]{Queloz2001, JenkinsTuomi2014, Santos2014, Diaz2018, Paneque2019}. All these activity indices are still used with iodine-cell spectrographs, since they fall outside the iodine forest.

Given that iodine provides a dense forest of lines from $\sim$5000\AA~to $\sim6300$\AA, every stellar line within this region is blended with several iodine lines, rendering the calculation of spectral indices and bisectors a useless task.

\begin{table*}
\centering
\caption{Instrumental parameters of echelle spectra.}
\label{tab:instruments}
\begin{tabular}{cccccc}
\hline\hline
\multicolumn{1}{c}{Instrument}& \multicolumn{1}{c}{Resolution} & \multicolumn{1}{c}{Total Number} & \multicolumn{1}{c}{Total Number} &\multicolumn{1}{c}{Chunks}& \multicolumn{1}{c}{Chunk width} \\ 
\multicolumn{1}{c}{}& \multicolumn{1}{c}{($\lambda /\Delta \lambda)$} & \multicolumn{1}{c}{of orders$^{a}$} & \multicolumn{1}{c}{of chunks}&\multicolumn{1}{c}{per order$^{b}$} & \multicolumn{1}{c}{(pixels)} \\\hline
PFS & 80,000 & 26 & 818 &  34&120 \\
HIRES & 60,000 & 14 & 718 &48 & 80 \\
UCLES & 50,000 & 20 & 944& 48& 50  \\ \hline\hline
\end{tabular}
\tablenotetext{a}{\footnotesize Effective number of orders of the echelle in the iodine region used for the computation of radial velocities.}
\tablenotetext{b}{\footnotesize For PFS, HIRES and UCLES, the first order is divided into 18, 31 and 32 chunks, respectively.}
\end{table*}

\begin{figure*}
\centering
\includegraphics[scale=1]{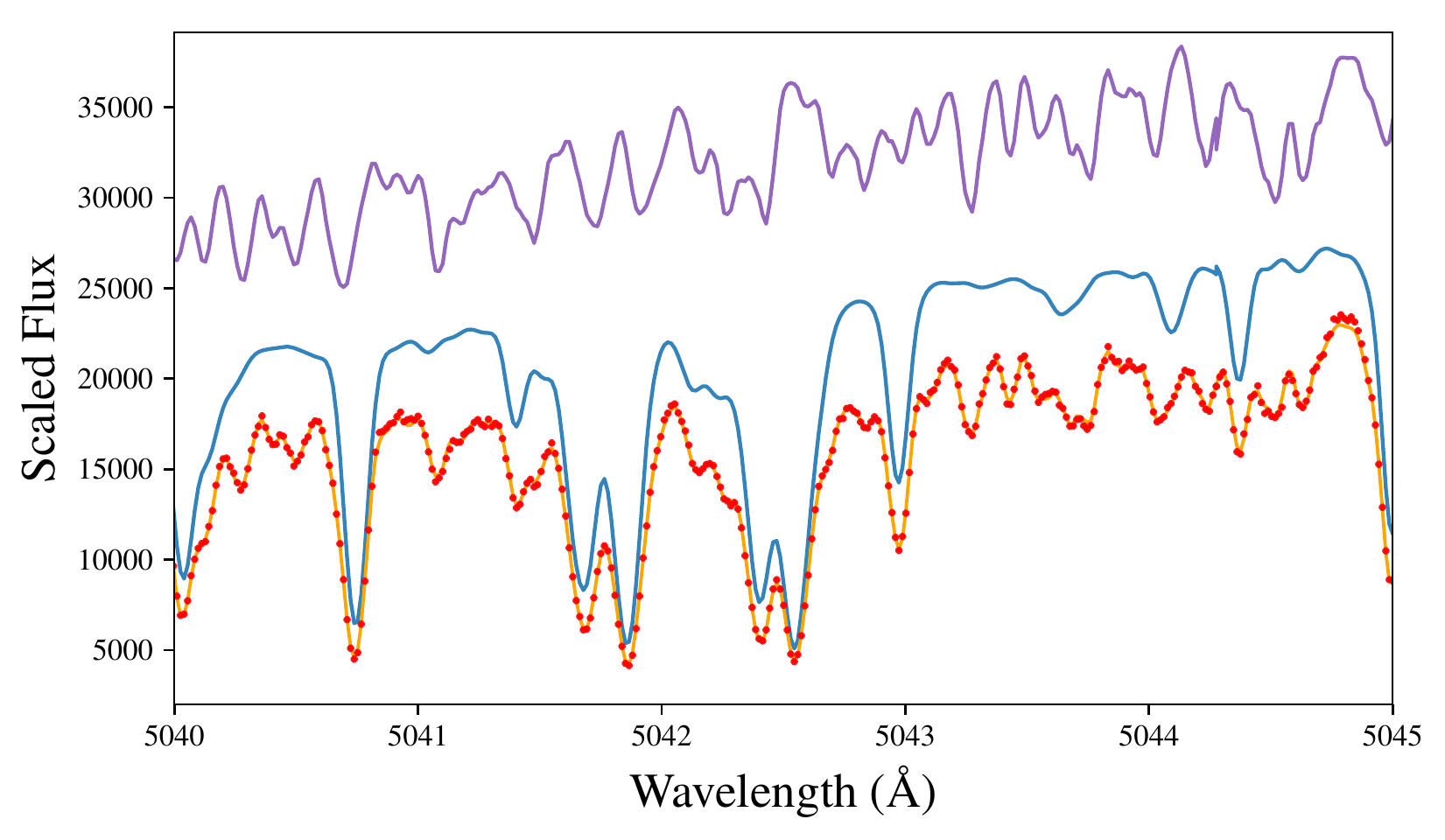}
\caption{{ Forward modelling process showing a 5\AA~portion for an observation of the star $\tau$ Ceti observed with PFS}. From top to bottom: Iodine spectrum convolved with the LSF ($T_{\rm I_{2}} \otimes$ LSF, purple),  intrinsic spectrum of the star, observed without iodine convolved with the LSF ($I_{\star}\otimes$LSF, blue), observation of the star through the iodine cell ($S_x$, red points), forward model of the observations ($S(x)$, orange). Note the scale in the y-axis, each spectrum is multiplied by some arbitrary factor for comparison only.}
\label{fig:mod}
\end{figure*}
\subsection{Fundamentals of Iodine Observations}

All the fundamentals of the iodine-cell method are described in \citet{Butler1996}, but in this section we recall the basic idea behind this technique. In particular, we rewrite the notation for an observed spectrum of a star through the iodine cell. That spectrum is modelled as follows

\begin{equation}\label{eq:fwdmodel}
S(x)=\kappa\,[T_{I_{2}}(\lambda(x))I_\star(\lambda(x)/(1-v/c))]\otimes\mathrm{LSF}(x)
\end{equation}
where the observation is modeled as the product of two functions: the intrinsic stellar spectrum, $I_\star$, and the transmission function of the iodine, $T_{I_{2}}$, and then convolved with the point spread function of the instrument. In our case, we work with extracted spectra so it is more accurate to use the term Line Spread Function (LSF) instead. The constant $\kappa$ is a normalization factor and $v$ corresponds to the Doppler shift, i.e, the radial velocity we will later fit as a free parameter. We note the use of a slightly different nomenclature to that in \citet{Butler1996}. Here we denote our forward model of the observations by $S(x)$ rather than $I_{obs}$. The observed spectrum, i.e., the actual data we record in the detector of the instrument is denoted by $S_x$. 

The transmission function of the iodine spectrum, $T_{I_{2}}$, is obtained by scanning the iodine cell itself using the Fourier Transform Spectrum (FTS) method described in \citet{Valenti1995}. The method relies on the acquisition of a high-resolution ($R\sim$10$^{6}$), high signal-to-noise (S/N$\sim$1000) spectrum of the iodine that will provide an extremely precise ($\sim$10$^{-8}$) vacuum wavelength scale.
 
The intrinsic stellar spectrum is obtained by observing the star without the iodine cell in place. However, this observed spectrum is not precisely $I_\star$, instead, this spectrum carries the smearing due to the instrumental profile, so $I_\star\otimes$LSF is obtained. 
 
In order to correct for the LSF smearing effect, a rapidly-rotating, bright (V$\sim$3 mag) B-type star is normally observed through the iodine cell before and after the observation of the stellar reference spectrum. 
Given the lack of spectral features for B-type stars, their spectrum is essentially continuum, they serve as incandescent lamps shining through the iodine absorption cell. 
Then a modified version \citep{Gilliland1992, Butler1996} of the original Jansson deconvolution procedure \citep{Jansson1984} allows this instrumental LSF to be deconvolved from the intrinsic stellar spectrum.

\begin{figure*}
\centering
\includegraphics[scale=0.59]{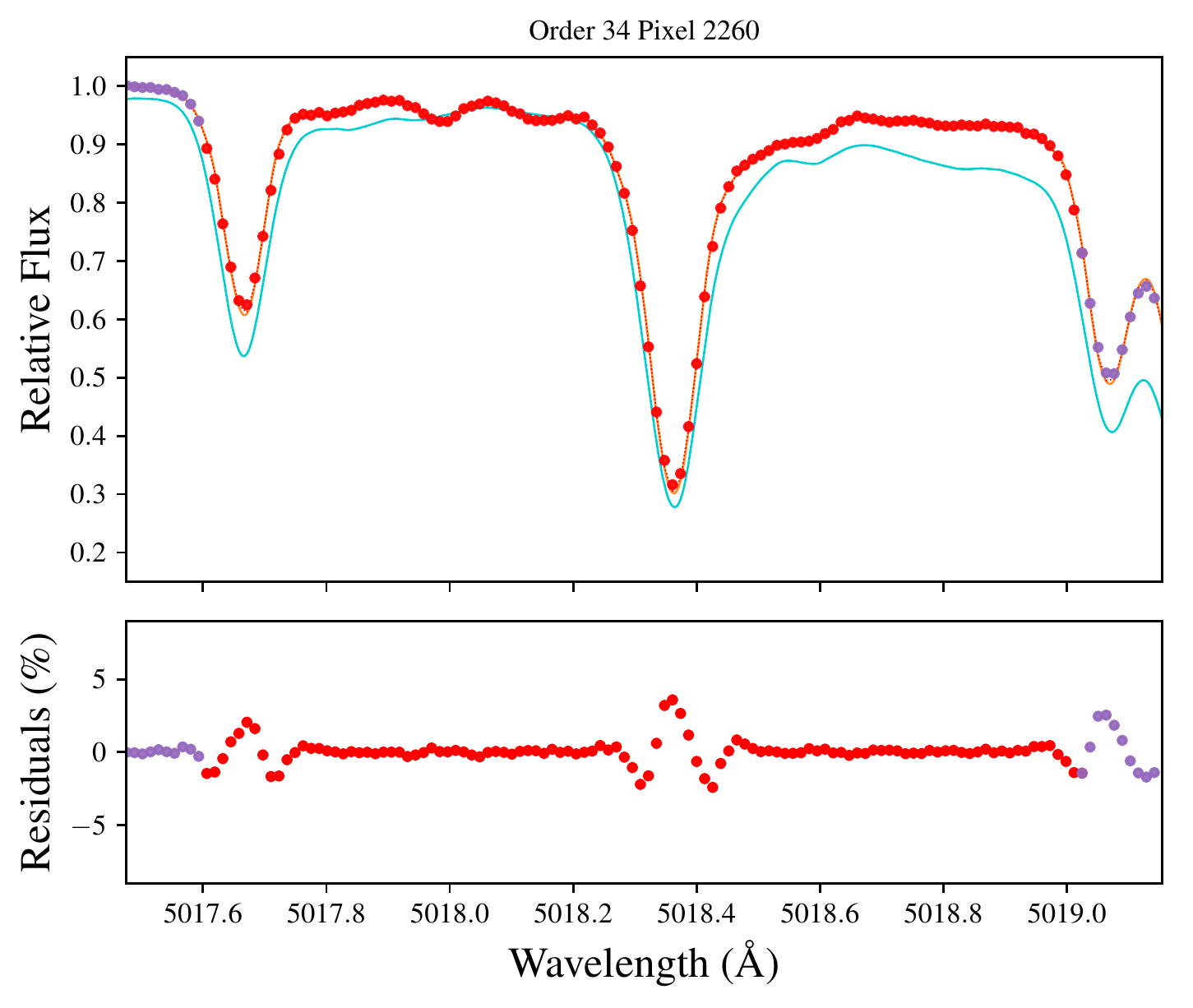}
\includegraphics[scale=0.59]{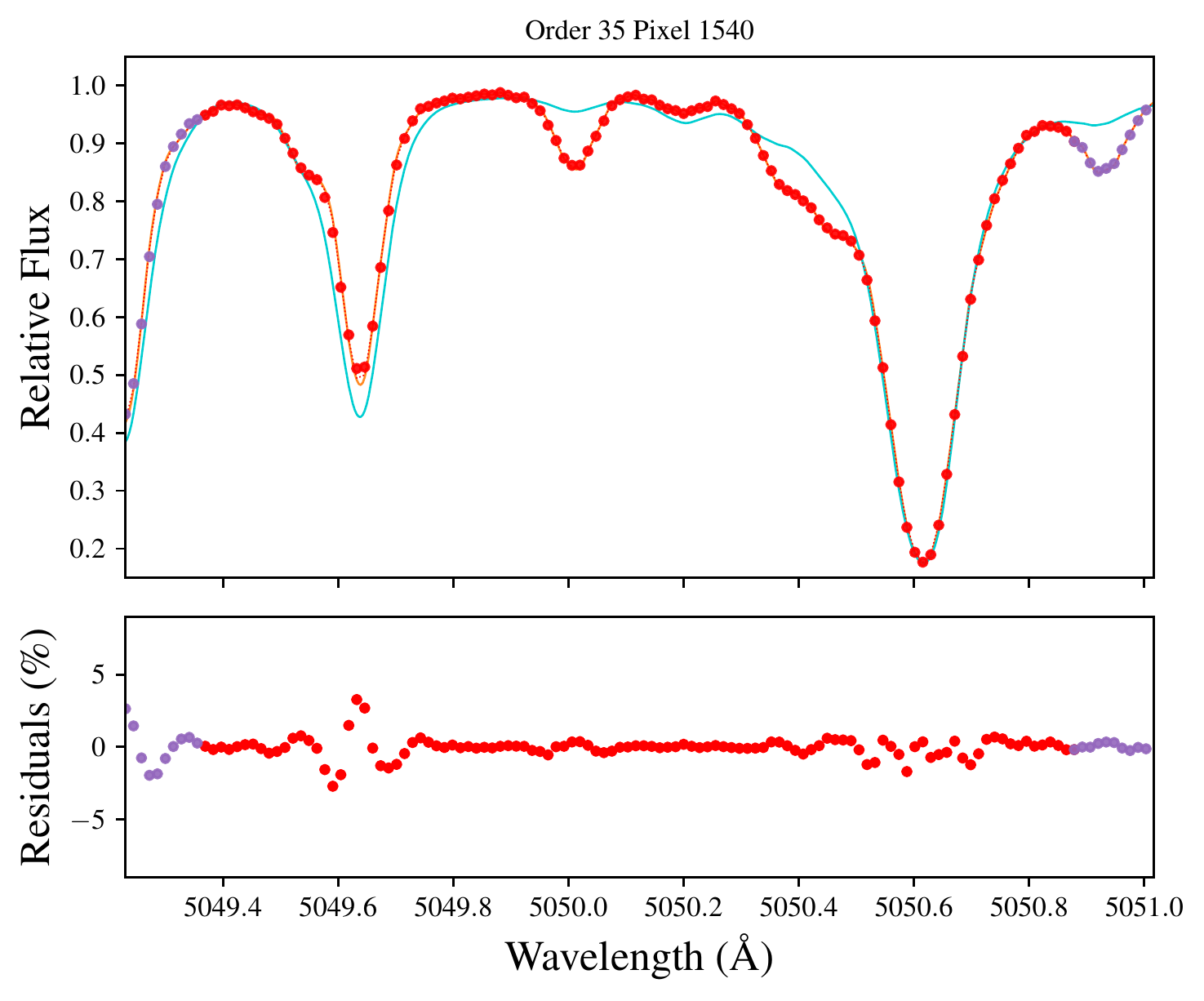}
\caption{Left Panel: Deconvolution of the stellar spectrum for the star $\tau$ Ceti. We show a portion consisting on 2\AA. Red points correspond to the observation. {Purple points show data from 10 pixels on adjacent chunks so the deconvolution routine can work properly at the edge of each chunk}. The deconvolved model of the observation is shown as a dashed-red curve (apparent on top of the solid orange curve). The deconvolved stellar spectrum is shown as a solid orange curve. The solar spectrum binned to the sampling of the observations is plotted in cyan as comparison. Bottom: Difference between the data and the deconvolved stellar spectrum, binned to the same sampling of the observations. Right Panel: Same as in left panel, but for another 2\AA~chunk.}
 \label{fig:deconv}
\end{figure*}

We generate a model of the observations, four times oversampled, by convolving the resultant deconvolved stellar spectrum with the instrumental profile and with the same sampling as the observations (see Figure \ref{fig:mod}). Figure \ref{fig:deconv} shows the deconvolution results for two chunks of spectrum for the star $\tau$ Ceti acquired with PFS. Red points corresponds to the observation of the star without the iodine cell in place. The deconvolved stellar spectrum model is shown with an orange solid curve. The purple points represent 10 pixels on each edge of a chunk. Due to the LSF, each pixel is the smeared average of the nearby pixels. In the center of a given chunk this is not a problem. At the edges, however, it is necessary to include some extra pixels from outside the chunk, for the deconvolution routine to be able to deal with the edge of the chunk. From the difference plots in the bottom panels of Figure \ref{fig:deconv} we note that the deconvolution is only very slightly different than the observation. The regions where the deconvolved spectrum is different compared with the observed data are mostly the depth of the lines and the difference is less than $\sim$5\%.

\subsection{Determining the LSF, wavelength solution and Doppler shift}

We implement the description of the LSF by a sum of Gaussians where the central one is fixed and the other are allowed to vary in amplitude, as explained in \citet{Valenti1995}. We define

\begin{equation}\label{eq:psf_desc}
LSF(x) = \sum_{m} G_{m}(x)
\end{equation}
where
\begin{equation}\label{eq:gauss_i}
G_{m}(x) = A_{m} \, e^ { ( x - x_{0,m})^{2}/2\sigma^{2}_{m}}
\end{equation}

The process of generating a LSF prescription involves finding a single fixed central Gaussian, $G_{1}$. The half-width of this central Gaussian, $\sigma_{1}$, is found by trial and error. Once the central Gaussian is chosen the additional Gaussians are placed at fixed positions, $x_{0,m}$, on either side of the central one such that their half-widths, $\sigma_{m}$ overlap, to enforce smoothness. The number of additional Gaussians, and their widths, are found by trial and error, and they are different for each instrument. The goal is to build a LSF prescription with the fewest number of free parameters, but with sufficient flexibility to deal with the variations due to the instrument and the input image on the entrance slit.

The wavelength scale is described by two parameters as follows 

\begin{equation}\label{eq:wav_mod}
{\lambda}_{i} = \lambda_{0} + b\, x_{i}
\end{equation}

where $\lambda_{0}$ refers to the wavelength zero point, $b$ corresponds to the linear dispersion and $i$ corresponds to the pixel index on the $j$-th chunk, $i=0,...,n_{pix}-1$, where $n_{pix}$ corresponds to the chunk width and $j=0,...,n_{chunk}-1$ (see Table \ref{tab:instruments}).

The initial guess for the wavelength scale is determined from an exposure of iodine and a quartz lamp as part of the nightly calibrations, but the final wavelength solution will be determined by a non-linear least squares minimization process that uses the Marquardt gradient-expansion technique \citep{Valenti1995, BevingtonRobinson2003}.
 
The radial velocity is defined by only one parameter

\begin{equation}\label{eq:z}
z = \frac{\Delta \lambda}{\lambda} = \frac{v}{c}
\end{equation}

The barycentric correction is calculated with our own package that makes use of the JPL ephemeris. This correction is subtracted from the derived Doppler shift.
 
In this way, 16 parameters are needed to fully describe our spectral modelling process: $\kappa$, $\lambda_{0}$, $b$, $v$ and $\sim$12 amplitudes for the Gaussians. We perform three iterations to derive the best-fit parameters that satisfy the  model (equations \ref{eq:wav_mod} to \ref{eq:z}). In the first pass, all the parameters are allowed to vary. In a second iteration, the linear dispersion is fixed. In a final pass, the linear dispersion and the LSF are fixed and the wavelength zero point and the Doppler shift are allowed to vary. The fixed LSF is a weighted average of LSF in a given chunk and the nearest adjacent chunks.

\section{Iodine-free spectra derivation}\label{sec:method}

The primary motivation for deriving iodine-free spectra is to recover the stellar spectrum so that stellar activity indicators from lines that fall within the iodine wavelength region can be measured. 
We have defined all the elements involved in the calculation of radial velocities from observations of a star acquired through the iodine cell. Now, we extend this idea to describe a straightforward approach to `remove' the iodine spectrum directly from each observation.

From equation \ref{eq:fwdmodel}, we recall that $S(x)$ represents the forward model of observed star through the iodine cell. 
Now, we define 
\begin{equation}\label{eq:modprima}
S_\star(\lambda) = \kappa \,I_\star(\lambda)\otimes \mathrm{LSF}
\end{equation}
where $I_\star(\lambda)$ is the intrinsic stellar spectrum, and $S_\star(\lambda)$ is what we would expect to observe if there was not any iodine.

Then, we introduce the following definition

\begin{equation}\label{eq:iodintensity} 
S_{I_{2}} = S(\lambda) - S_\star(\lambda)
\end{equation}
where $S_{I_{2}}$ corresponds to the intensity of the iodine spectrum in the model of the data.

From equations \ref{eq:fwdmodel}, \ref{eq:modprima} and \ref{eq:iodintensity} we can write 
\begin{equation}
S_{\star,x} = S_\star(x) - S_{I_{2}} 
\end{equation}
So, the iodine-free spectrum can be described by
\begin{equation}\label{eq:i2free}
S_{\star,x} = S_{x} - S(\lambda) + S_\star(\lambda)
\end{equation}

{Another way of understanding this method is by considering $S_{x}$ - $S(\lambda)$, this corresponds to the data minus the forward model of the observation. This expression can also be understood as the residuals between the data and the model. Then, $S_{\star} + (S_{x} - S(\lambda))$ will represent the template spectrum, without iodine, with the observed LSF plus the residuals between the data and the model (see Figure \ref{fig:zoomchunks}). }

In order to derive the iodine-free spectra we perform the following steps.

\begin{enumerate}
\item First, the parameters that describe the LSF are defined: the pixels at which the Gaussians are centered, $x_{0, m}$, and the widths, $\sigma_{m}$, with $m=1,...,13$ for the instruments listed in Table \ref{tab:instruments}.  We then retrieve the information from the deconvolved stellar spectrum (template) of the star of interest and the information is stored in our database that consists of an \texttt{IDL} data structure. Each one of these structures contain all the observations for that particular star, the name of the original files, the best-fit parameters from the modelling process. All the steps are carried out relative to the template, chunk by chunk. There are 818 chunks in the 26 iodine orders of the PFS spectra, for instance (see Table \ref{tab:instruments}).

\item We start on the first pixel from the first order of the deconvolved stellar spectrum and we generate a wavelength scale based on the parameters from the template: $\lambda_{0}$ and $b$ from equation \ref{eq:wav_mod}. From the observation, we read the radial velocity from the data structure. This, by definition, corresponds to the shift between the observed spectrum compared with the template, as we do not compute absolute radial velocities.

\item Next, we read the iodine FTS spectrum atlas and we bin it to the same resolution of the observation. After this, we construct a weighted average LSF within a pre-specified domain (adjacent chunks) on the echelle format of the spectrograph. This averaged LSF is then used to convolve the binned iodine spectrum, the template spectrum and the forward model of the observations.

\item The iodine intensity in the observation is obtained by subtracting the template plus iodine and LSF from the forward model (equation \ref{eq:iodintensity}). Finally, we obtain the iodine-free spectrum by subtracting the iodine intensity from the observation, as shown in equation \ref{eq:i2free}. 

\item This process is repeated for every consecutive chunk, then by `pasting' each chunk we generate a reconstructed iodine-free spectrum for each one of the orders. These products are saved and the process is repeated for the next observation of the star.\\
\end{enumerate}

\begin{figure*}
\centering
\includegraphics[scale=0.5]{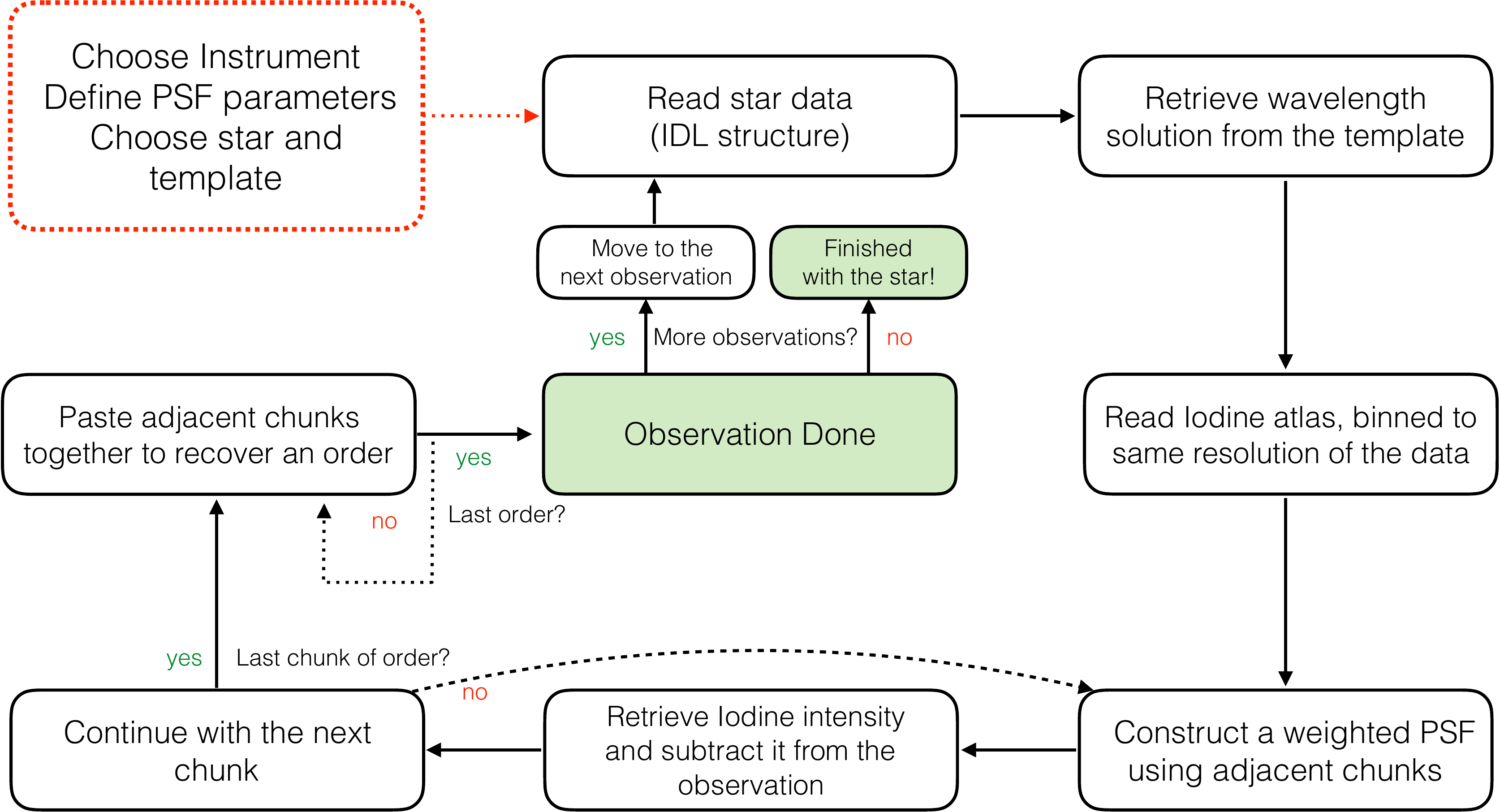}
\caption{Flowchart showing the main steps in the code for the derivation of the clean spectra.}
\label{fig:flowchart}
\end{figure*}

The implementation of the code is written in \texttt{IDL}, and the motivation to do so is to interact with the existing libraries and codes that our group currently uses for Doppler analysis. The iodine-free code is currently working on the instruments that we describe in the following subsections. The iodine-free derivation process is analogous for these instruments with minor differences that take into account each instrument's parameters (see Table \ref{tab:instruments}). Figure \ref{fig:flowchart} shows a simplified flowchart with the main steps of the code.

\subsection{PFS}
The Planet Finder Spectrograph \citep[PFS,][]{Crane2006,Crane2008,Crane2010} is a temperature-controlled, high-resolution echelle spectrograph mounted at the Magellan II/Clay-6.5m telescope at Las Campanas Observatory in Chile. The main purpose of this spectrograph is to search for exoplanets using the iodine cell method. 
The portion of the spectrum that is used to compute radial velocities consists of 26 orders covering the iodine region from $\sim$5011\AA~to $\sim$6337\AA. Each image has a format of 4096x4096 pixels, with a pixel size of 15 $\mu$m. Science observations with PFS typically make use of a 0.5" slit that delivers a resolving power of $\sim$80,000. For the acquisition of the template, stellar spectra are acquired using a narrower 0.3" slit, achieving a higher resolution of $\sim$130,000. 
For the computation of the radial velocities, each order is divided into 34 smaller chunks of 120 pixels width that represent $\sim$2\AA~of spectra. 

\subsection{HIRES}
The High Resolution Echelle Spectrometer \citep[HIRES,][]{Vogt1994} is mounted on the 10m Keck I telescope at the W. M. Keck Observatory in Mauna Kea, Hawai'i. The spectrograph delivers a resolution of $\sim$65,000. The HIRES detector is composed of a 3-chip CCD mosaic covering the 300-1100 nm range.  We refer to these chips as `red', `green' and `blue'. For RV measurements, the iodine spectrum is collected in the `green' CCD. A spectrum obtained through the iodine cell consists of 14 orders between $\sim$4460 \AA~and $\sim$6500 \AA~. In the procedure where the radial velocities are computed, each order in the echelle format is divided into 32 smaller chunks, each of 80 pixels width that are equivalent to $\sim$2\AA~ in the spectra. 

\subsection{UCLES}
The University College London Echelle Spectrograph \citep[UCLES,][]{Diego1990} was mounted on the 3.9m Anglo-Australian Telescope at the Australian Astronomical Observatory. The instrument offers a resolving power of $R\sim$50,000 when operated in its 31 lines/mm mode and making use of a 1 arcsec slit. 
The UCLES detector is a 2K$\times$4K CCD that covers the wavelength range between 300-1100 nm. The iodine region consists of 20 orders, where each order is divided into 48 chunks of 55 pixels width, equivalent to $\sim$2\AA~ in the spectra.
Recently decomissioned, this instrument was used to carry out the Anglo-Australian Planet Search (AAPS) project for 18 years. We derived iodine-free spectra from the data available from the AAPS survey for the star $\tau$ Ceti as well.

\section{Results}
\begin{figure*}
\centering
\includegraphics[scale=1]{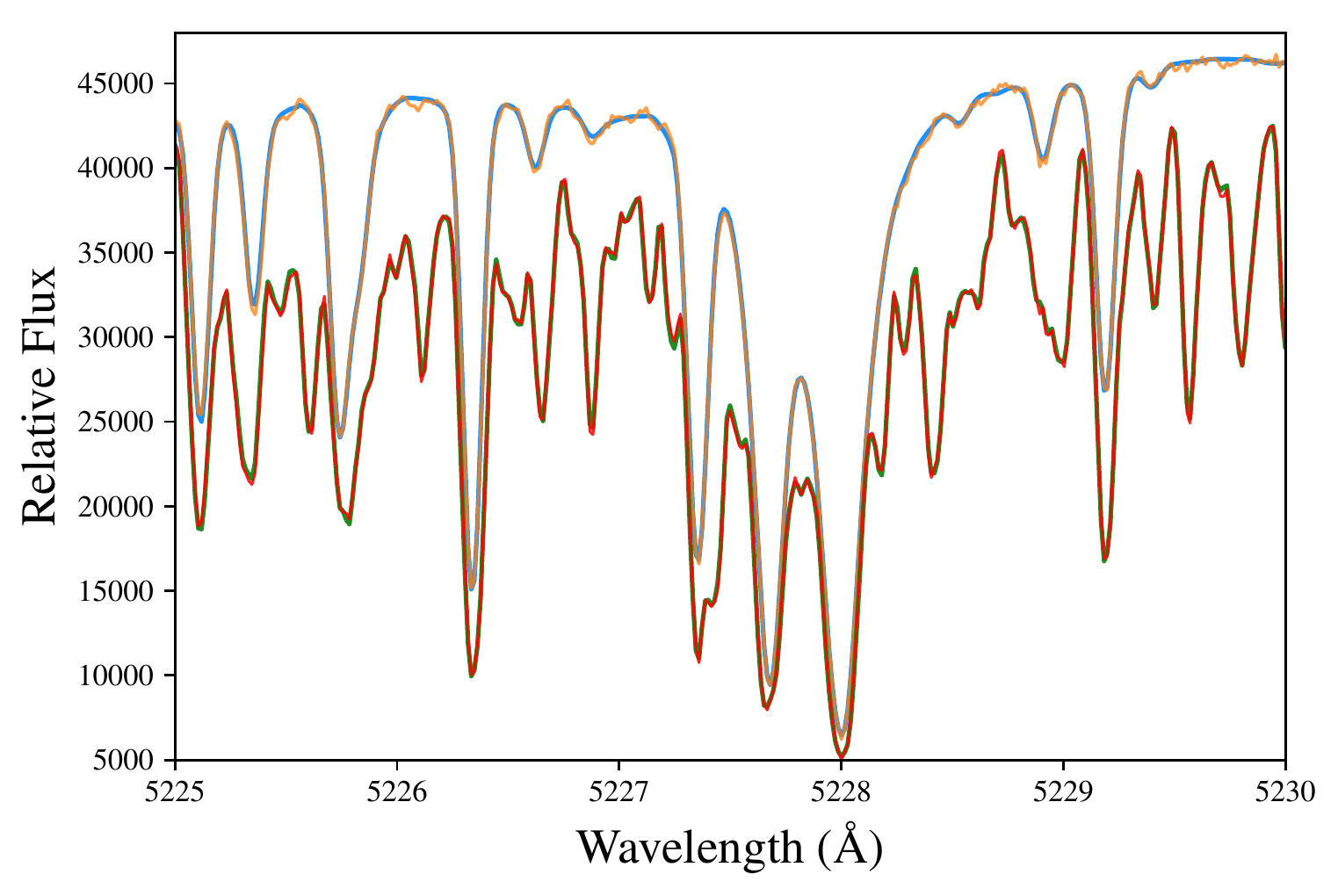}
\includegraphics[scale=1]{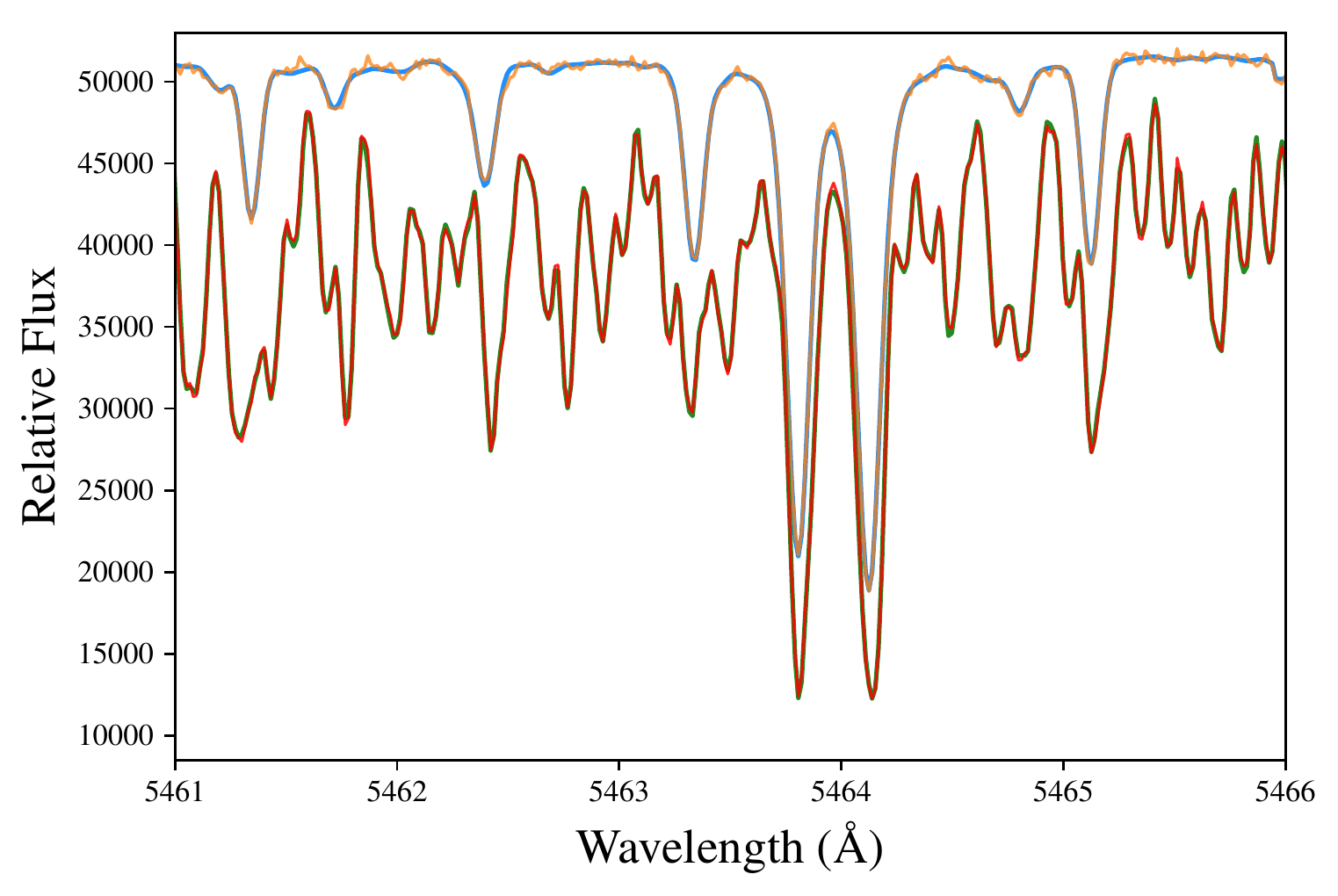}
\caption{5\AA\ of PFS spectra for orders 40 and 45. The red spectrum, $S_{x}$ shows the observation of the star $\tau$ Ceti taken through the iodine cell. The green spectrum represents the forward model of the observations, $S(x)$. The blue spectrum corresponds to the intrinsic stellar spectrum, without iodine, convolved with the LSF, denoted by $S_\star(\lambda)$. The orange spectrum on top of the template represents the iodine-free spectrum, $S_{\star,x}$ derived using the method described in Section \ref{sec:method}.}
\label{fig:zoomchunks}
\end{figure*}

\begin{figure*} 
\includegraphics[scale=0.34]{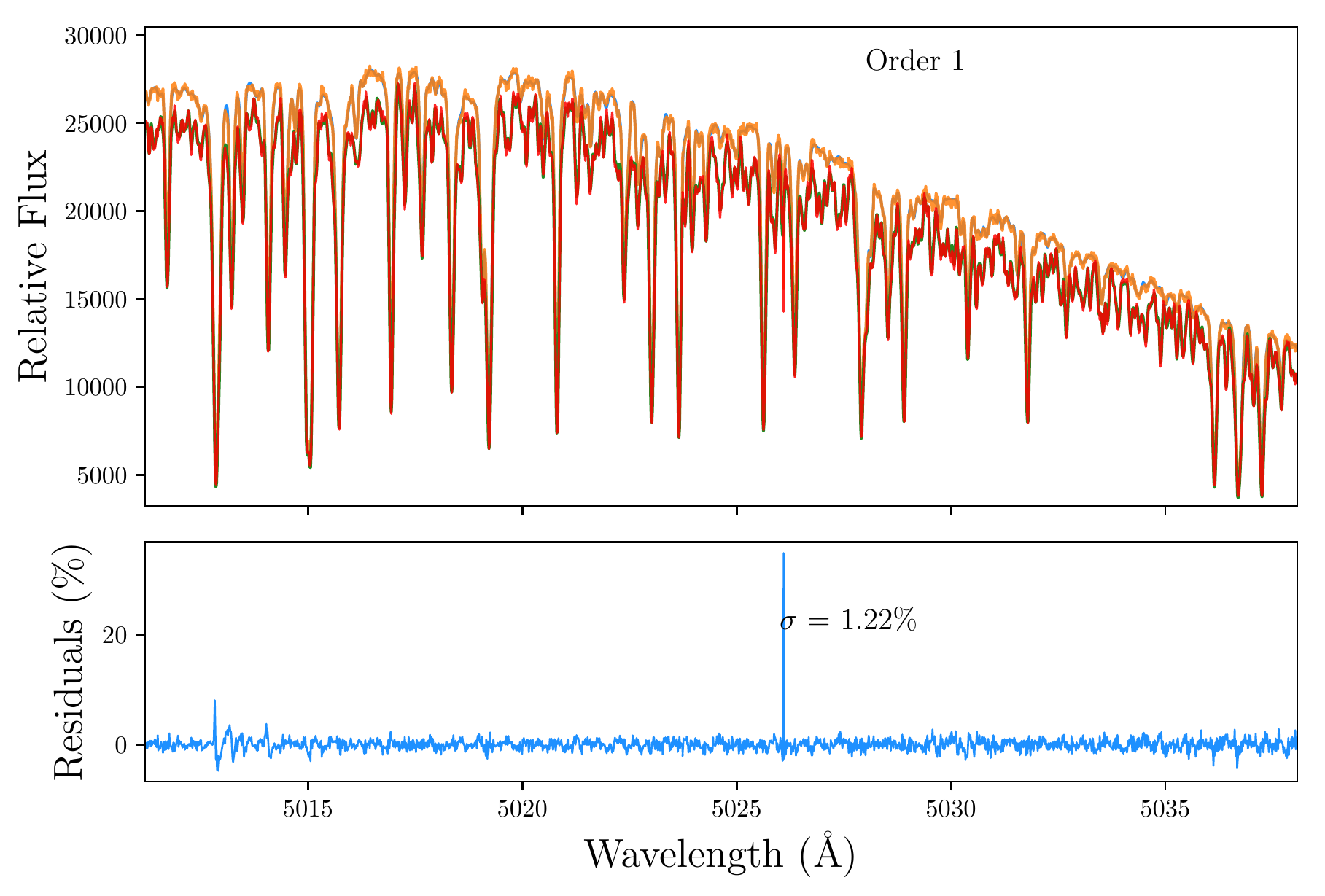}
\includegraphics[scale=0.34]{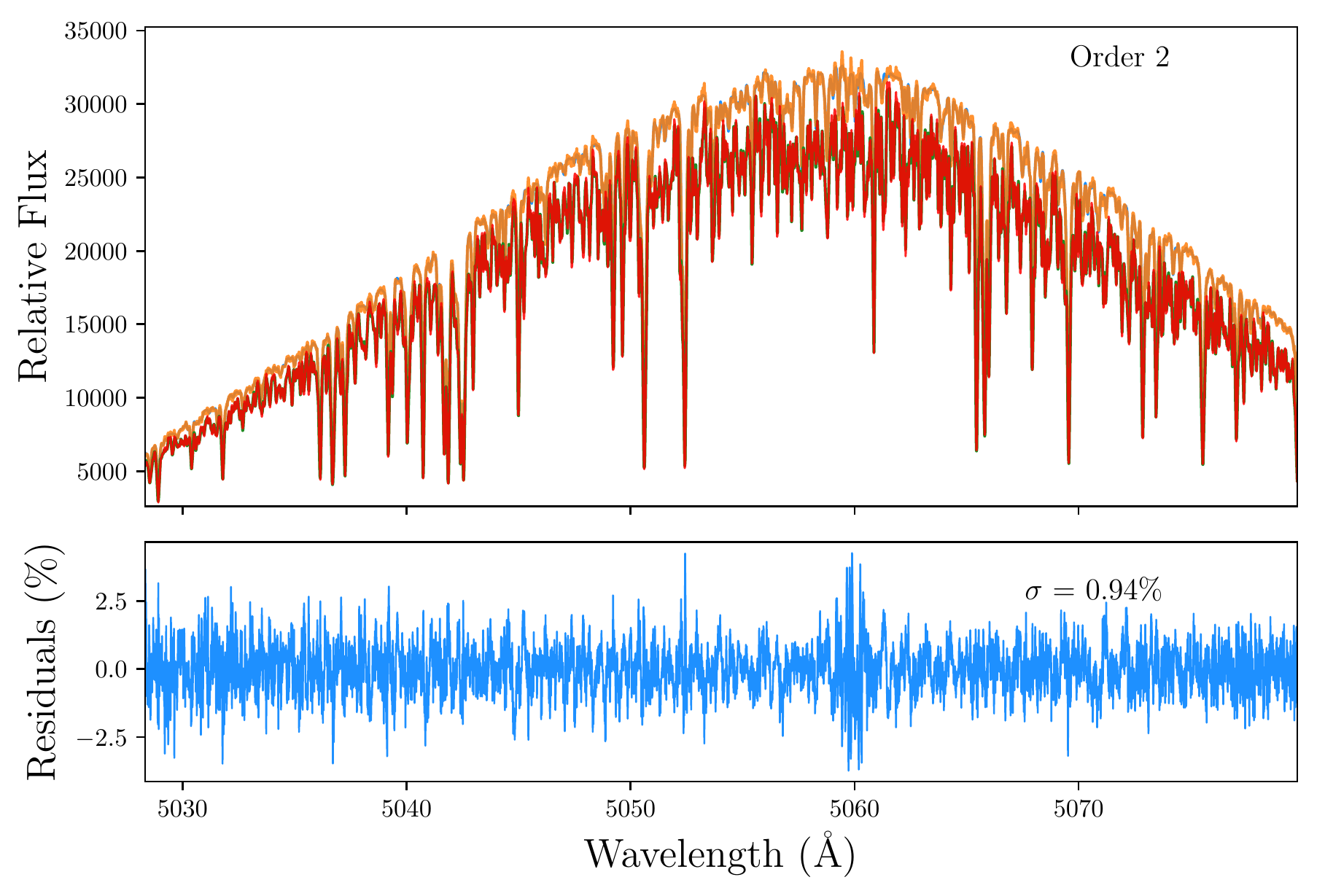}
\includegraphics[scale=0.34]{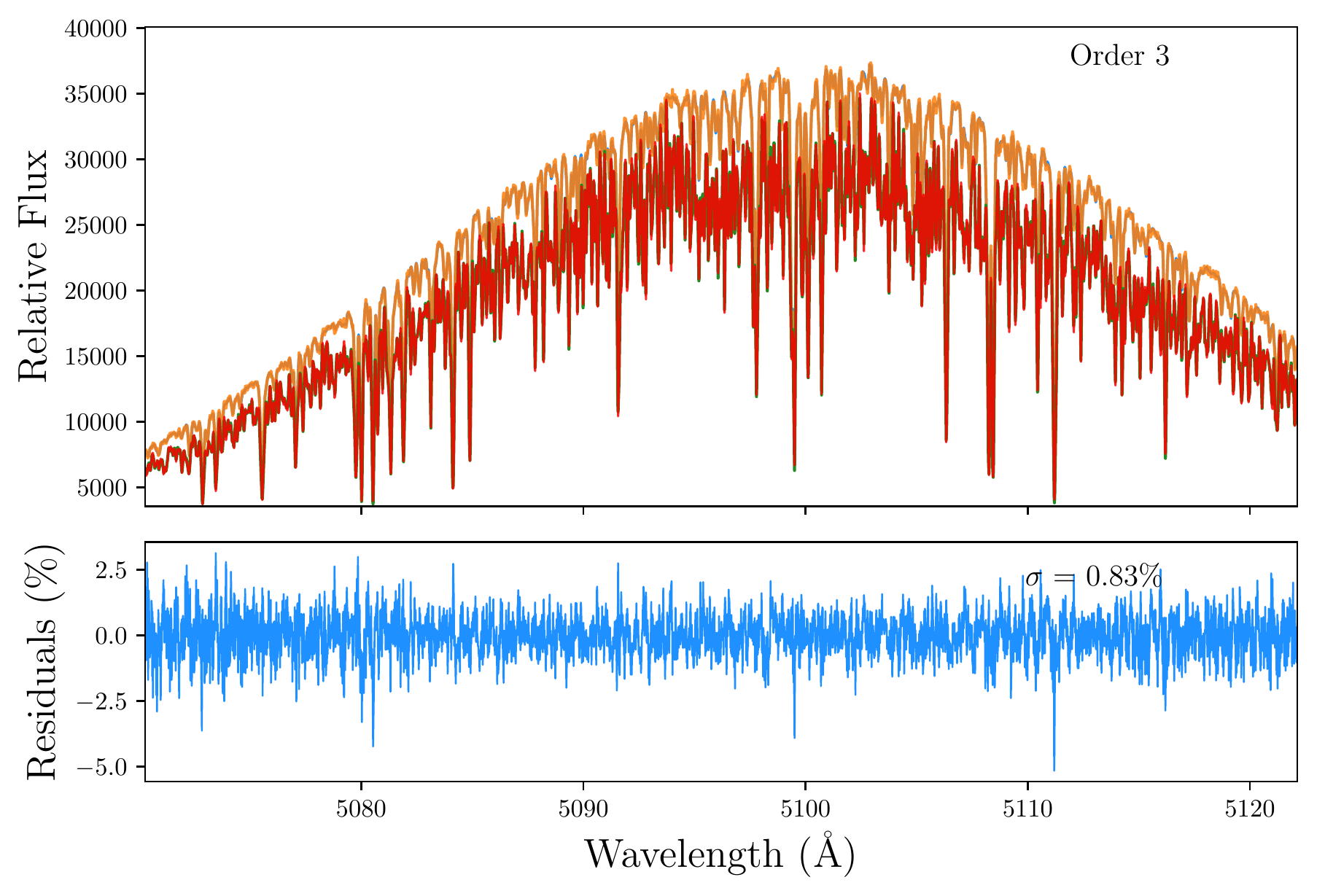}
\includegraphics[scale=0.34]{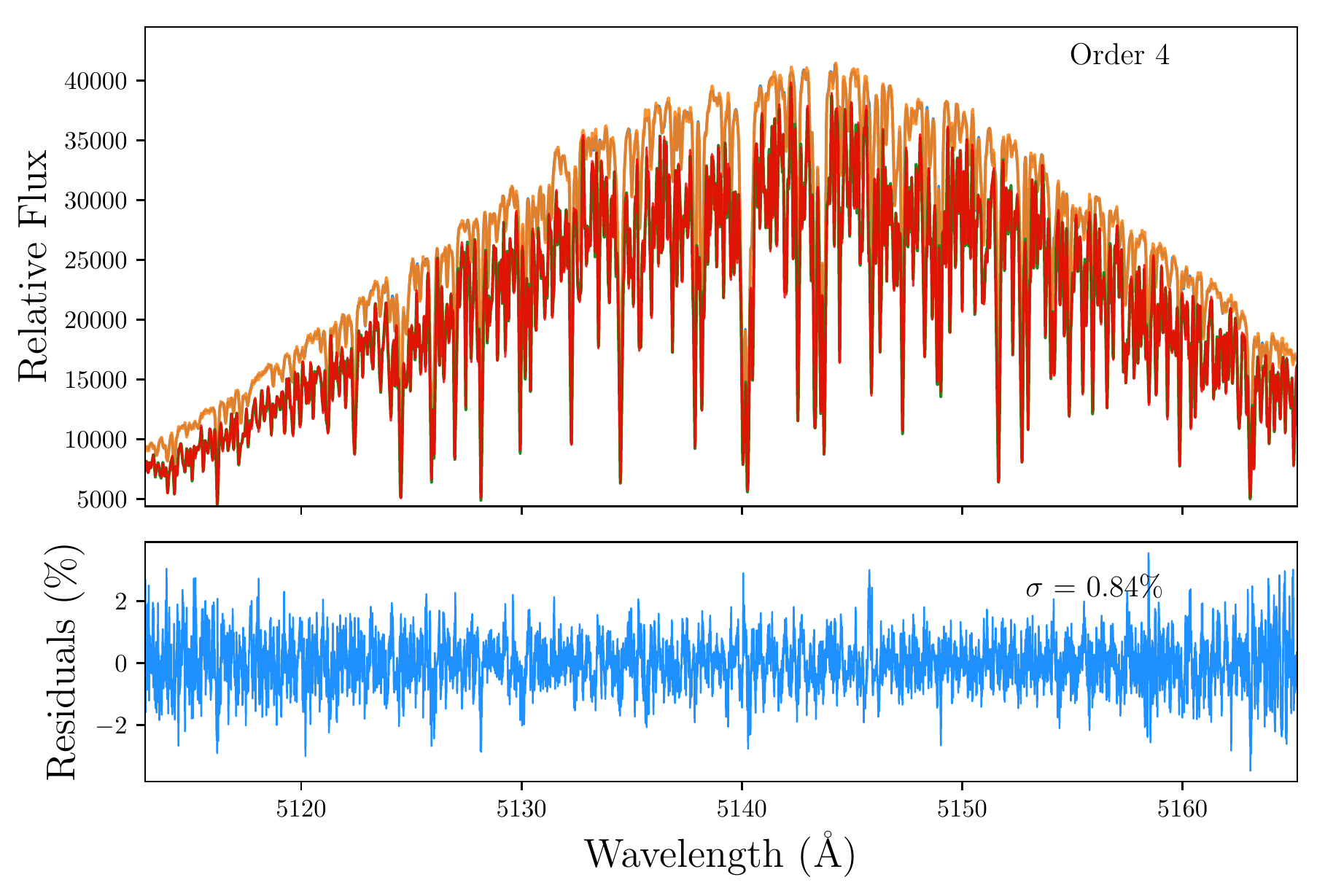}
\includegraphics[scale=0.34]{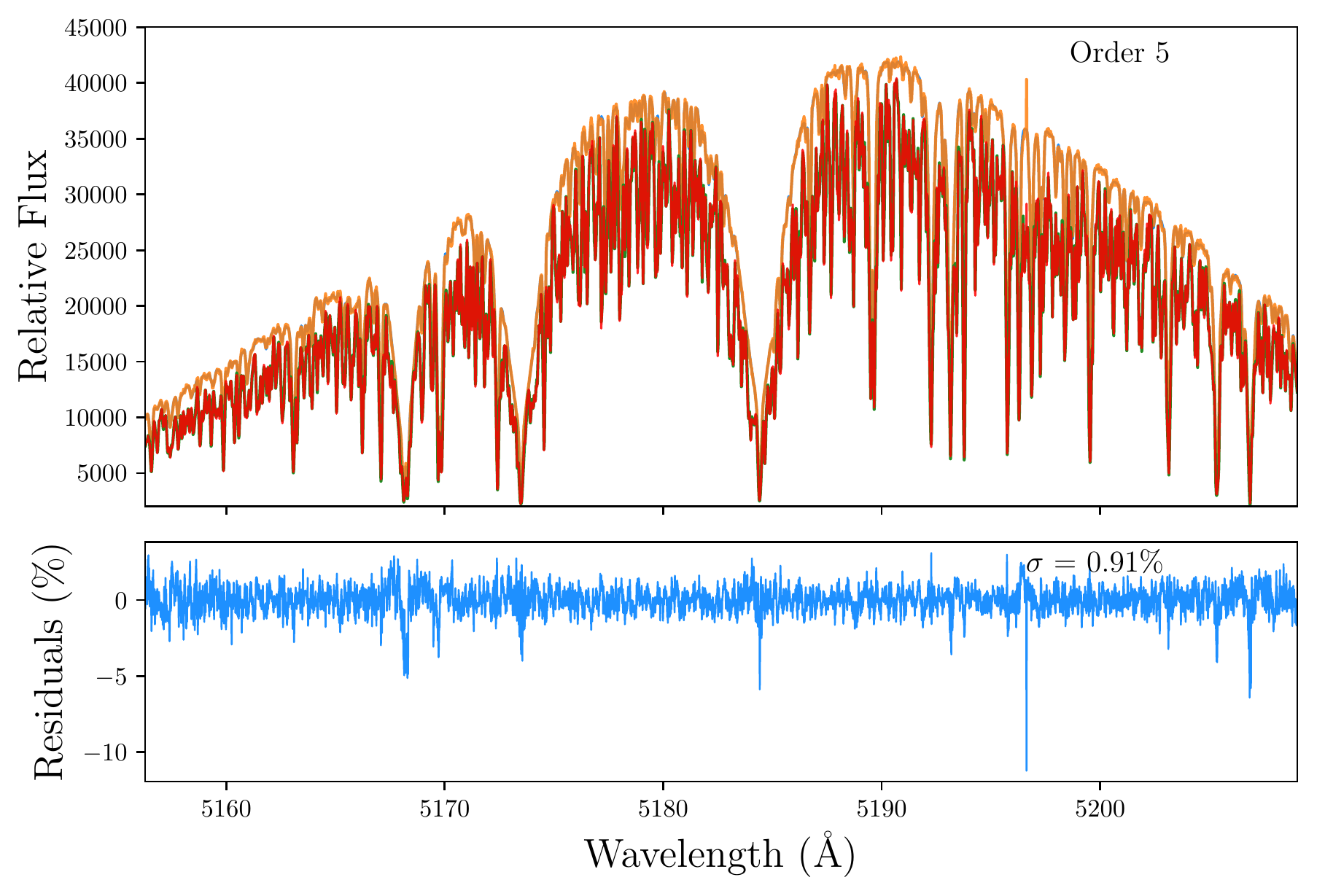}
\includegraphics[scale=0.34]{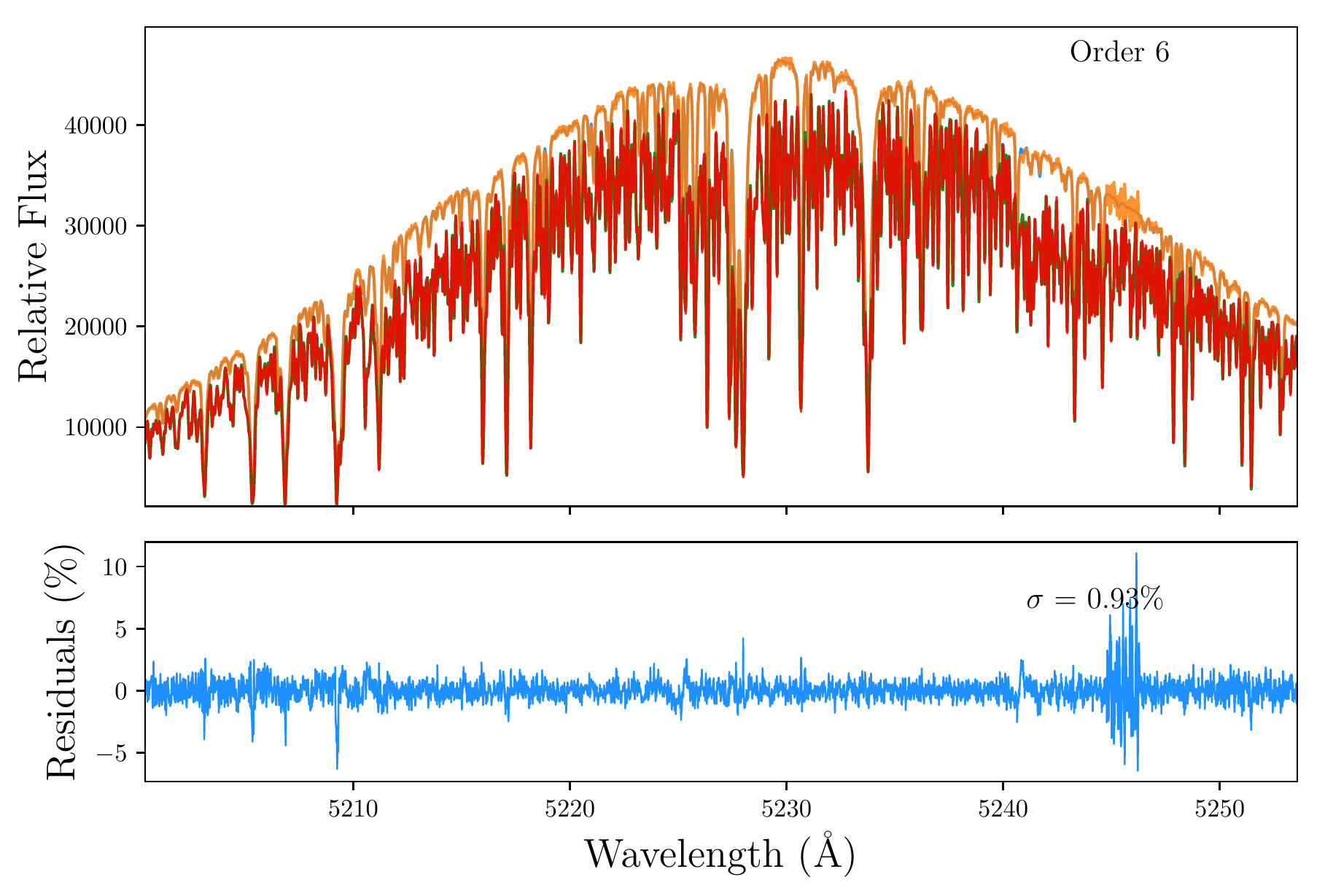}
\includegraphics[scale=0.34]{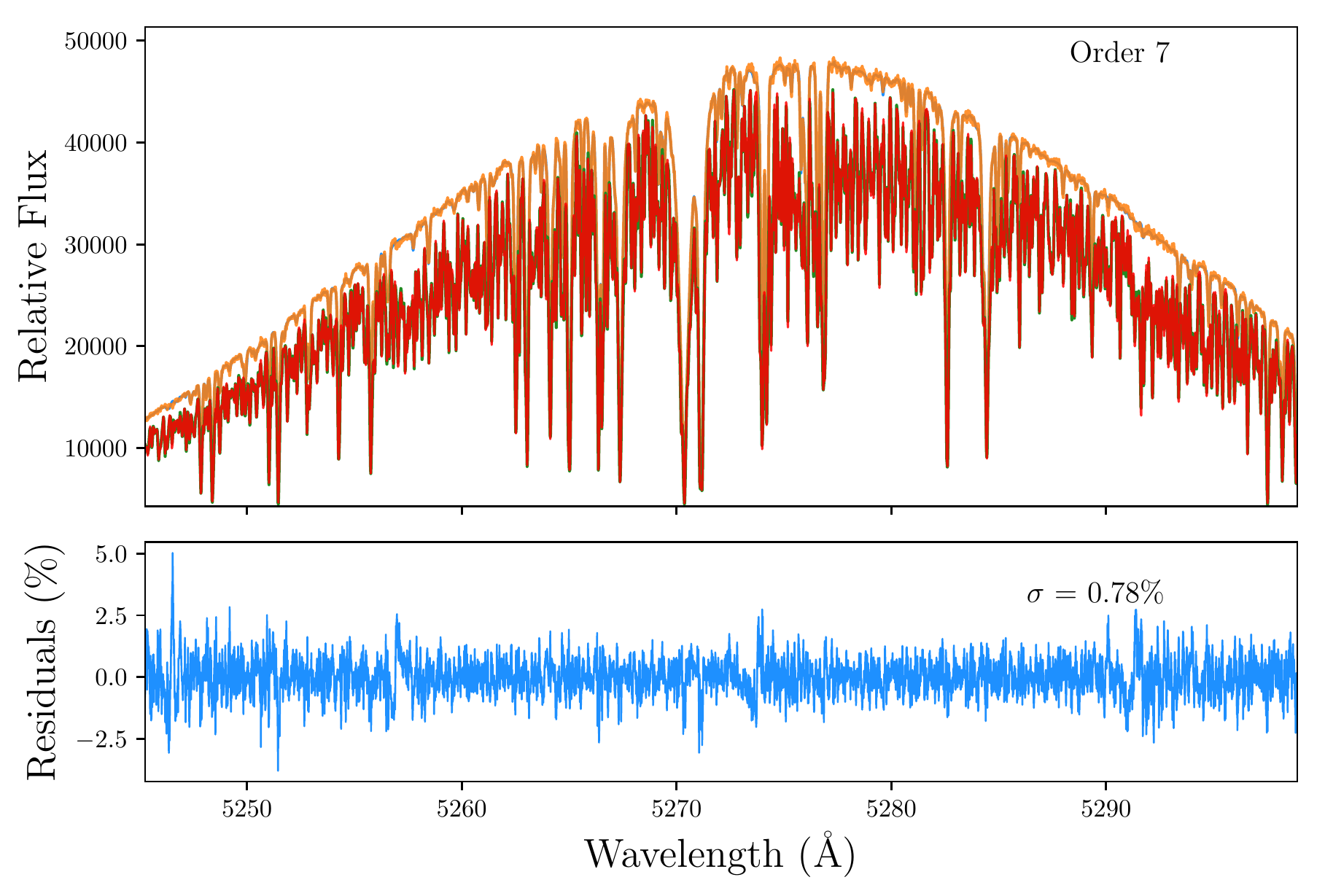}
\includegraphics[scale=0.34]{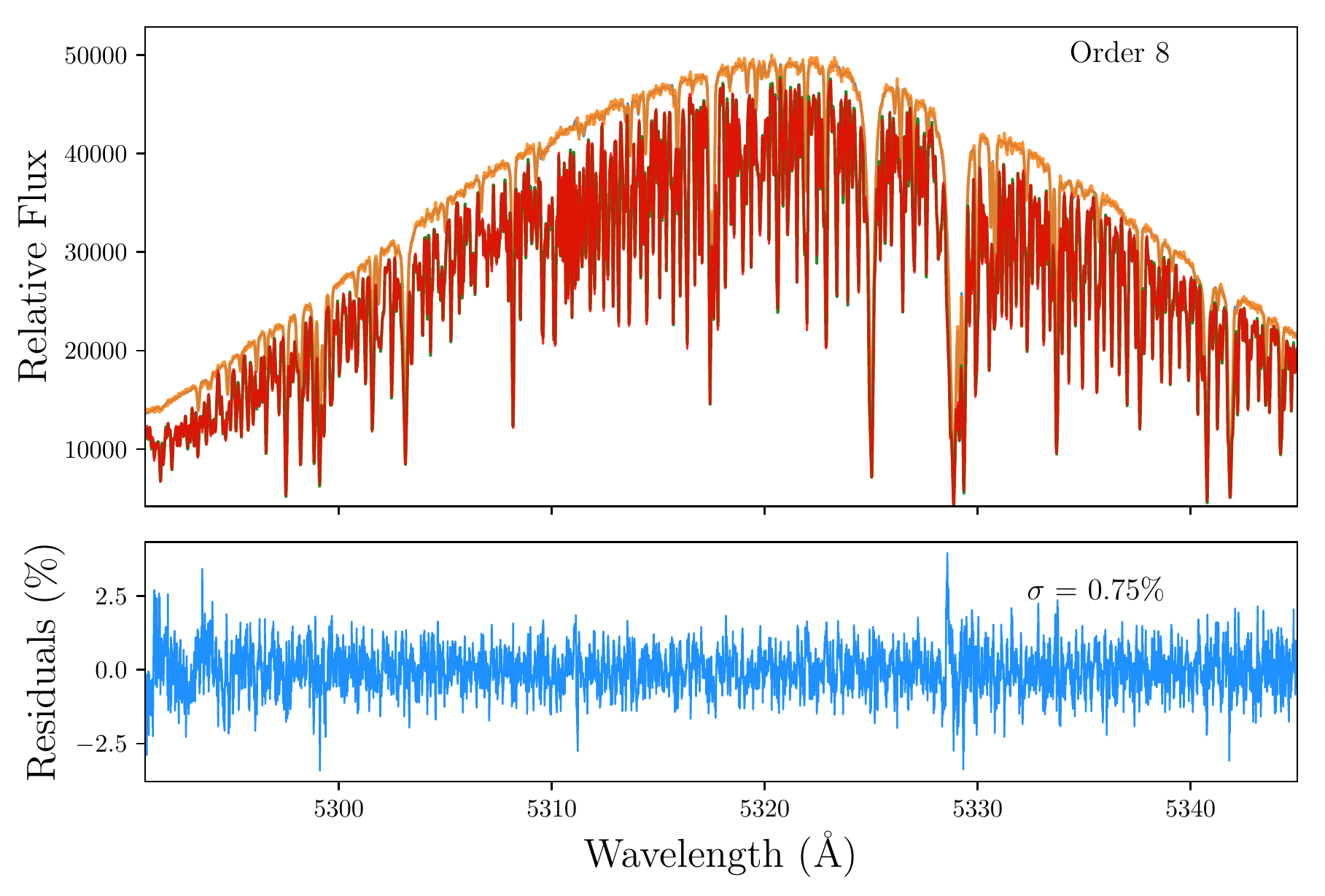}
\includegraphics[scale=0.34]{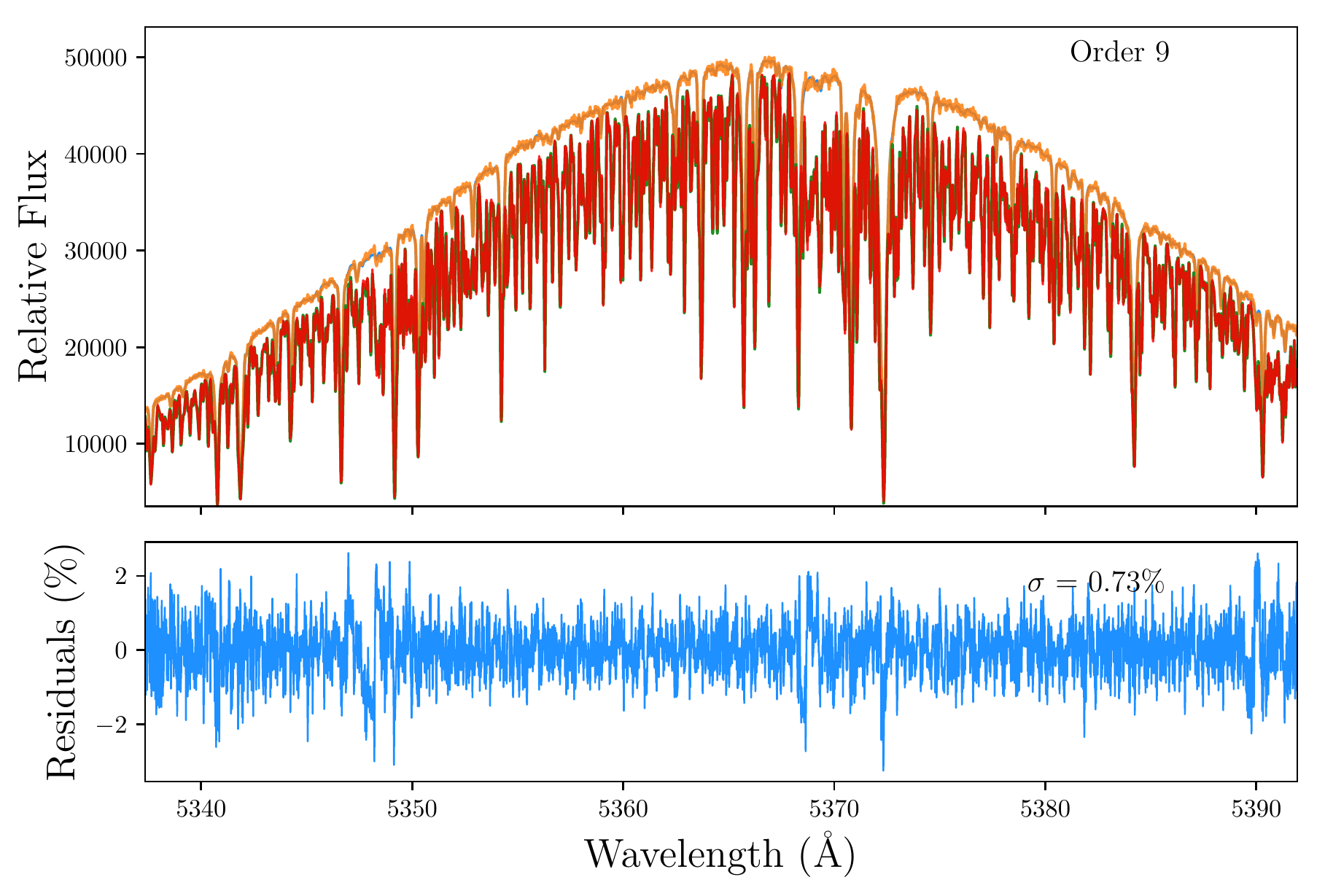}
\includegraphics[scale=0.34]{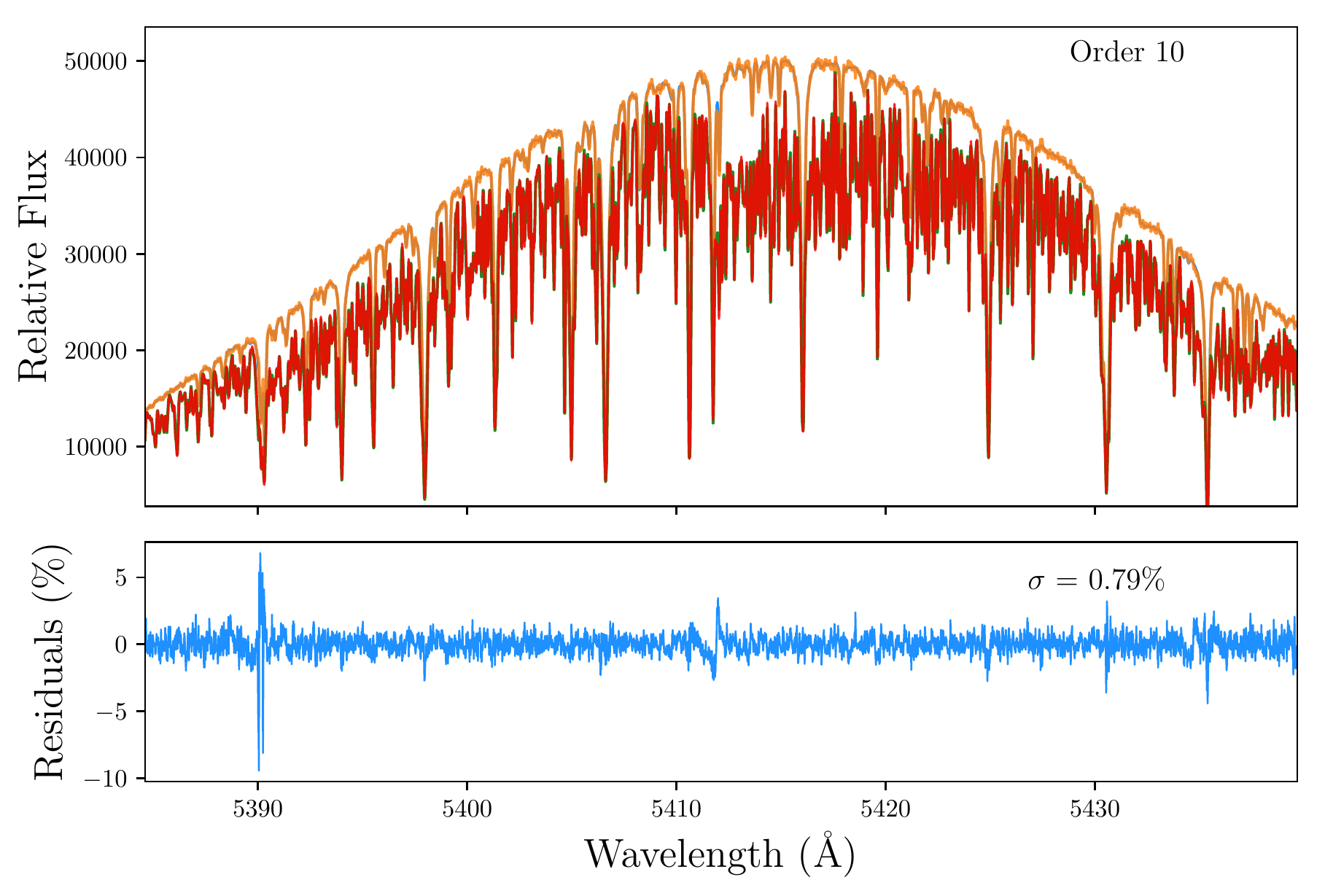}
\includegraphics[scale=0.34]{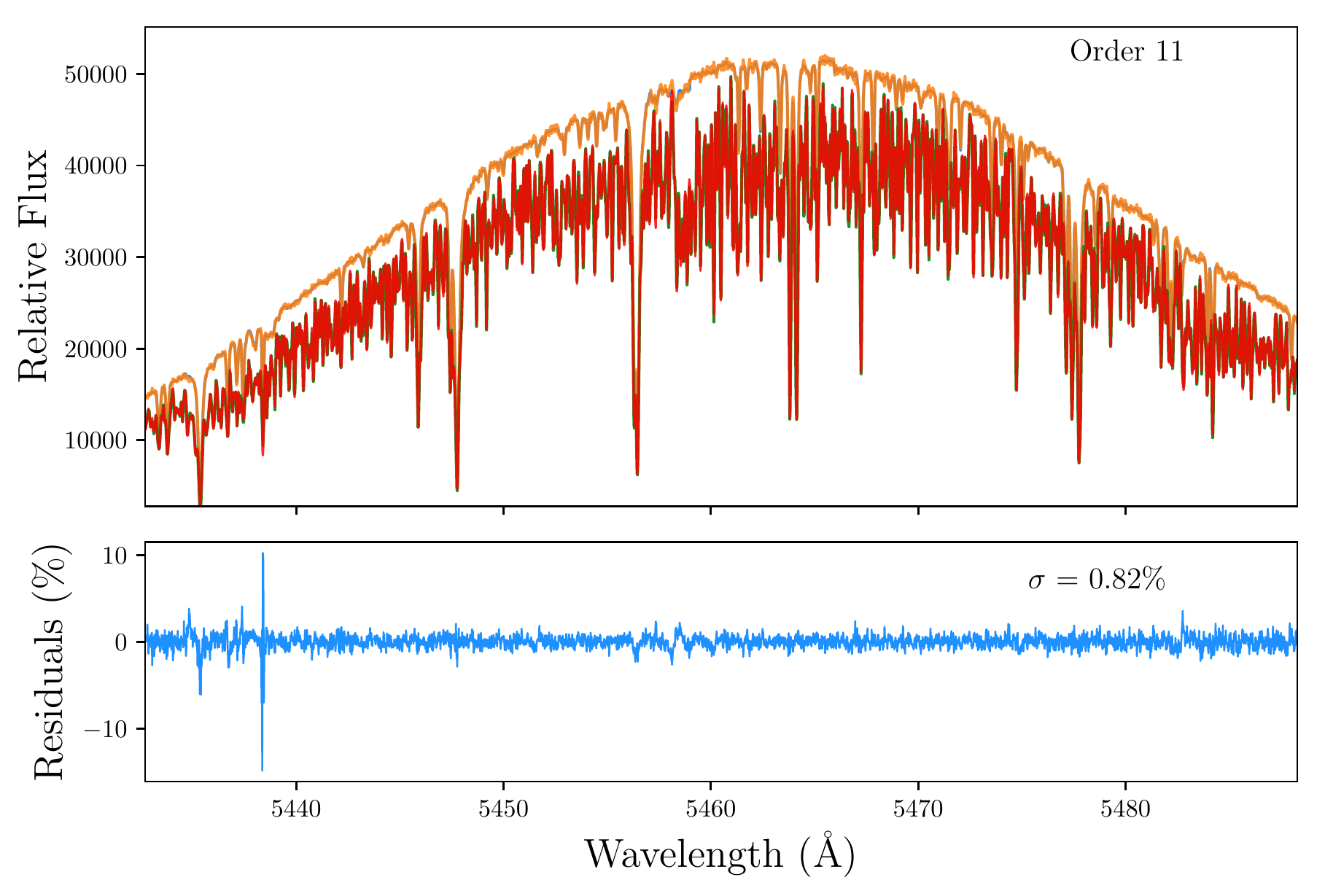}
\includegraphics[scale=0.34]{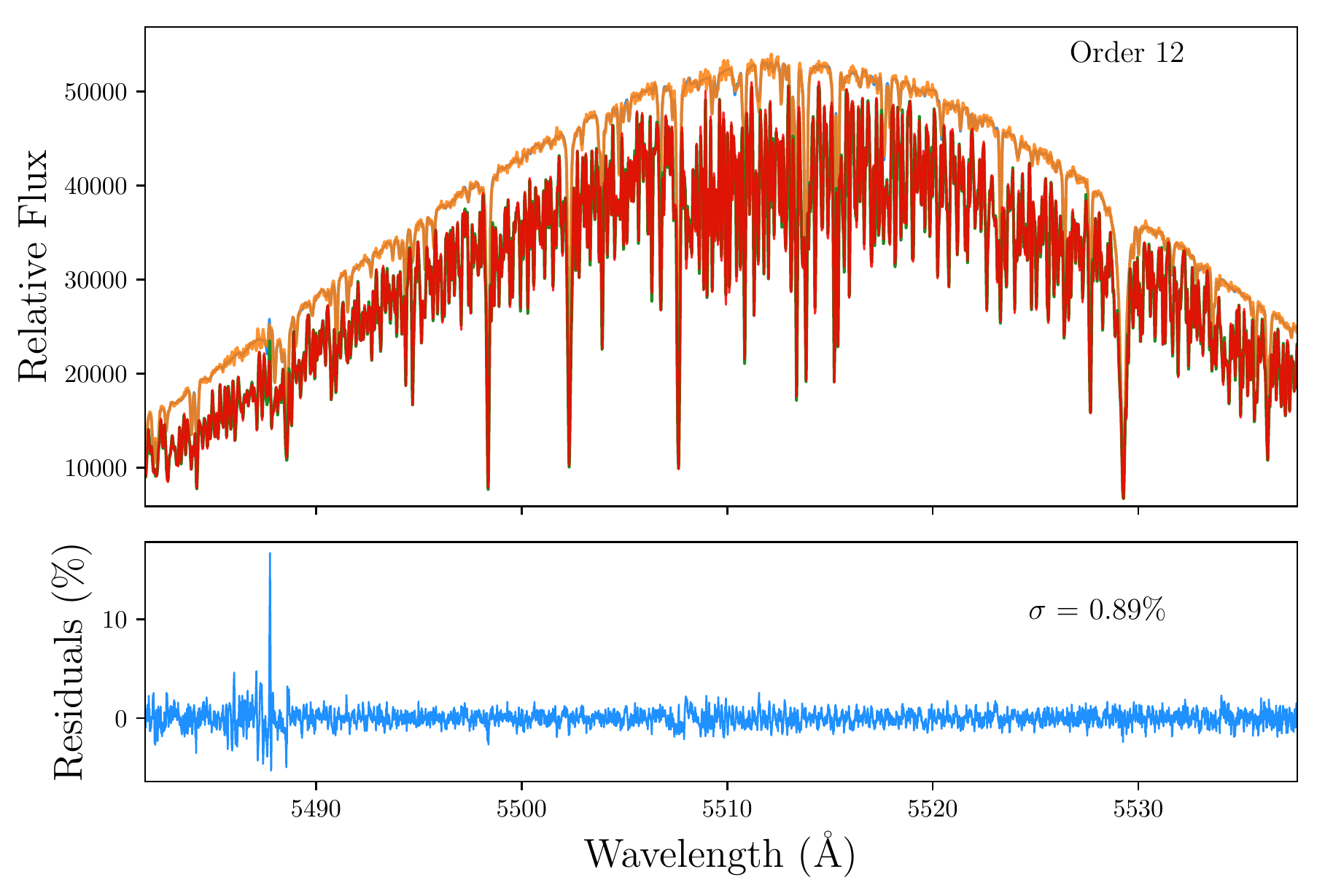}
\includegraphics[scale=0.34]{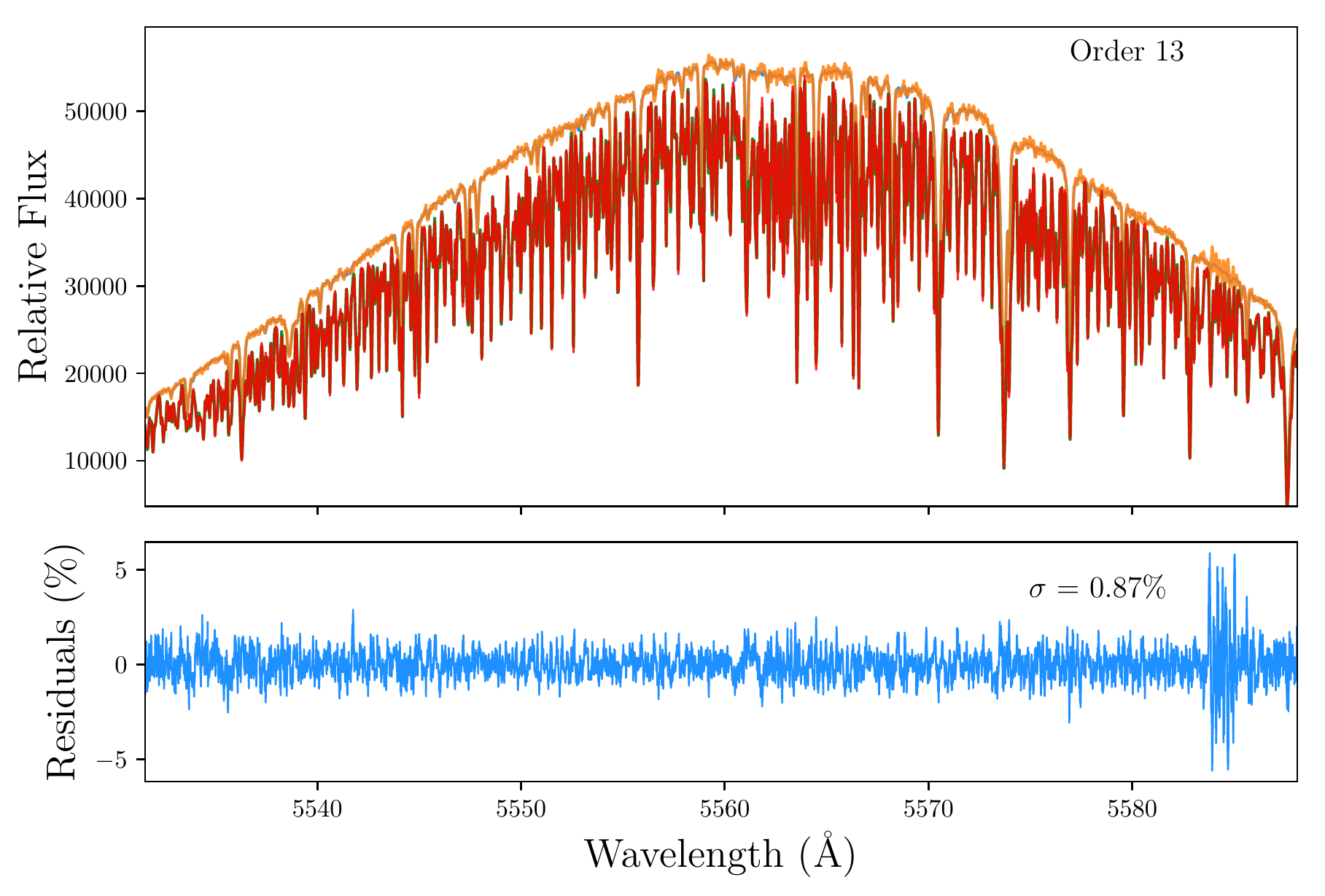}
\includegraphics[scale=0.34]{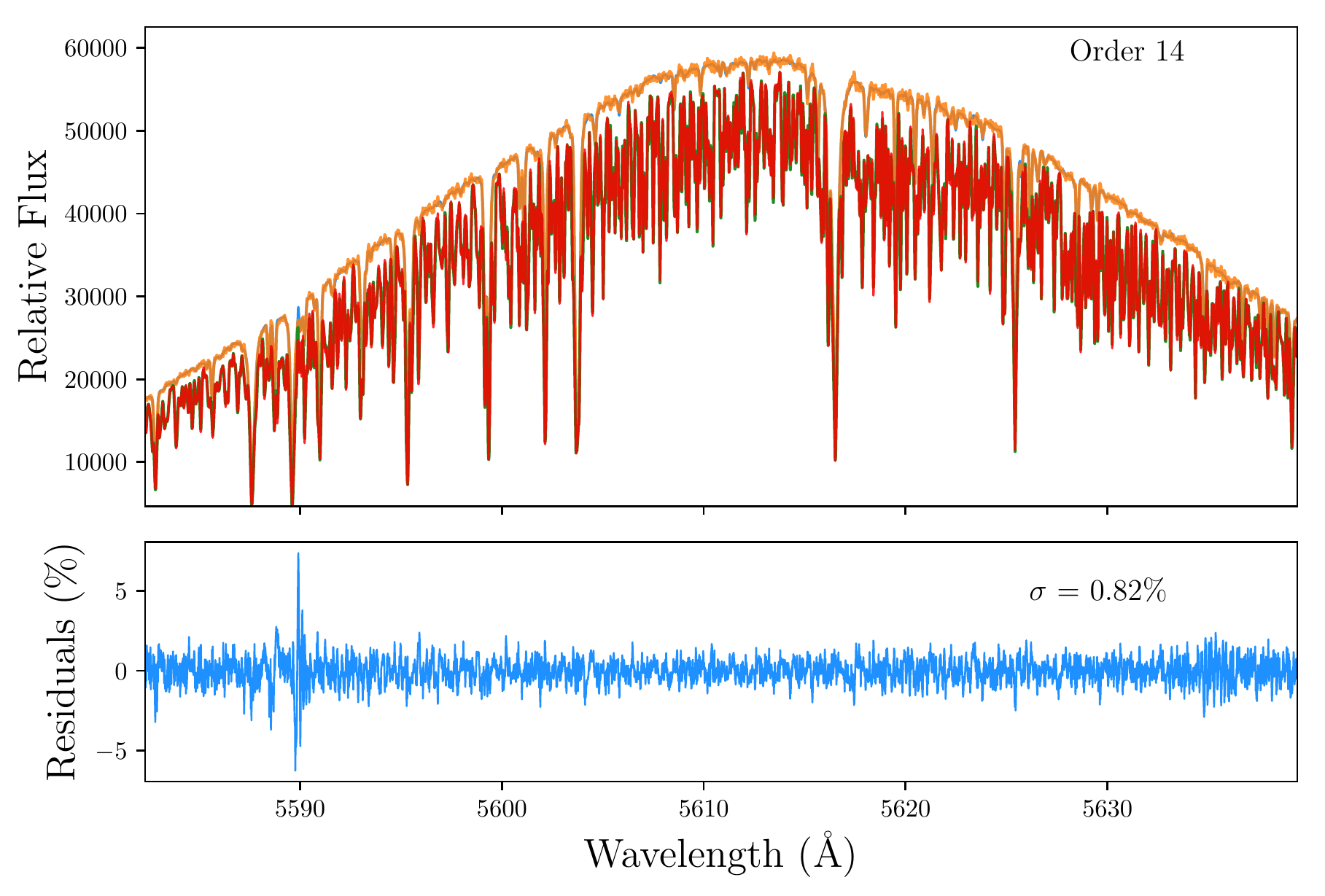}
\includegraphics[scale=0.34]{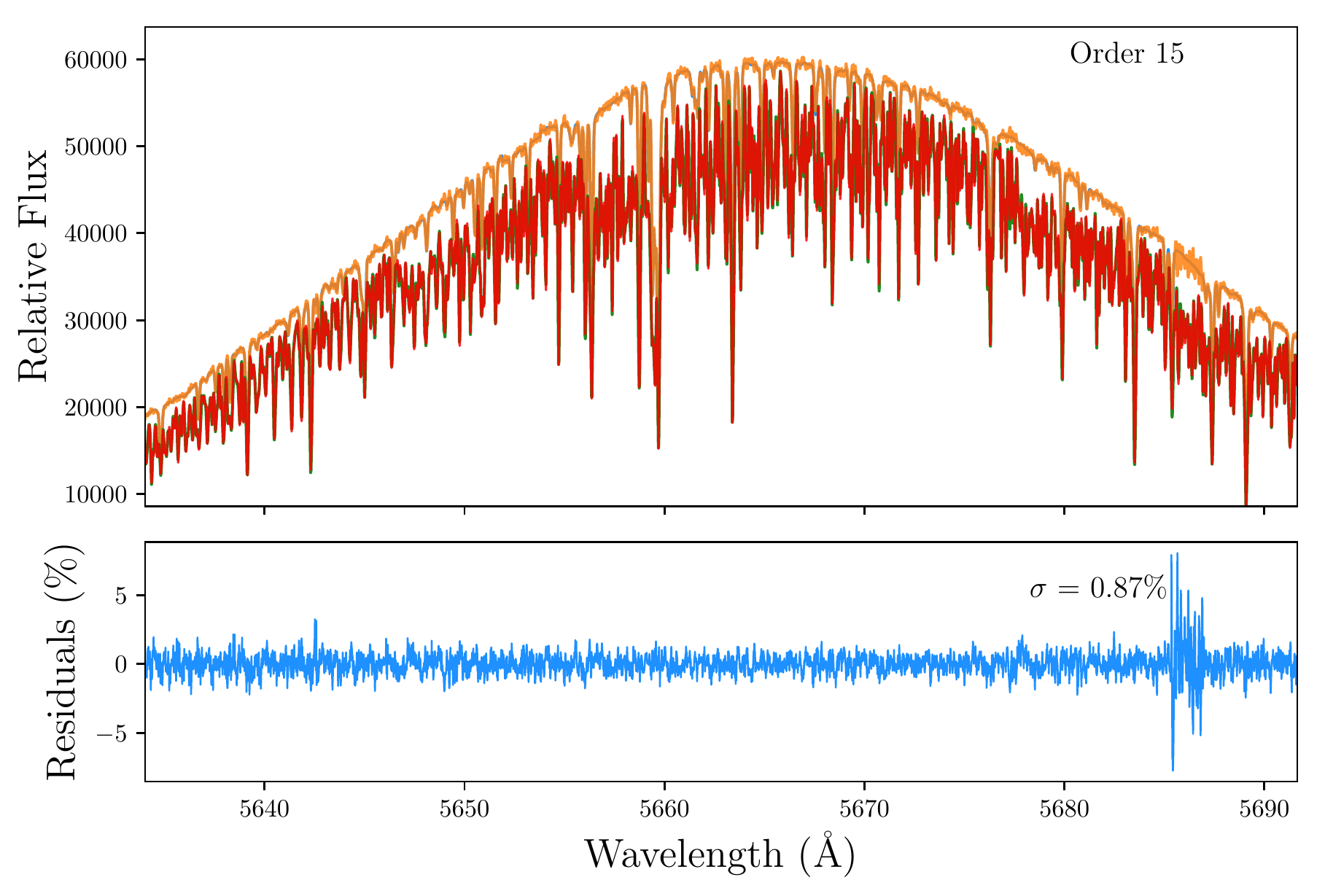}
\caption{Results of the iodine-free derivation for the star $\tau$ Ceti. We show the 26 orders from the iodine region of PFS. On each plot, we show the observation through iodine (red), the template observation without iodine (blue), the forward model of the observations through iodine (green) and the recovered iodine-free spectrum (orange). Bottom panels on each plot show the level of precision of the recovered spectrum compared with the template (in $\%$). }
\label{fig:plot_orders_pfs1}
\end{figure*}

In figure \ref{fig:zoomchunks} we show two 5\AA~ portions of the spectra acquired with PFS where the iodine-free spectrum (orange line) is shown on top of the template spectrum of the star $\tau$ Ceti and the original observation through the iodine cell (red). In figures \ref{fig:plot_orders_pfs1}, \ref{fig:plot_orders_hires} and \ref{fig:plot_orders_ucles1} we show the iodine-free derived spectra for each order, for observations of the same star acquired with PFS at Magellan, HIRES at Keck and UCLES at the AAT, respectively. This star has a magnitude of V=3.5~mag, providing excellent, high signal to noise spectra for testing purposes. At this time, we have collected 205 observations of $\tau$ Ceti, where the median signal to noise for observations is $\sim$192 per pixel, with a typical exposure time of $\sim$10s. The total computation time needed for one observation consisting of 25 orders divided into 818 chunks is less than one second per chunk, totaling less than 2 minutes to derive and save an iodine-free spectrum in the same format as the observation: an \texttt{IDL} structure that can be easily read in both \texttt{IDL} and \texttt{Python} for further analysis, or indeed a range of other programming languages. Each one of these data structures contains the wavelength scale, template, observation, model, iodine spectrum and iodine-free derived spectrum.

From figures \ref{fig:plot_orders_pfs1} to \ref{fig:plot_orders_ucles1} (bottom panels) we also show the residuals after subtracting our derived iodine-free spectrum from the original iodine-free template observation of $\tau$ Ceti, as a function of chunk, $(S_x-S(x)) / S(x)$. The overall scatter is found to be at the $\sim$1.3-1.5\% level, obtained by computing the RMS for the resultant residual spectrum for a given order. For each order, $\sigma$ represents the scatter of the residual spectrum to highlight the cases where the deconvolution fails due to cosmetics and the iodine free spectrum carries these high-frequency-like patterns for some chunks (e.g., orders 39, 46, 48). When this happens, the whole chunk is not affected, i.e. these patterns are not visible across the whole chunk, but on localized regions that span 20-50 pixels across a few columns in the CCD. We note here that during the raw reduction process of the images we perform cosmic ray removal. However, cosmics and outliers do remain in the spectra. Since each observation produces a velocity for each chunk, we are left with $\sim$700 velocities for each observation. We recall that in our Doppler analysis procedure, weights are defined by the estimated variance of each chunk. The RV for a given observation is the weighted mean of the chunks. The chunks that produce velocities at the tail of the distribution are automatically discarded from the velocity analysis. The final error is computed by taking into account these weights as discussed in \citet{Butler1996}.

Throughout the process of developing this algorithm, we encountered a number of artifacts in each of the spectra, artifacts that must be precisely removed. In particular, if we look at the PFS instrument, located, for example, at $\sim$5245\,\AA~in order 39, there is a defect that needs to be dealt with in a future revision of code. Checking the raw images and comparing them with a smoothed version of a flat field image, we identified bad regions that span several pixels wide and across a few columns on the detector for the orders where we see large residuals after our procedure. These defects are not smoothed away after performing flat fielding, and then when the deconvolution routine tries to reproduce the observed spectrum, it still carries these bad regions into the template by adding systematics that are not stellar features. These problematic regions of the spectra were never fixed since the radial velocity code performs an iterative rejection of deviant chunks based on high $\chi^{2}$ values for each chunk. In this way, a chunk where the fit of the model is not good, will be removed and will not be considered in the final computation of the velocities, meaning they will not affect the precision too negatively.
Despite the latter, trying to overcome and fix these erroneous regions will be important to deal with in the future.
In total we found 15 parts of the HIRES detector that introduced noisy features, along with $\sim$30 chunks for PFS and $\sim$10 for UCLES.

\renewcommand{\thefigure}{\arabic{figure} (Cont.)}
\addtocounter{figure}{-1}

\begin{figure*} 
\includegraphics[scale=0.34]{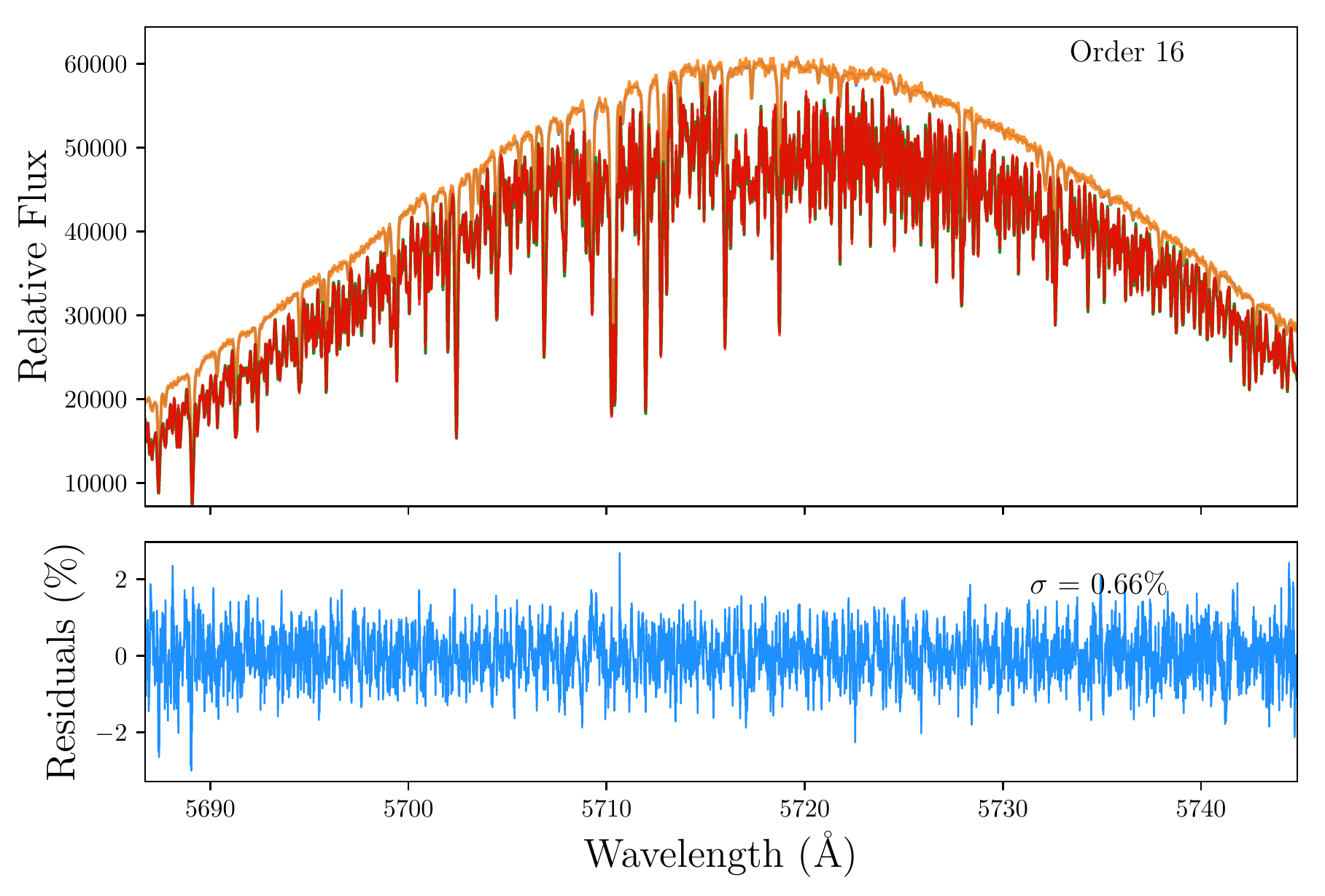}
\includegraphics[scale=0.34]{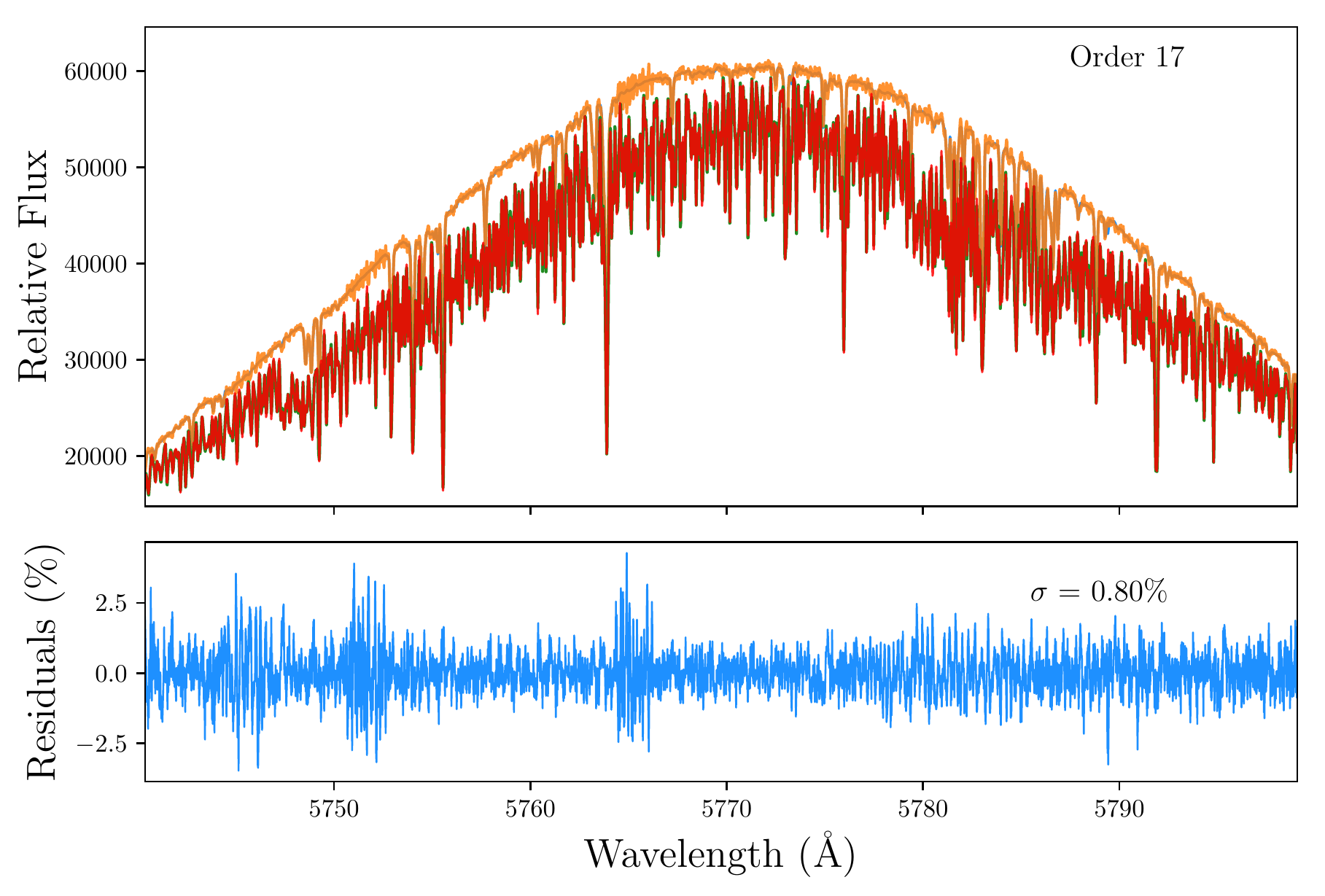}
\includegraphics[scale=0.34]{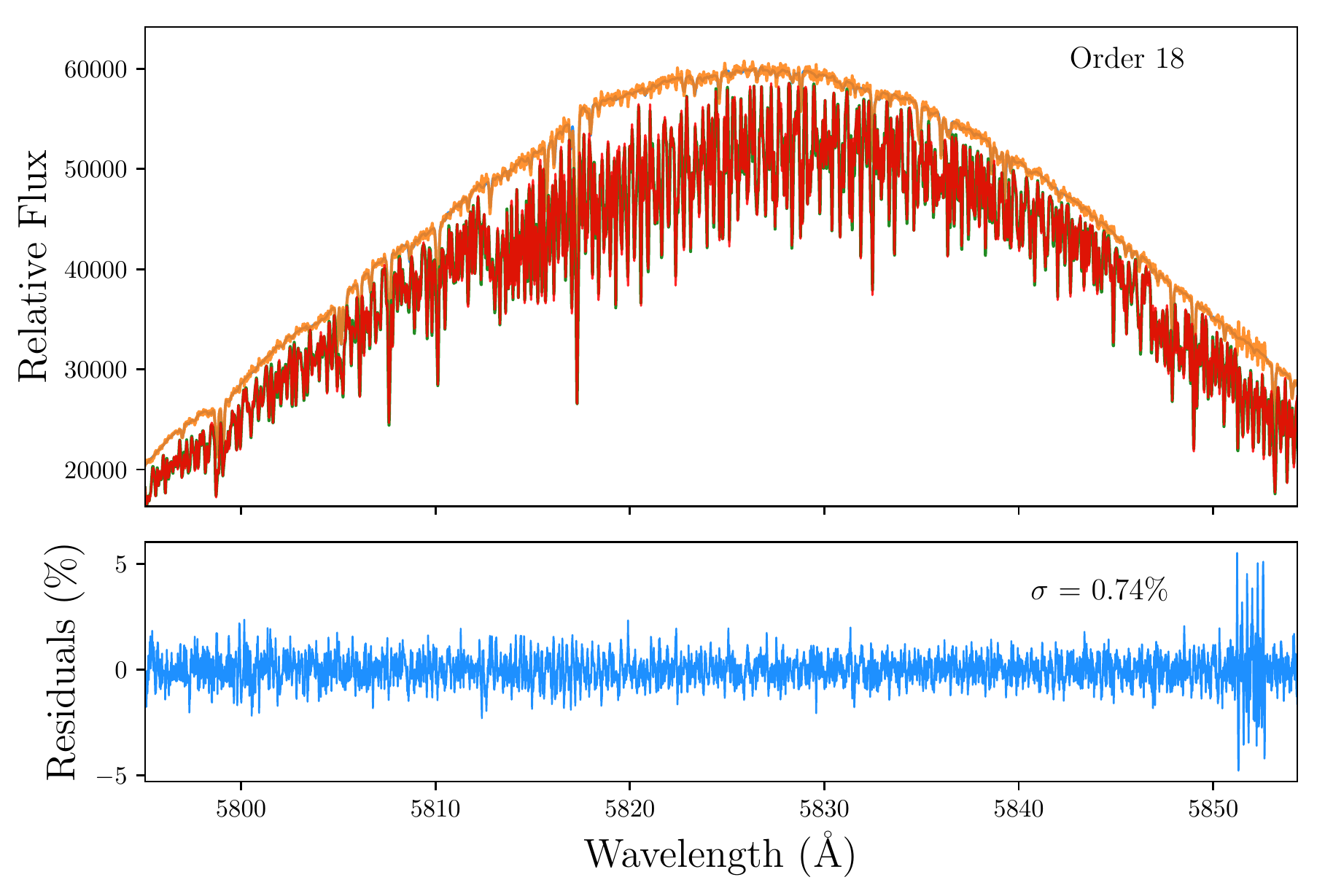}
\includegraphics[scale=0.34]{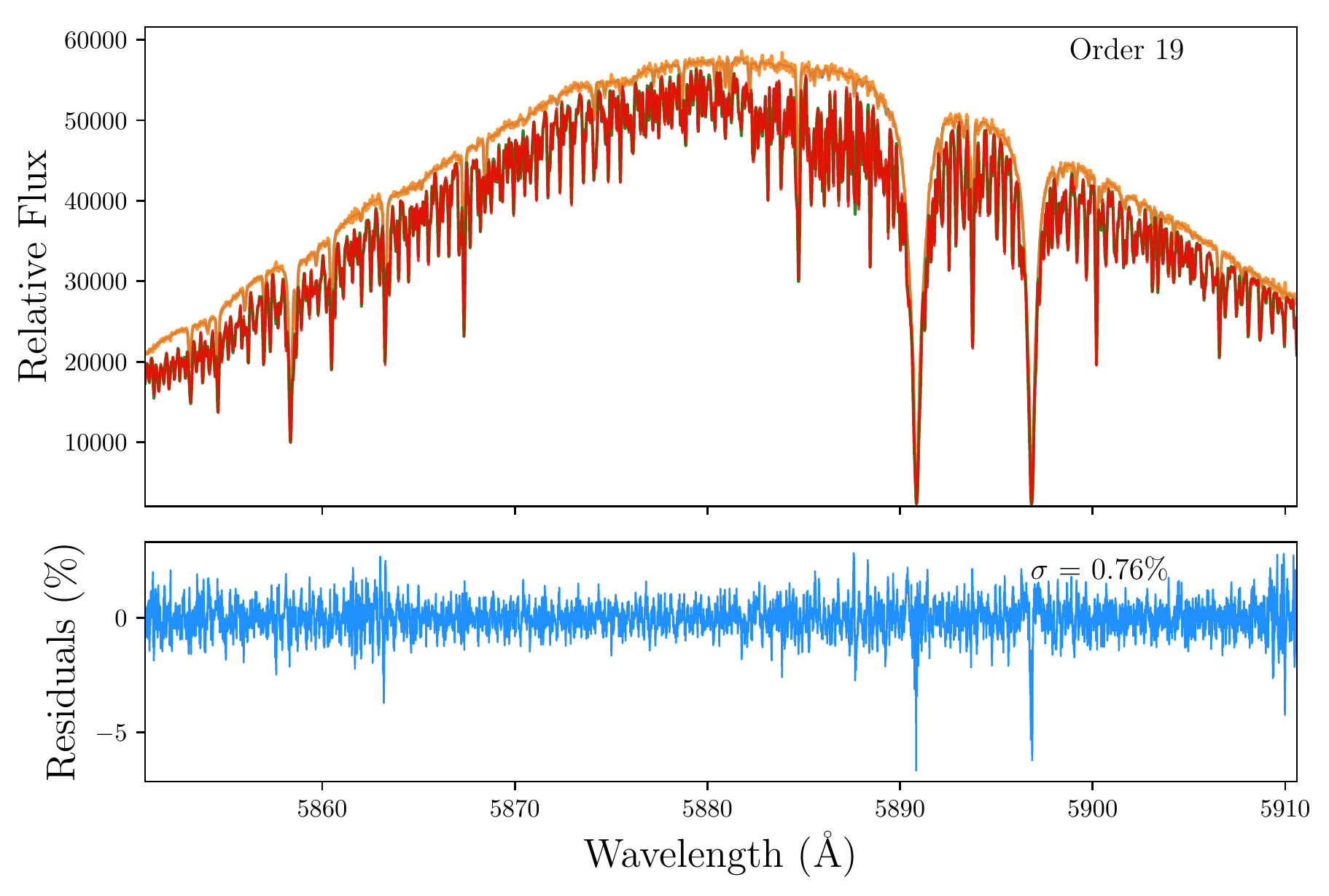}
\includegraphics[scale=0.34]{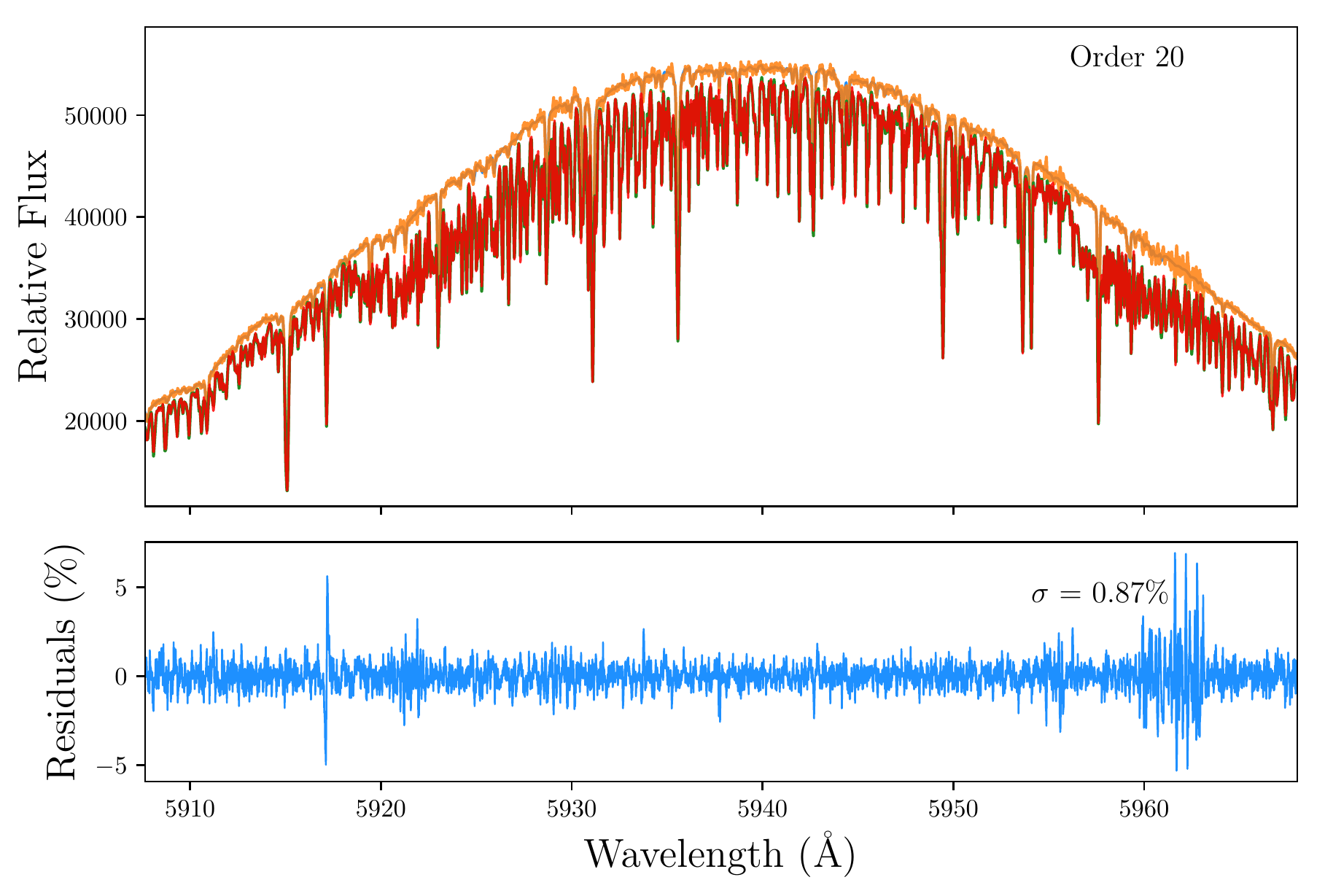}
\includegraphics[scale=0.34]{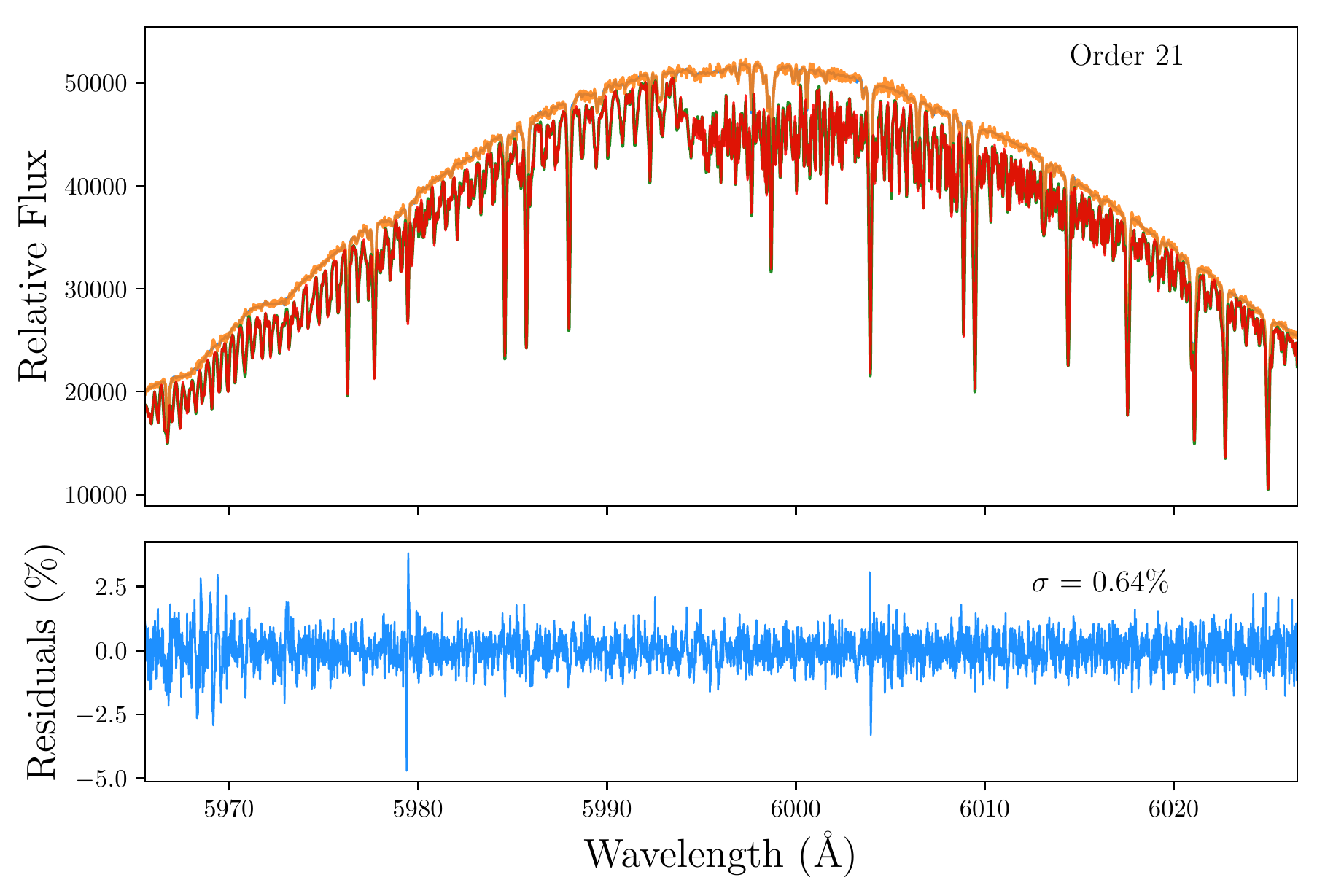}
\includegraphics[scale=0.34]{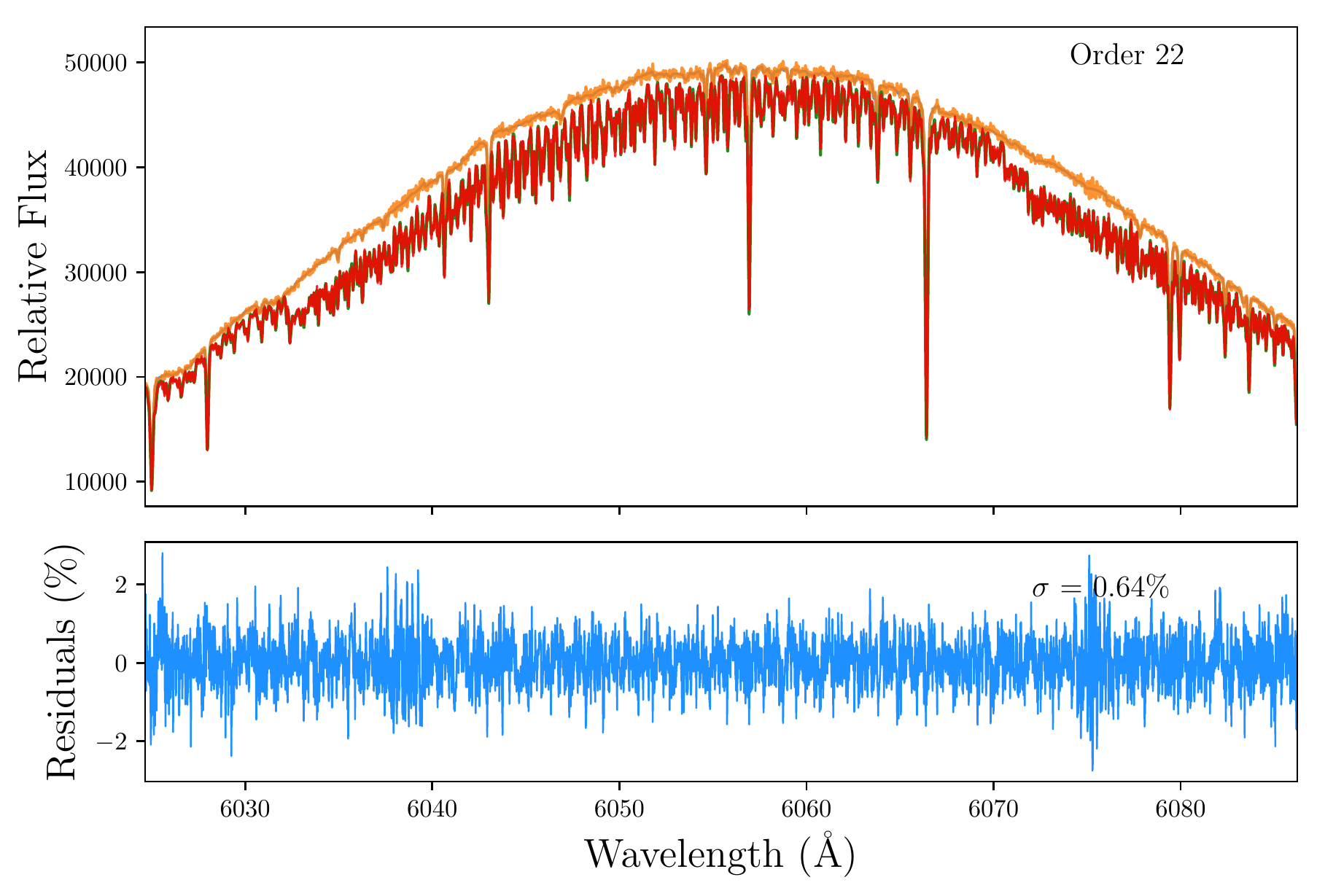}
\includegraphics[scale=0.34]{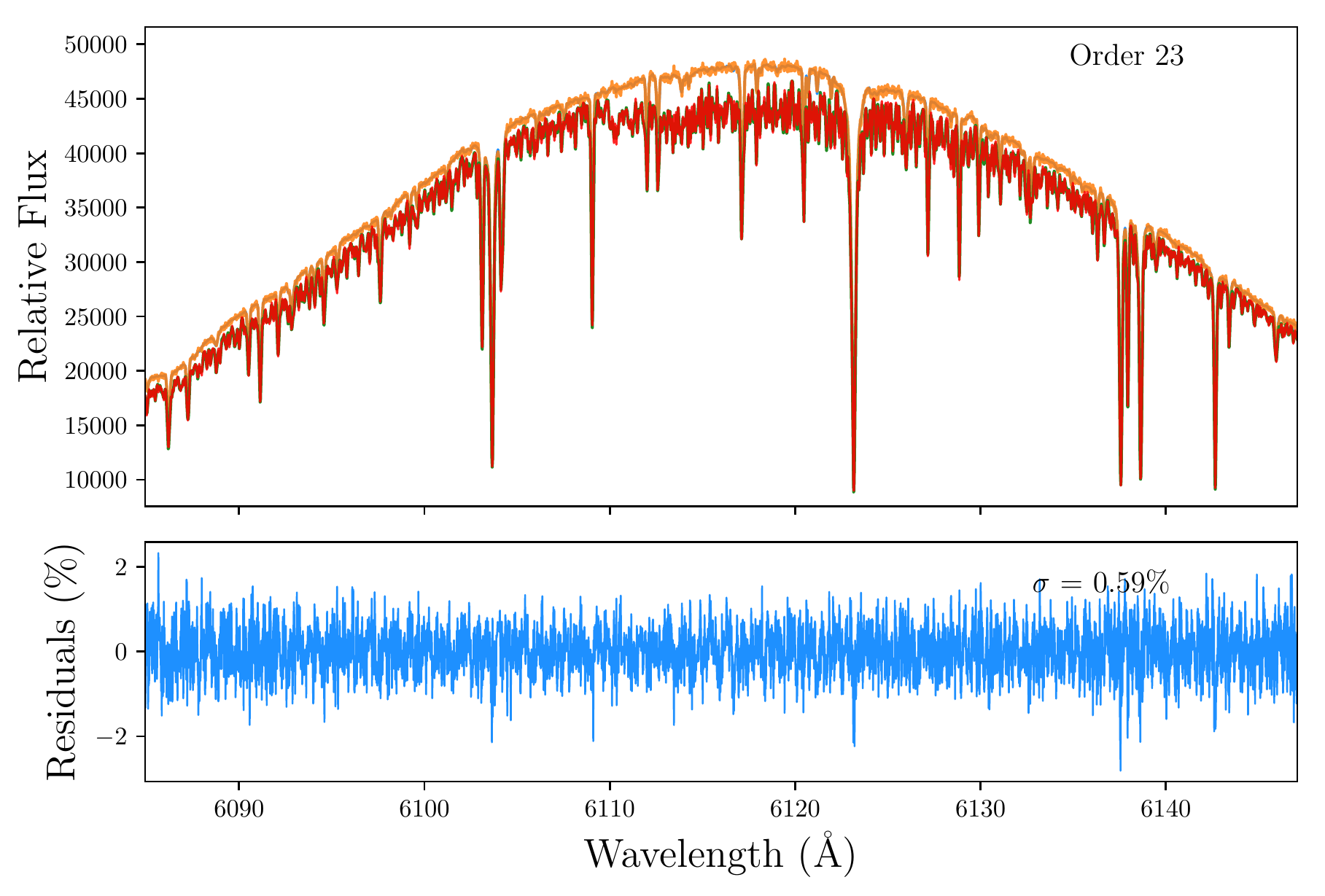}
\includegraphics[scale=0.34]{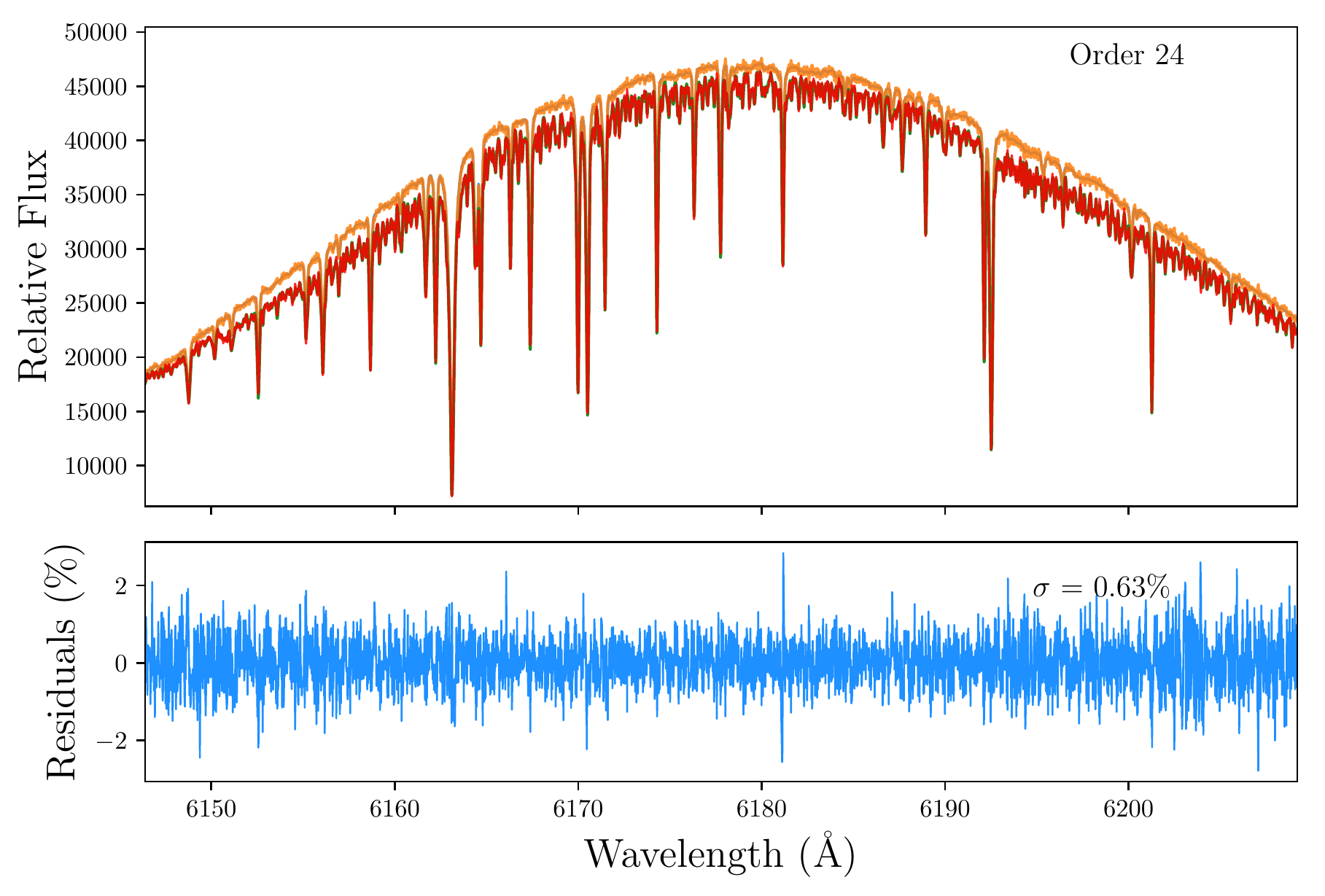}
\includegraphics[scale=0.34]{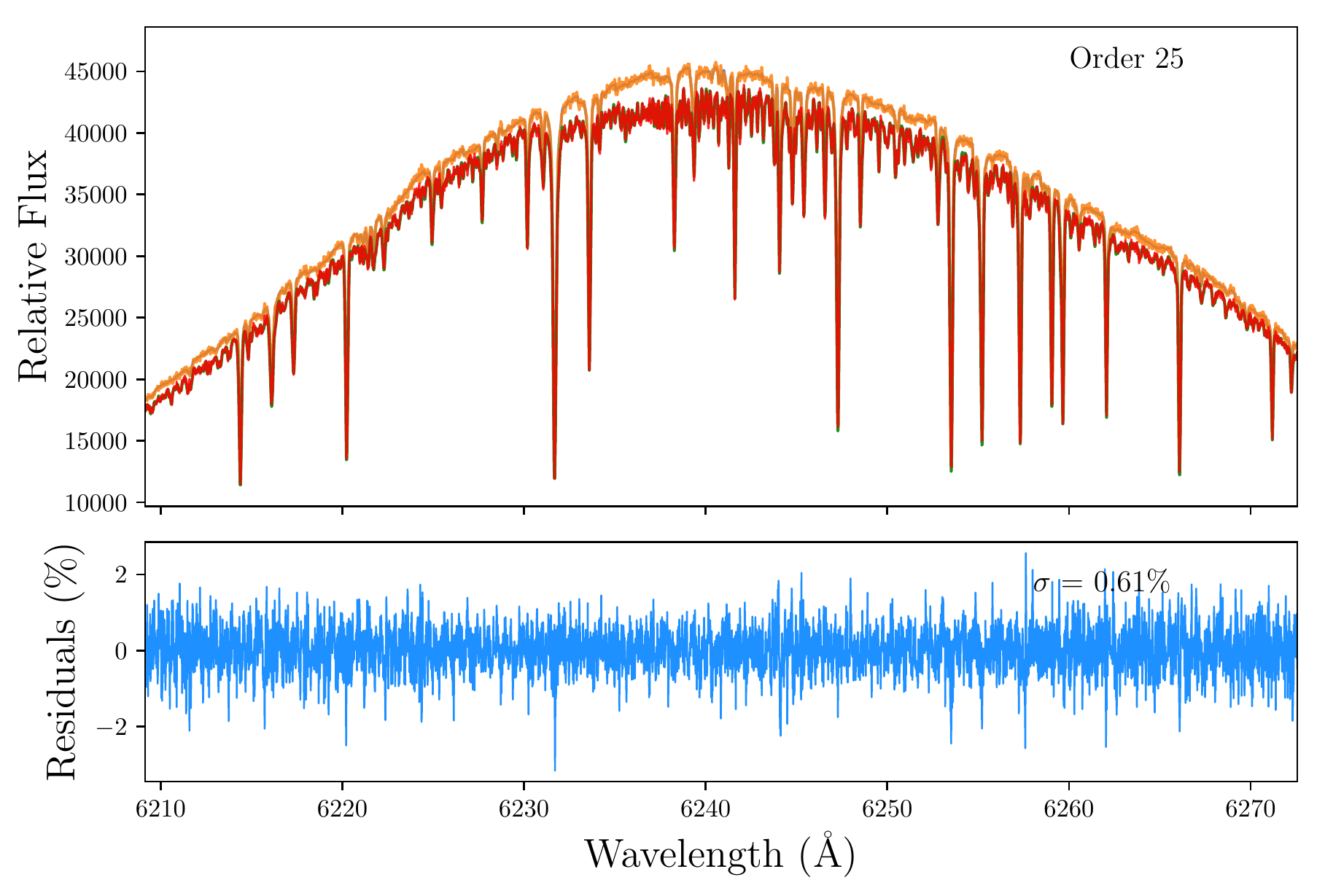}
\includegraphics[scale=0.34]{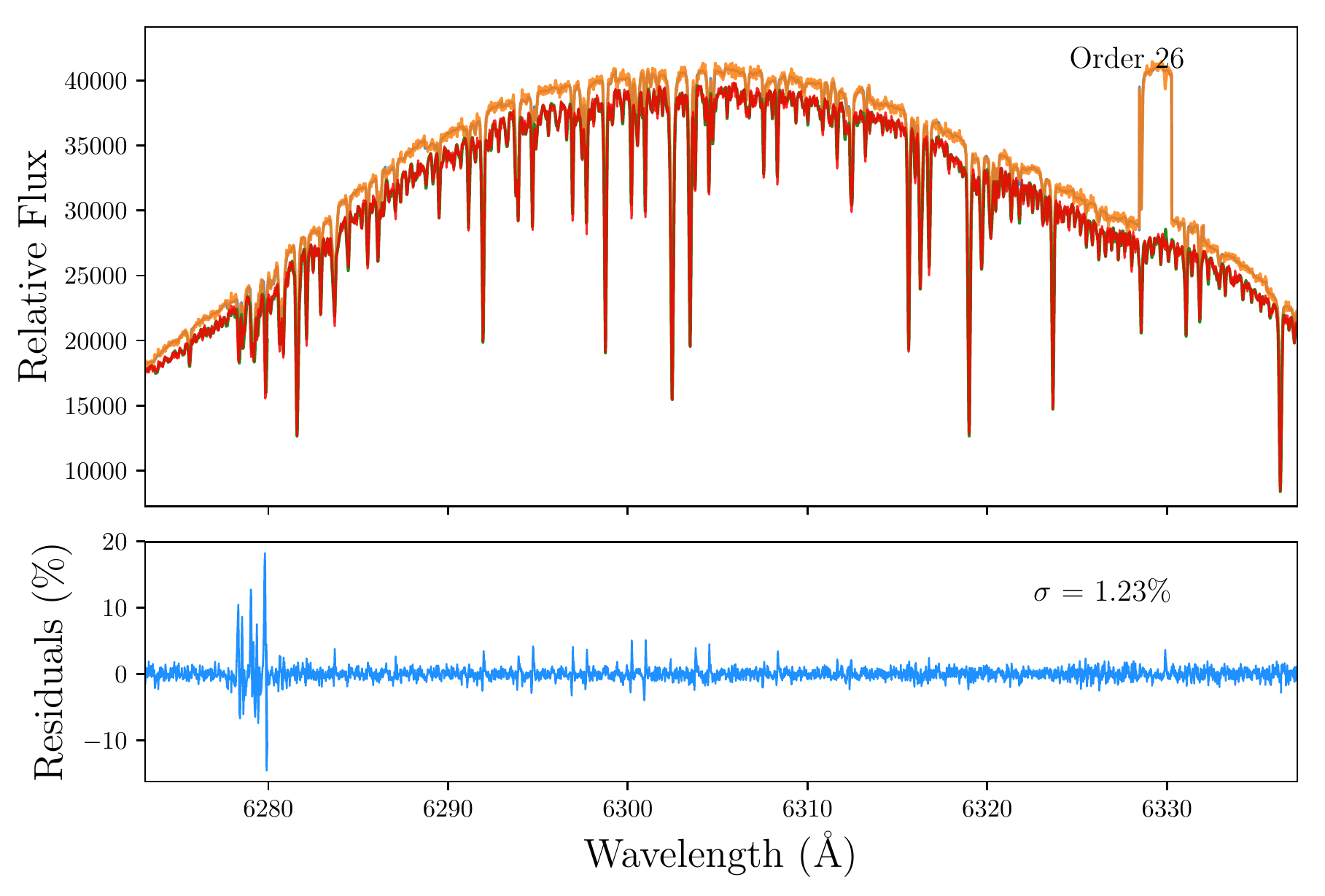}
\caption{Results of the iodine-free derivation for the star $\tau$ Ceti. We show the 26 orders from the iodine region of PFS. On each plot, we show the observation through iodine (red), the template observation without iodine (blue), the forward model of the observations through iodine (green) and the recovered iodine-free spectrum (orange). Bottom panels on each plot show the level of precision of the recovered spectrum compared with the template (in $\%$). }
\label{fig:plot_orders_pfs2}
\end{figure*}

\renewcommand{\thefigure}{\arabic{figure}}
\begin{figure*} 
\includegraphics[scale=0.34]{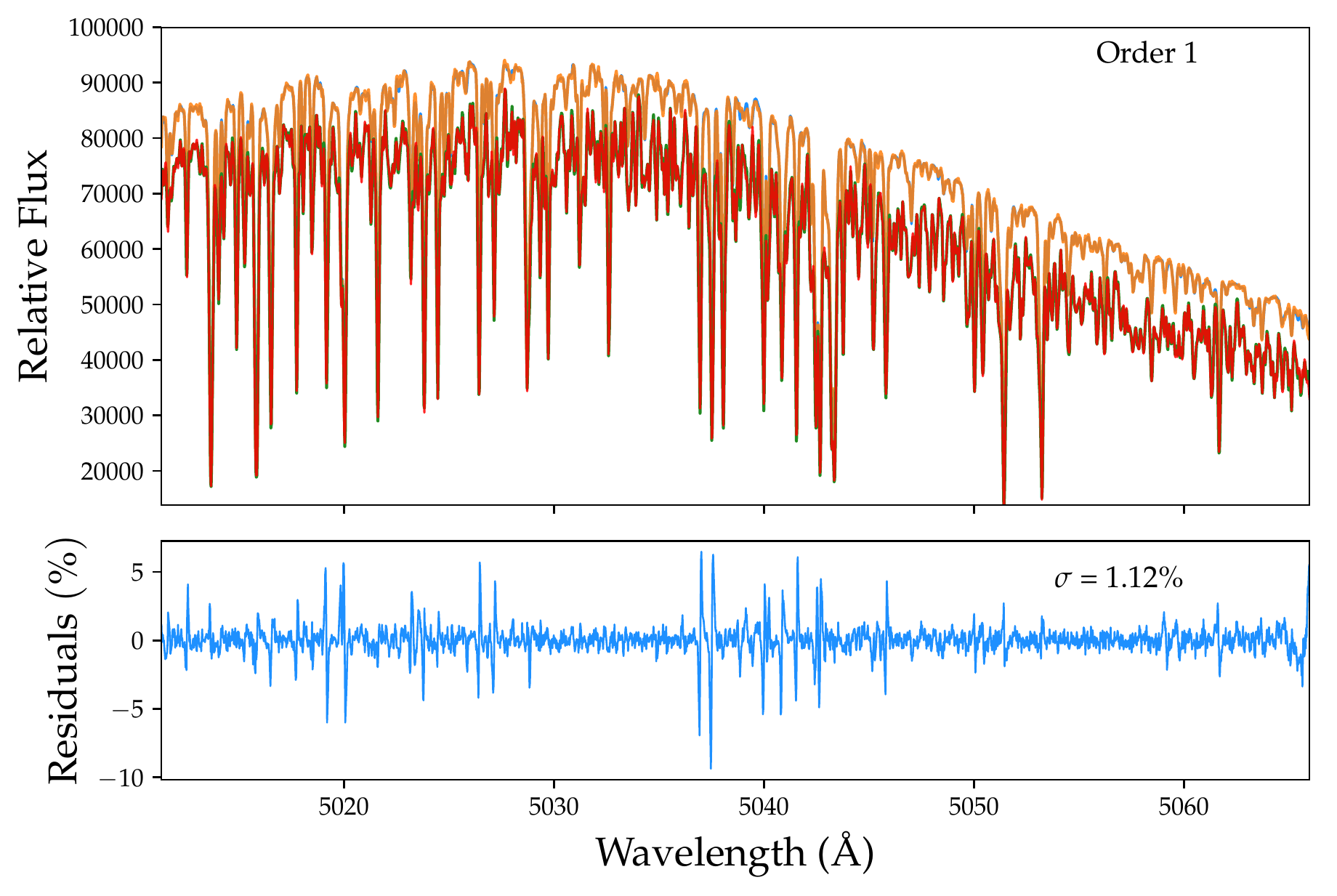}
\includegraphics[scale=0.34]{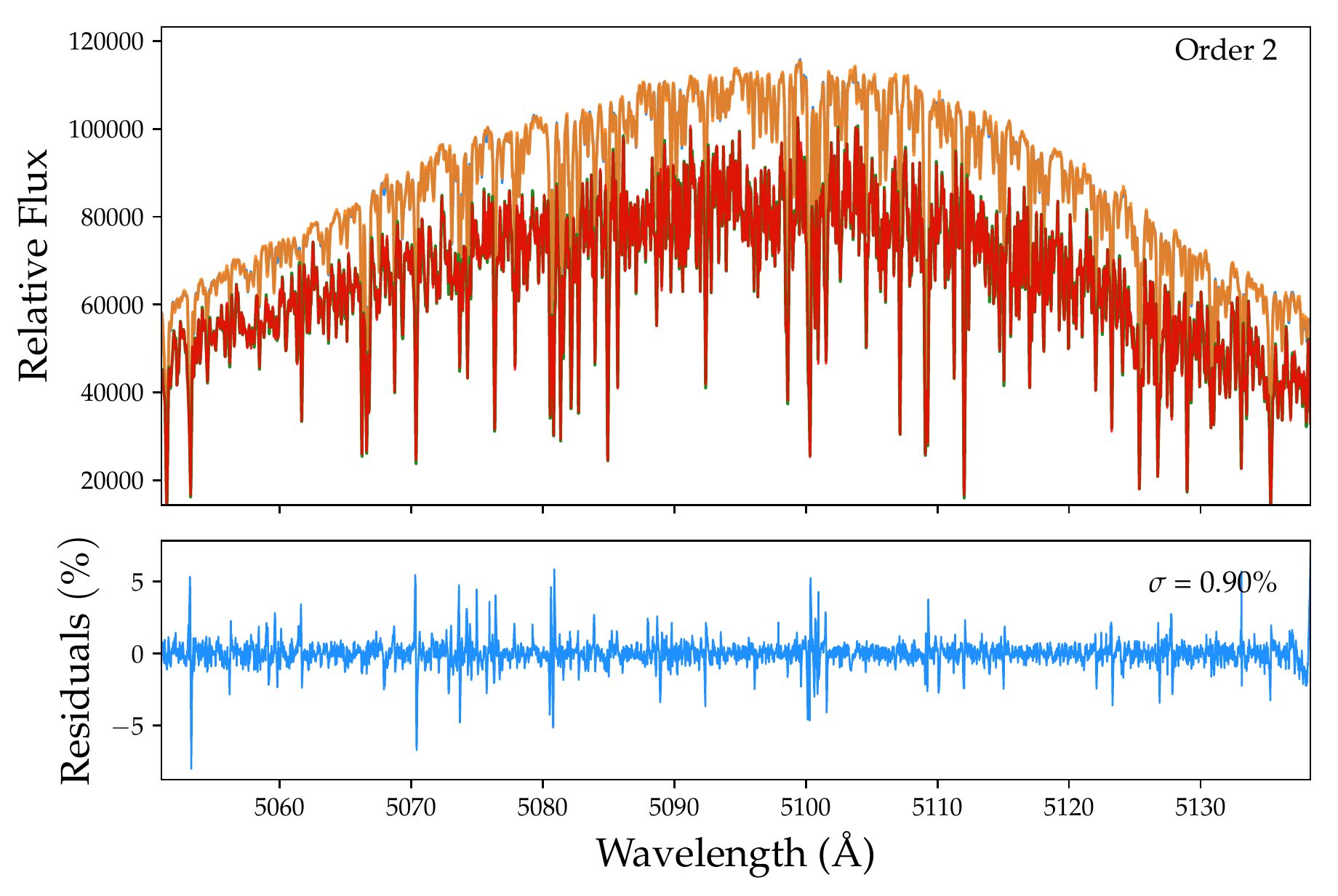}
\includegraphics[scale=0.34]{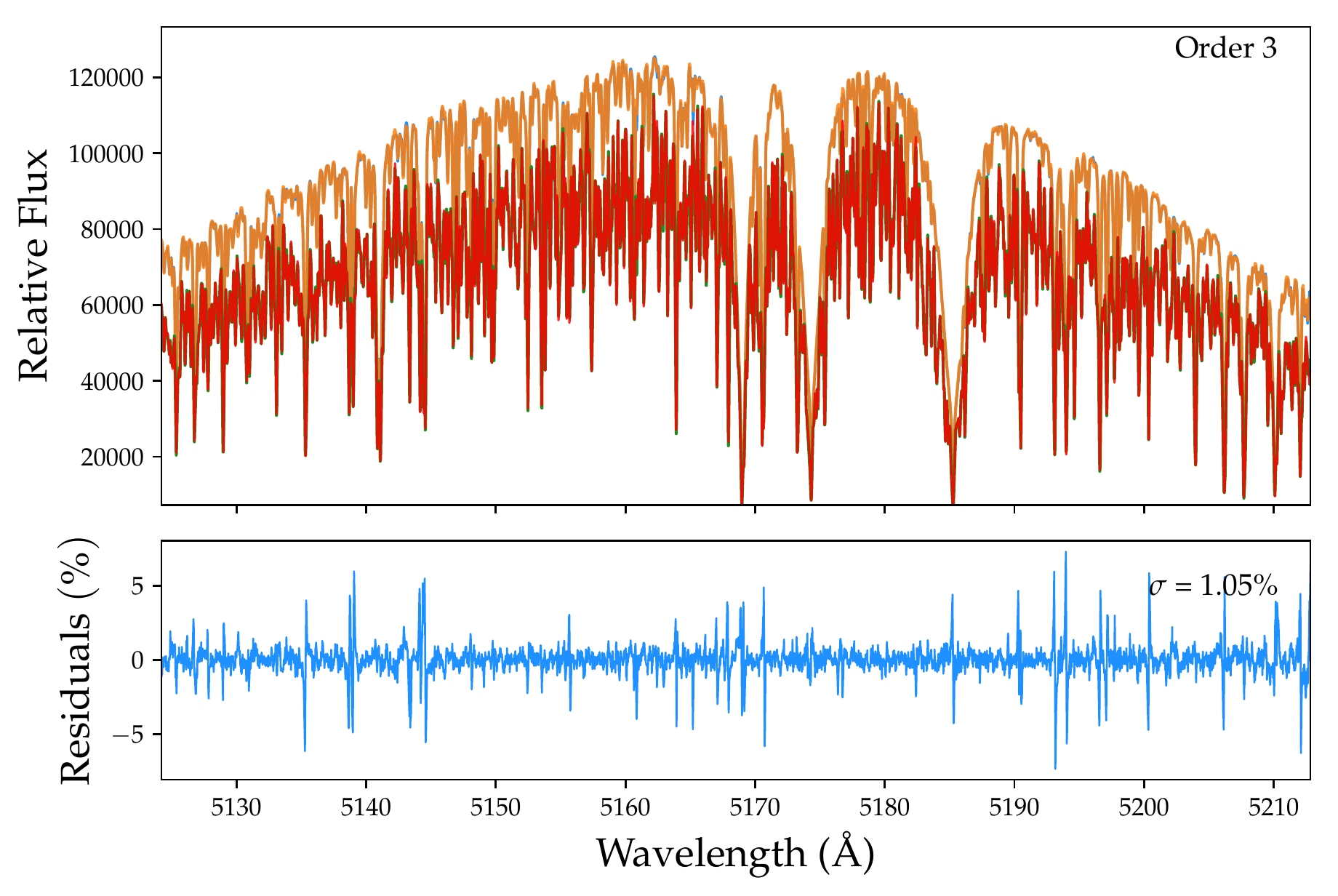}
\includegraphics[scale=0.34]{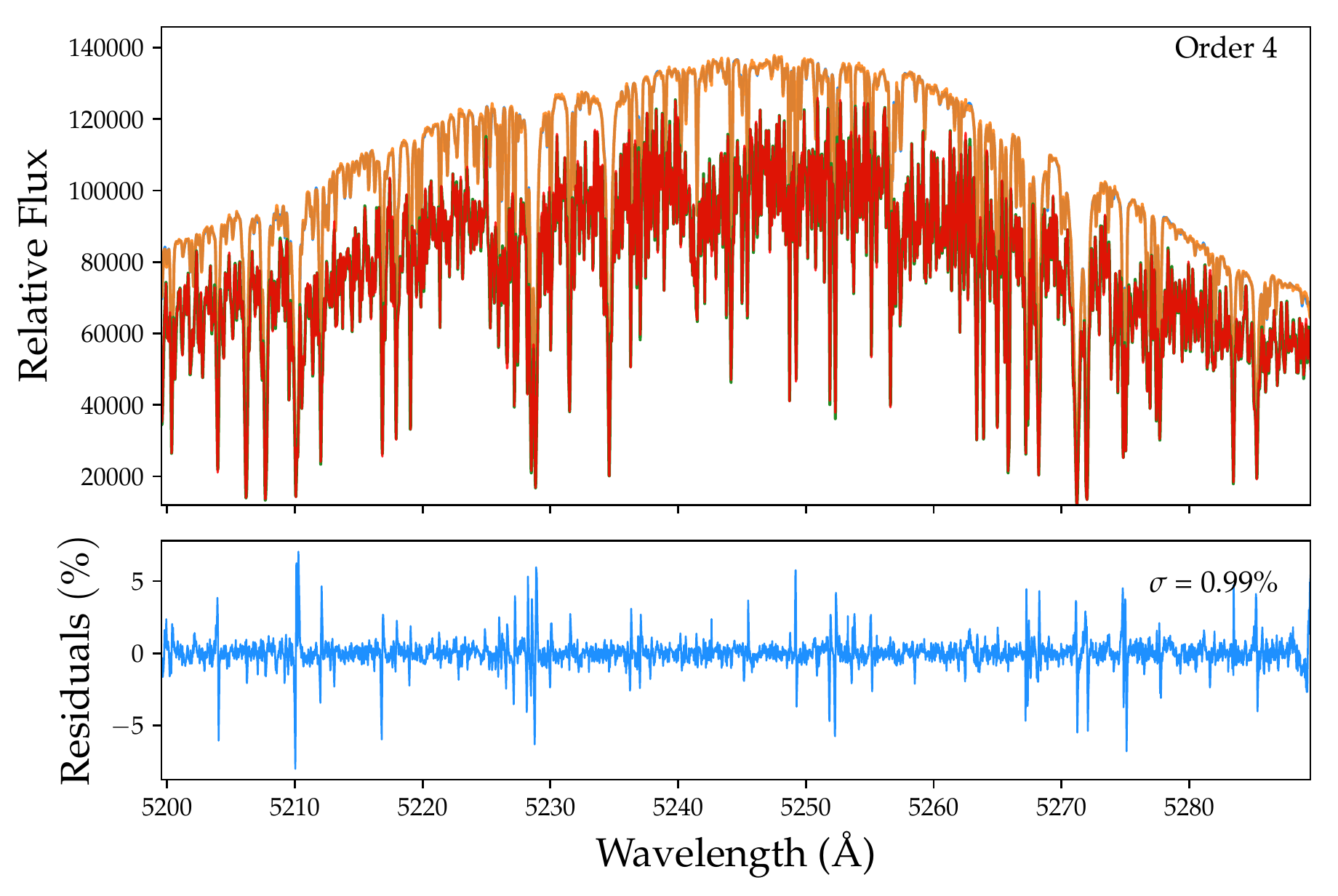}
\includegraphics[scale=0.34]{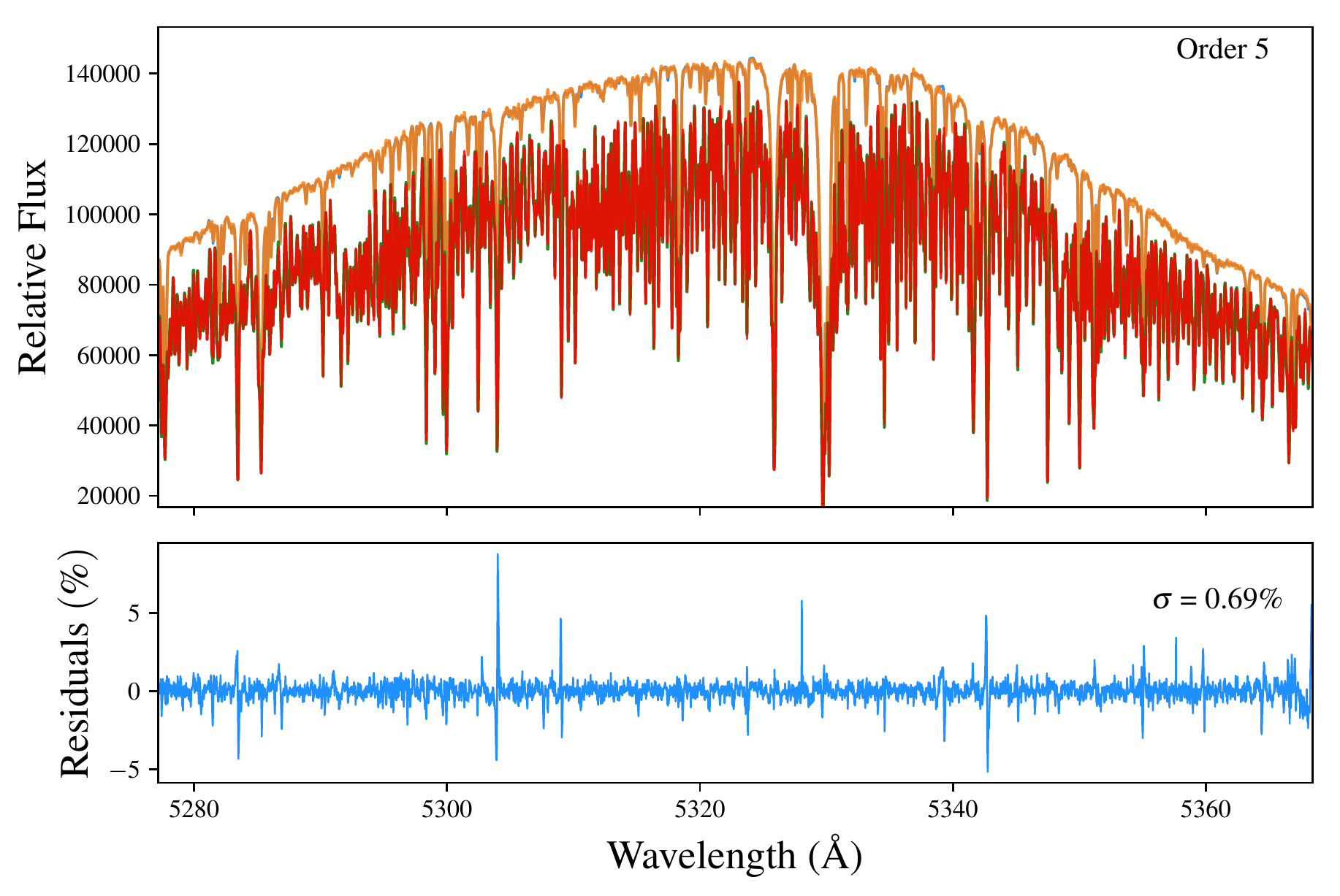}
\includegraphics[scale=0.34]{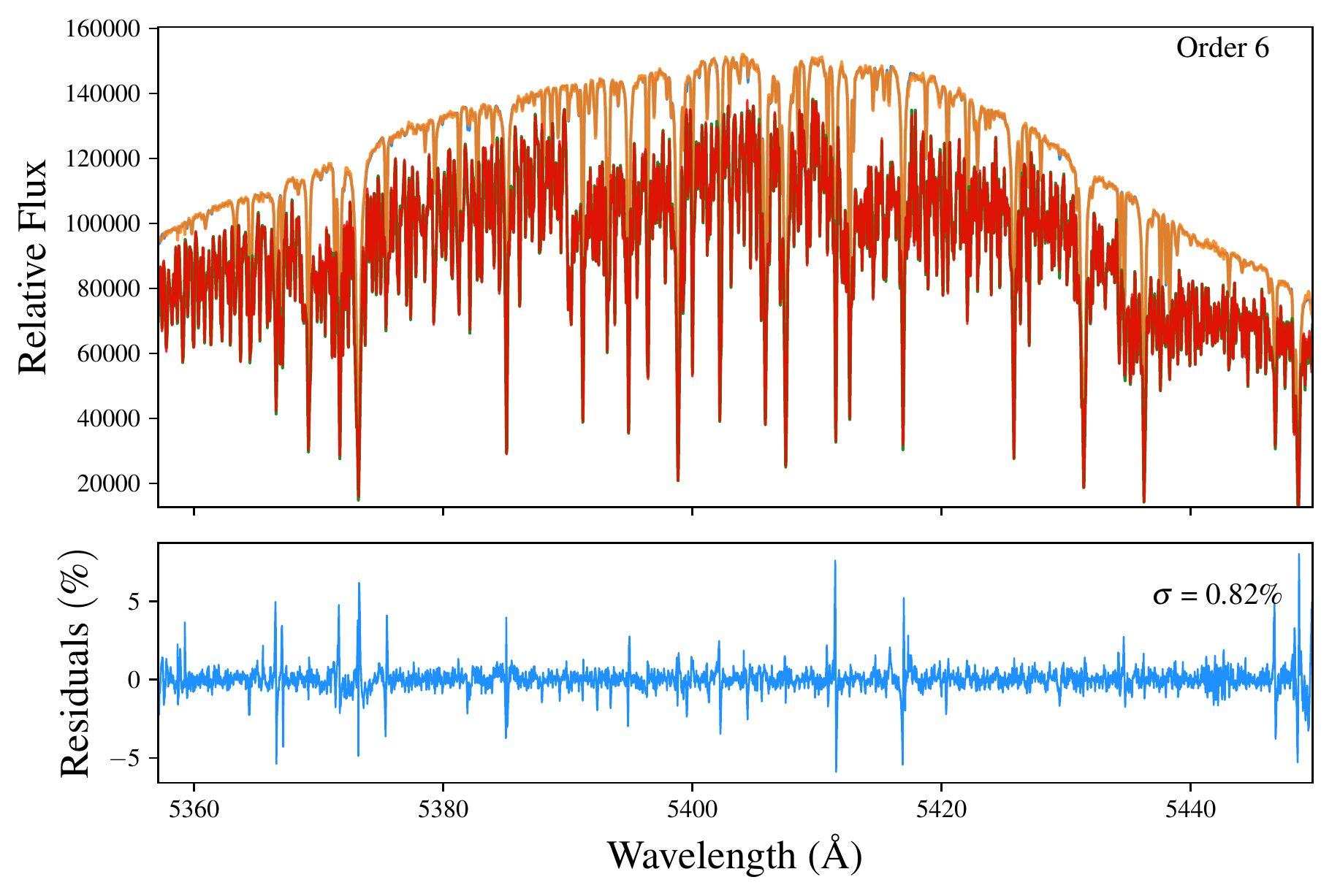}
\includegraphics[scale=0.34]{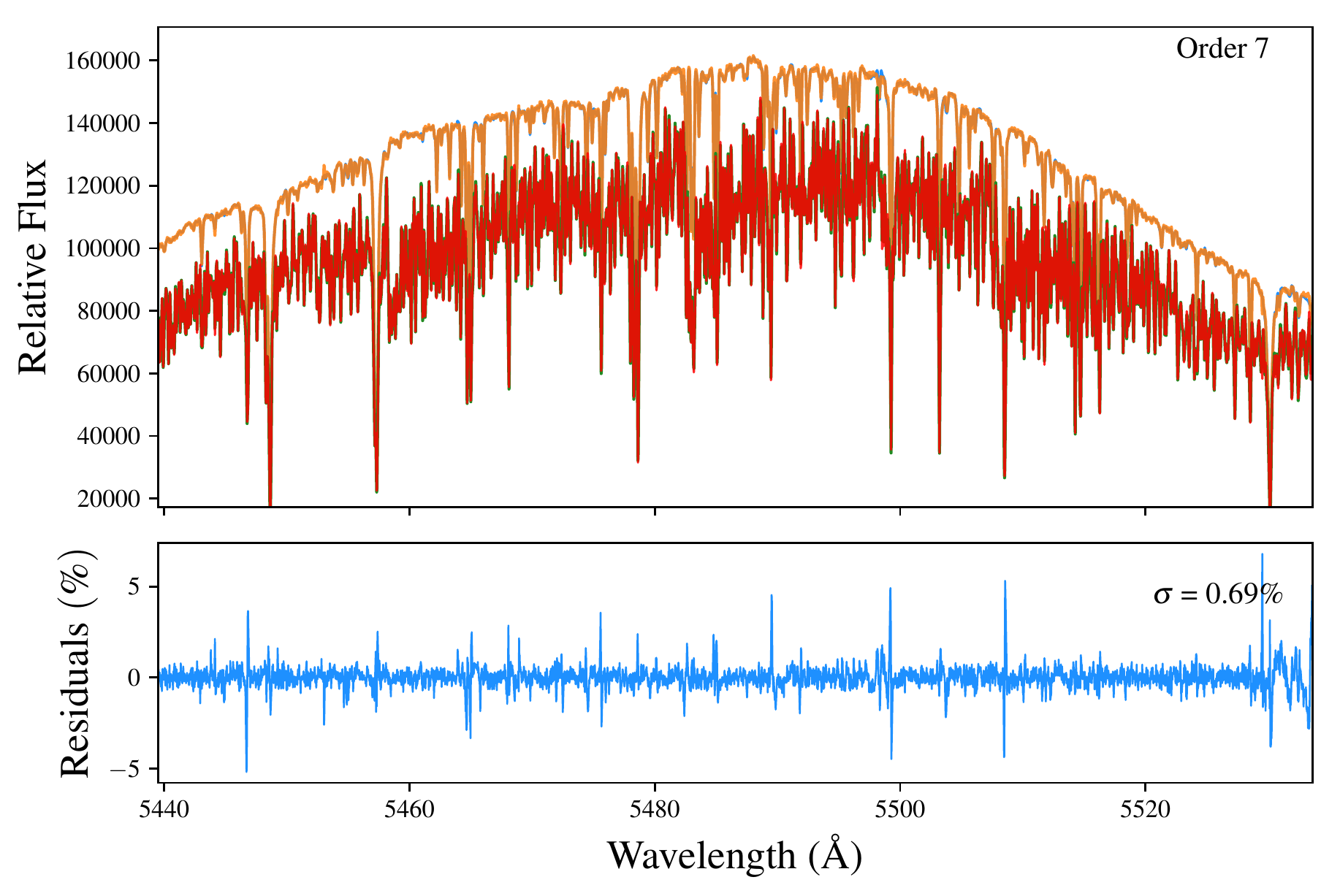}
\includegraphics[scale=0.34]{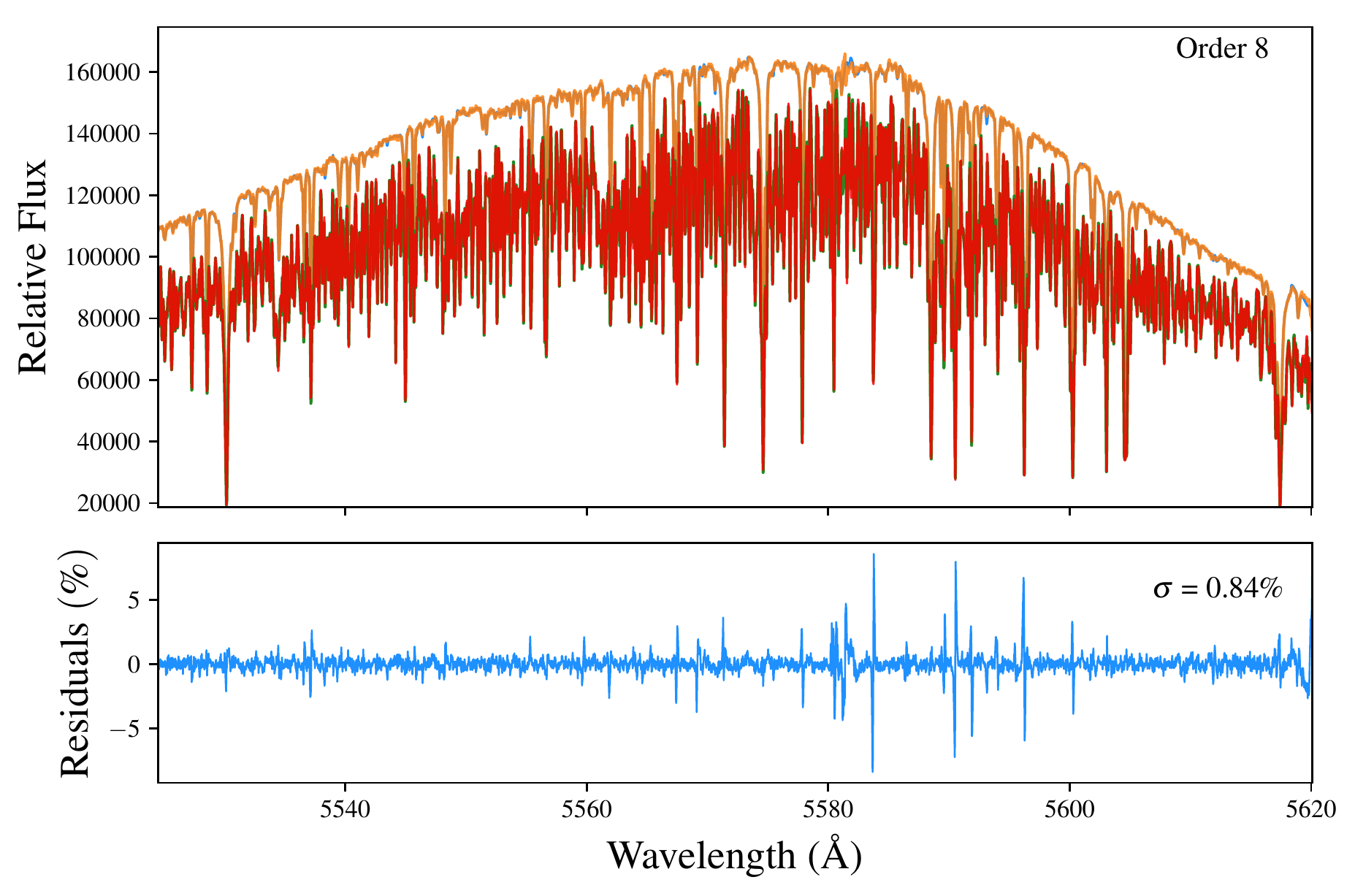}
\includegraphics[scale=0.34]{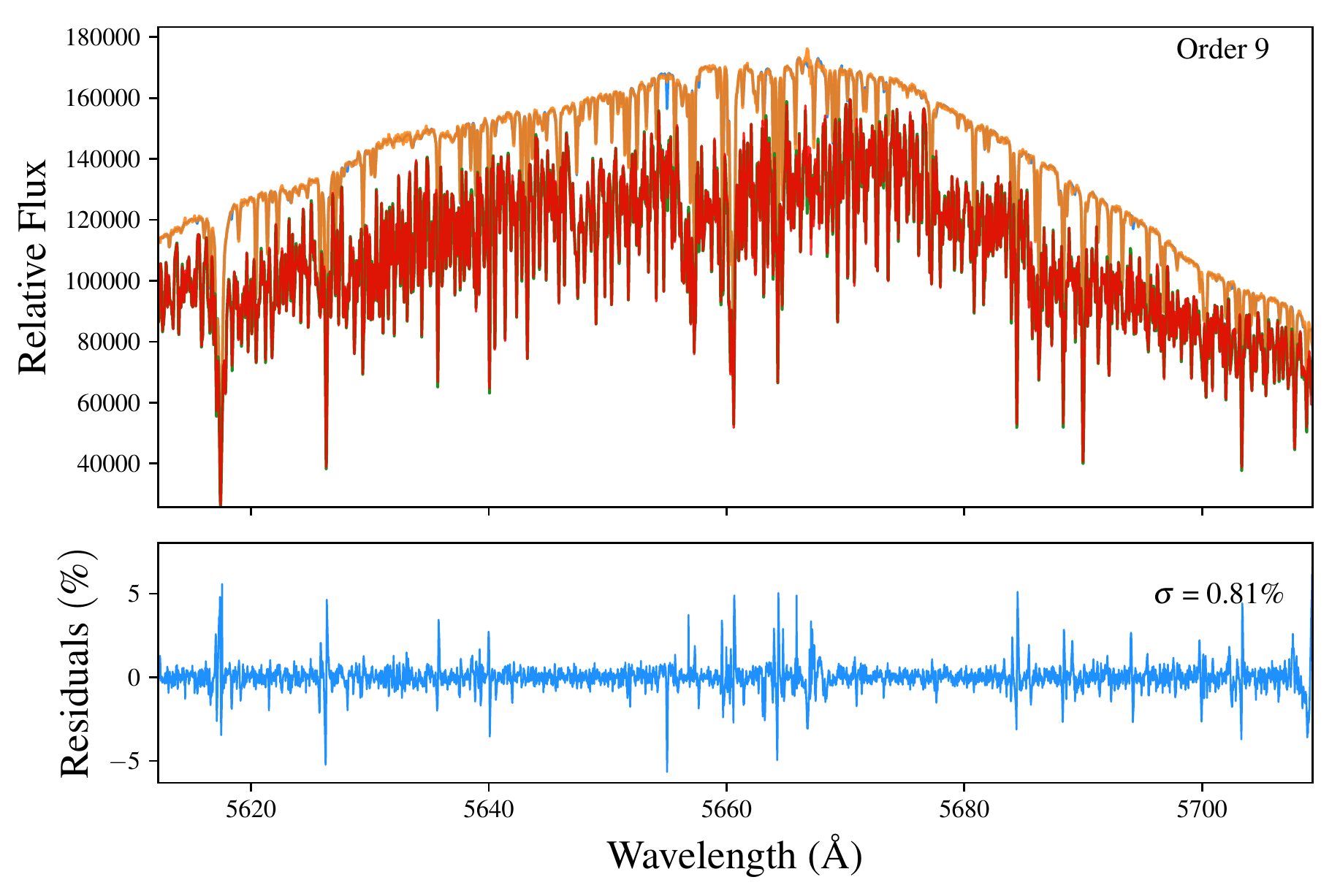}
\includegraphics[scale=0.34]{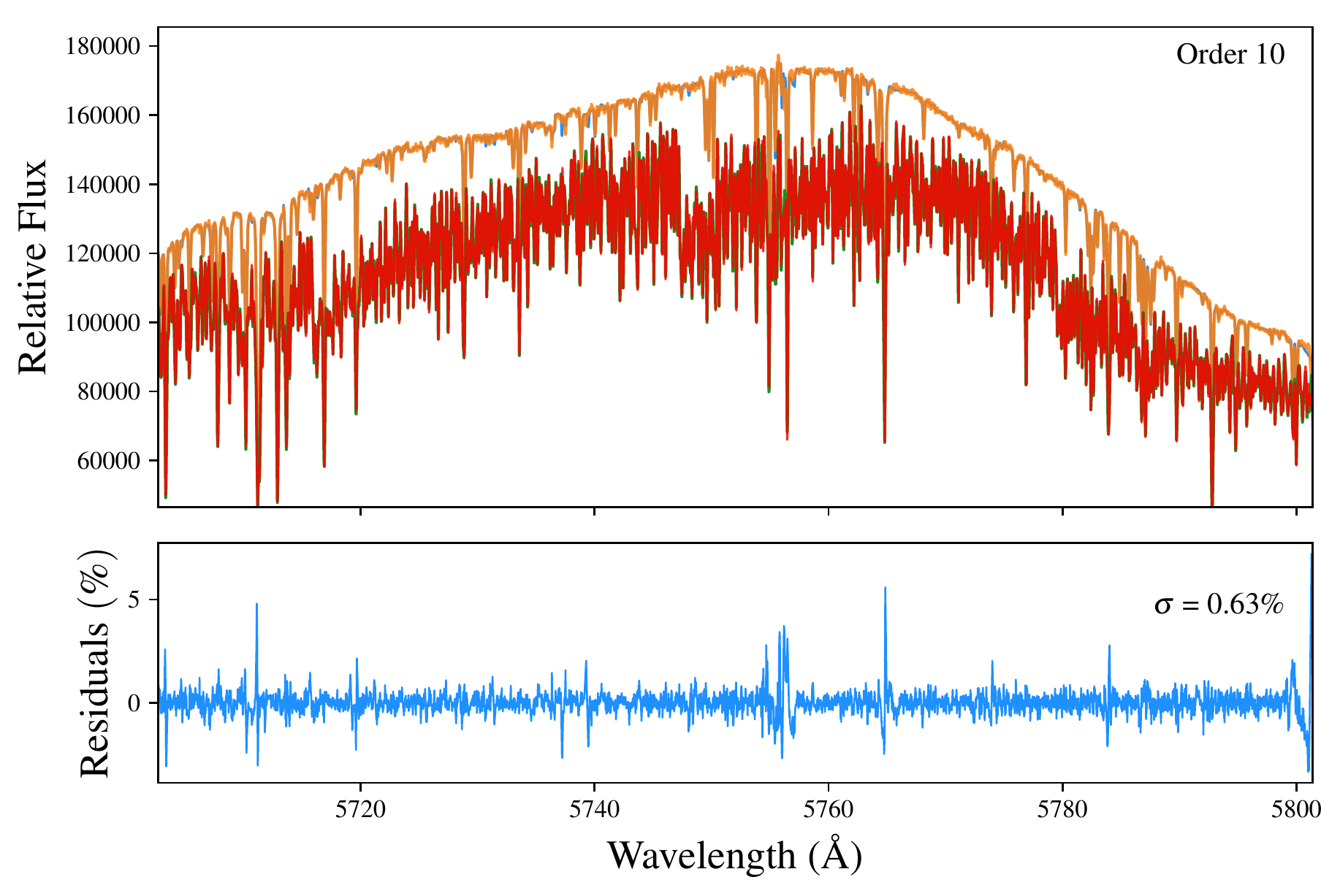}
\includegraphics[scale=0.34]{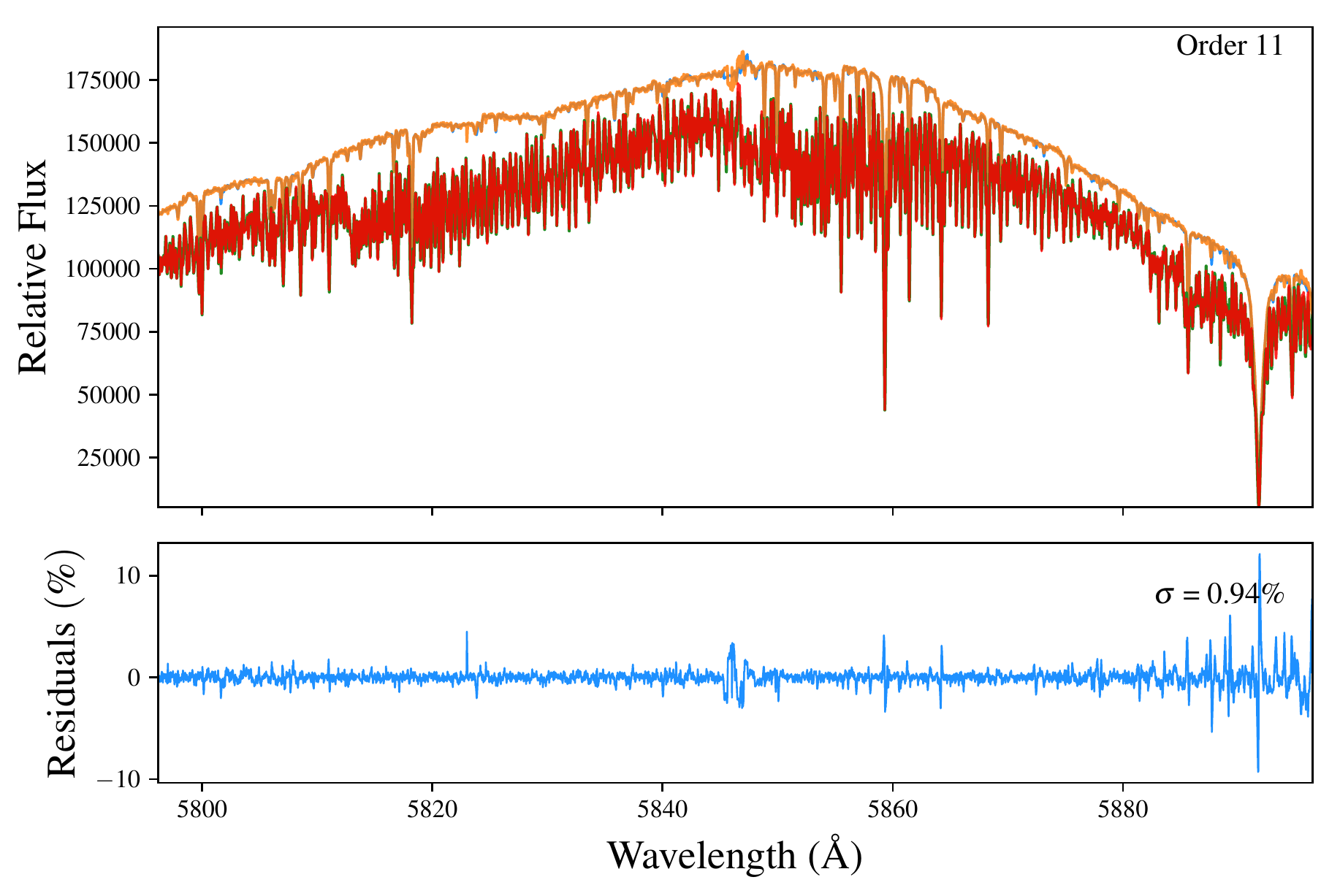}
\includegraphics[scale=0.34]{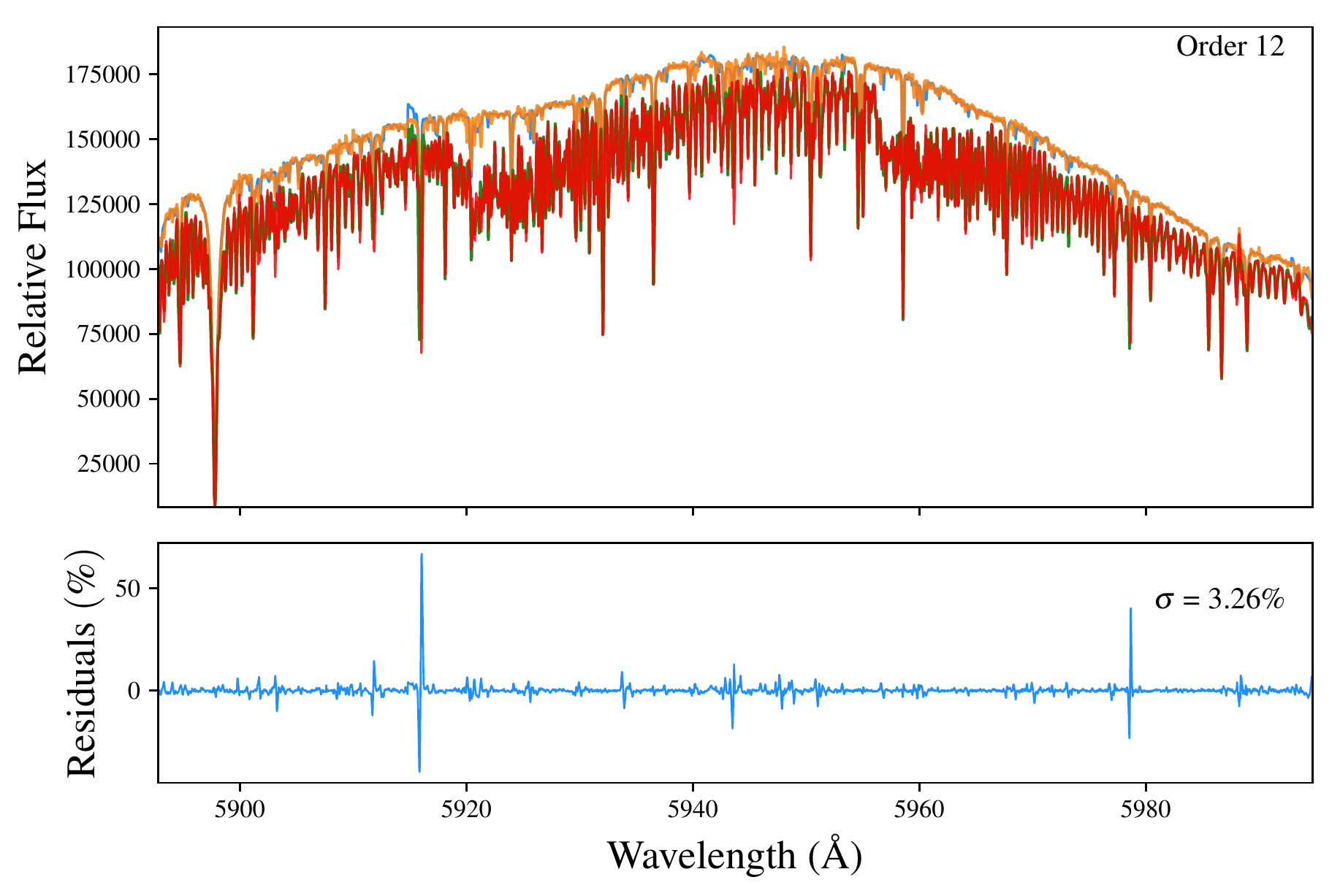}
\hspace{2cm}\includegraphics[scale=0.34]{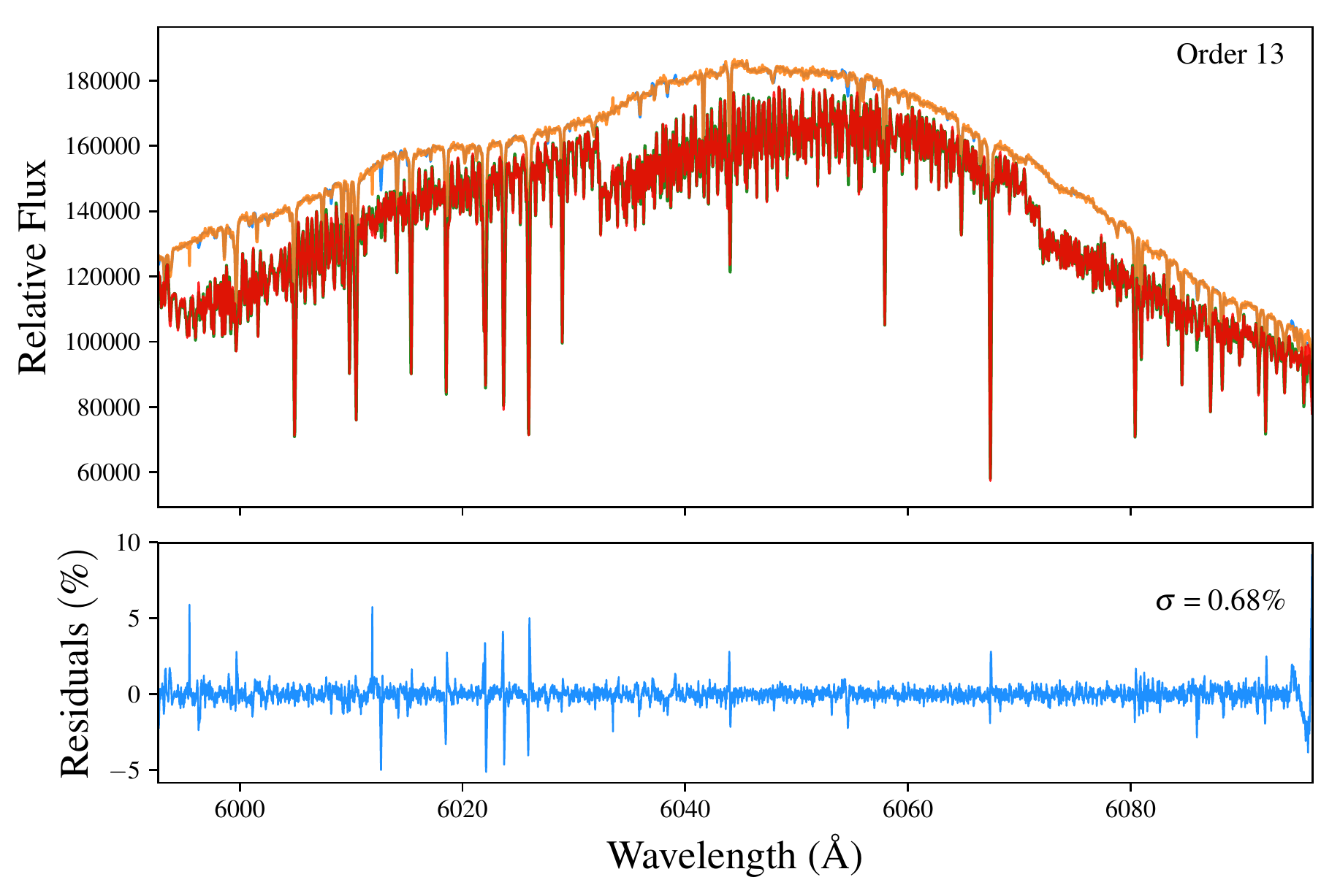}
\includegraphics[scale=0.34]{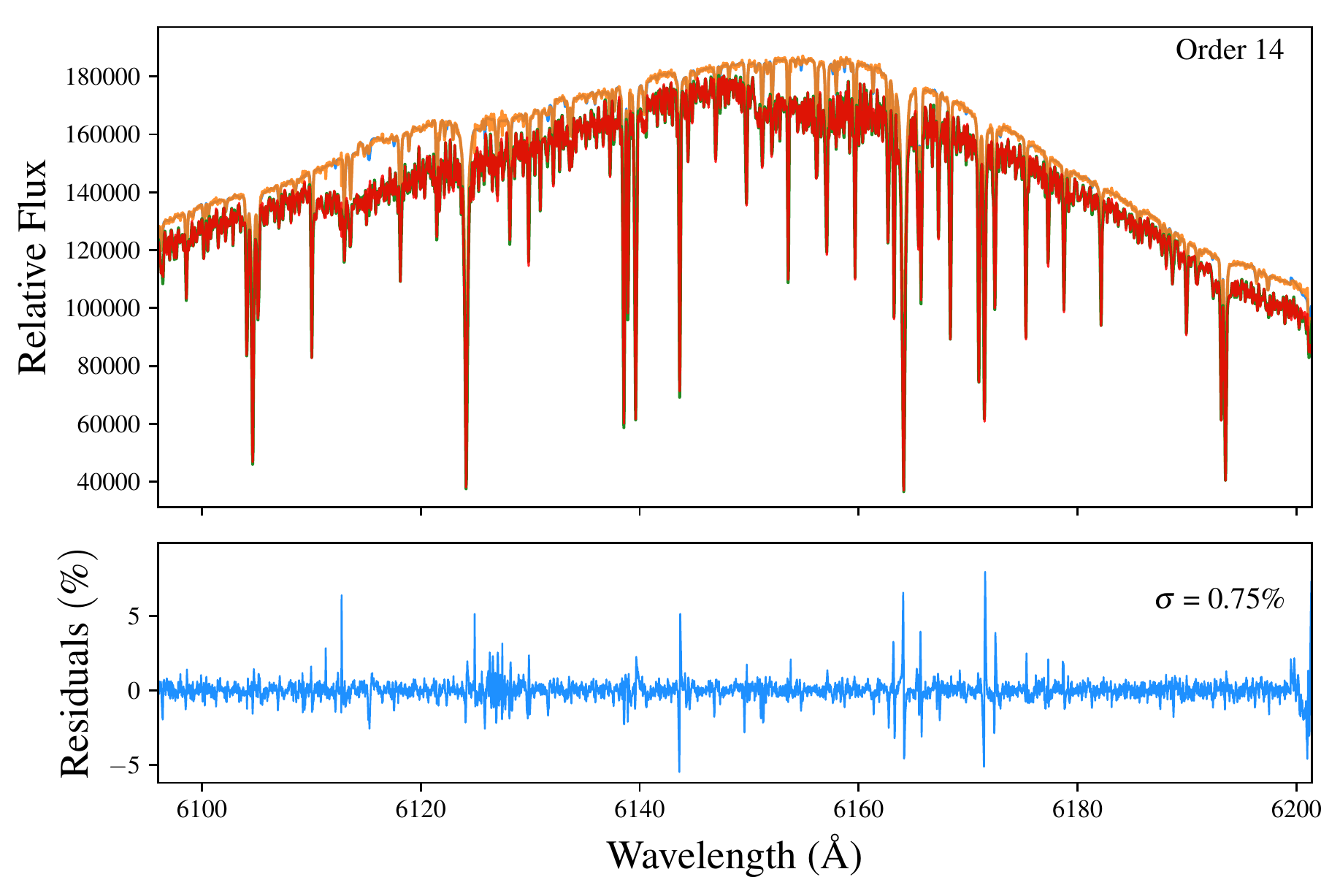}
\caption{Results of the iodine-free derivation for the star $\tau$ Ceti. We show the 14 orders from the iodine region of HIRES. On each plot, we show the observation through iodine (red), the template observation without iodine (blue), the forward model of the observations through iodine (green) and the recovered iodine-free spectrum (orange). Bottom panels on each plot show the level of precision of the recovered spectrum compared to the template (in $\%$).}
\label{fig:plot_orders_hires}
\end{figure*}

\begin{figure*} 
\includegraphics[scale=0.34]{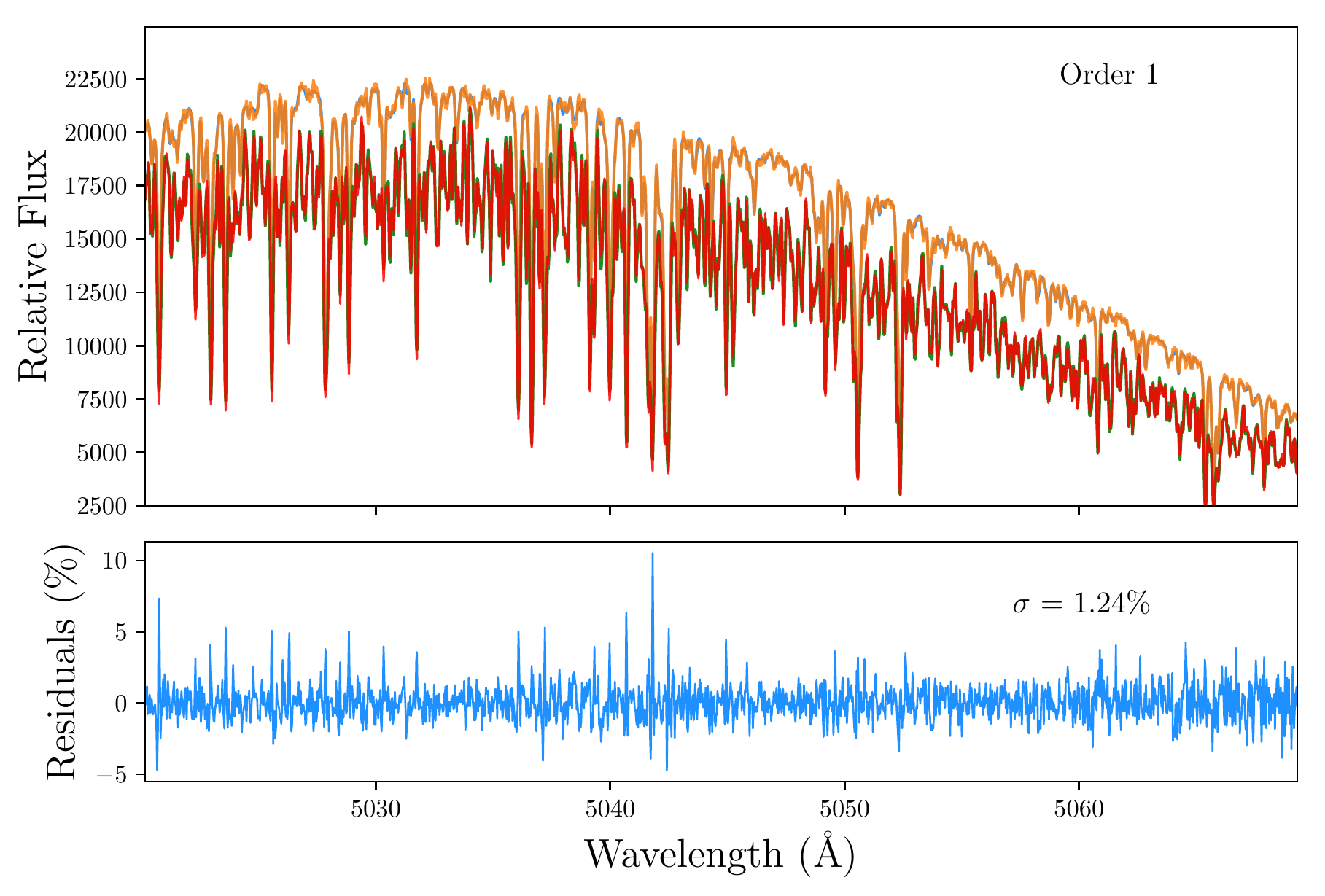}
\includegraphics[scale=0.34]{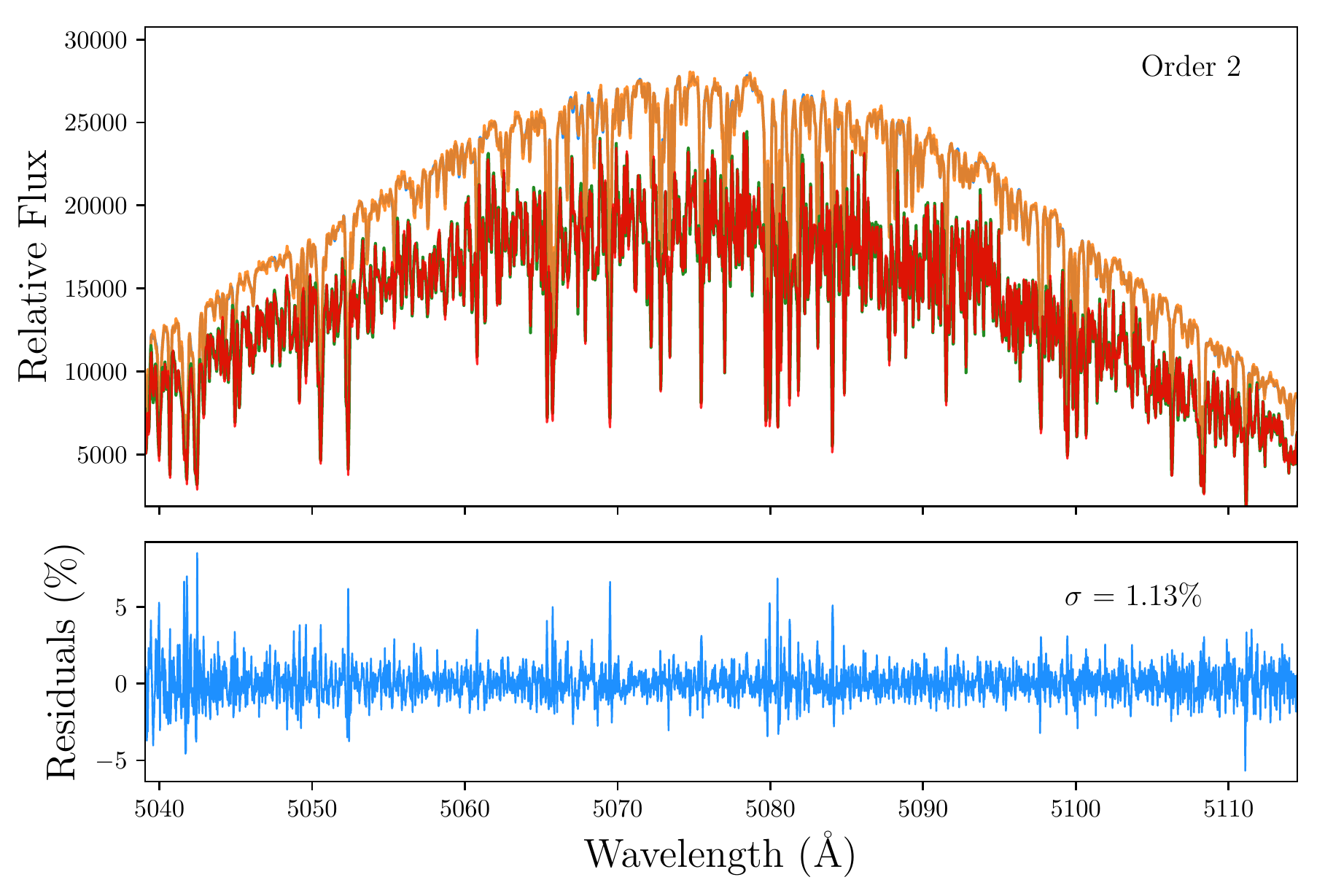}
\includegraphics[scale=0.34]{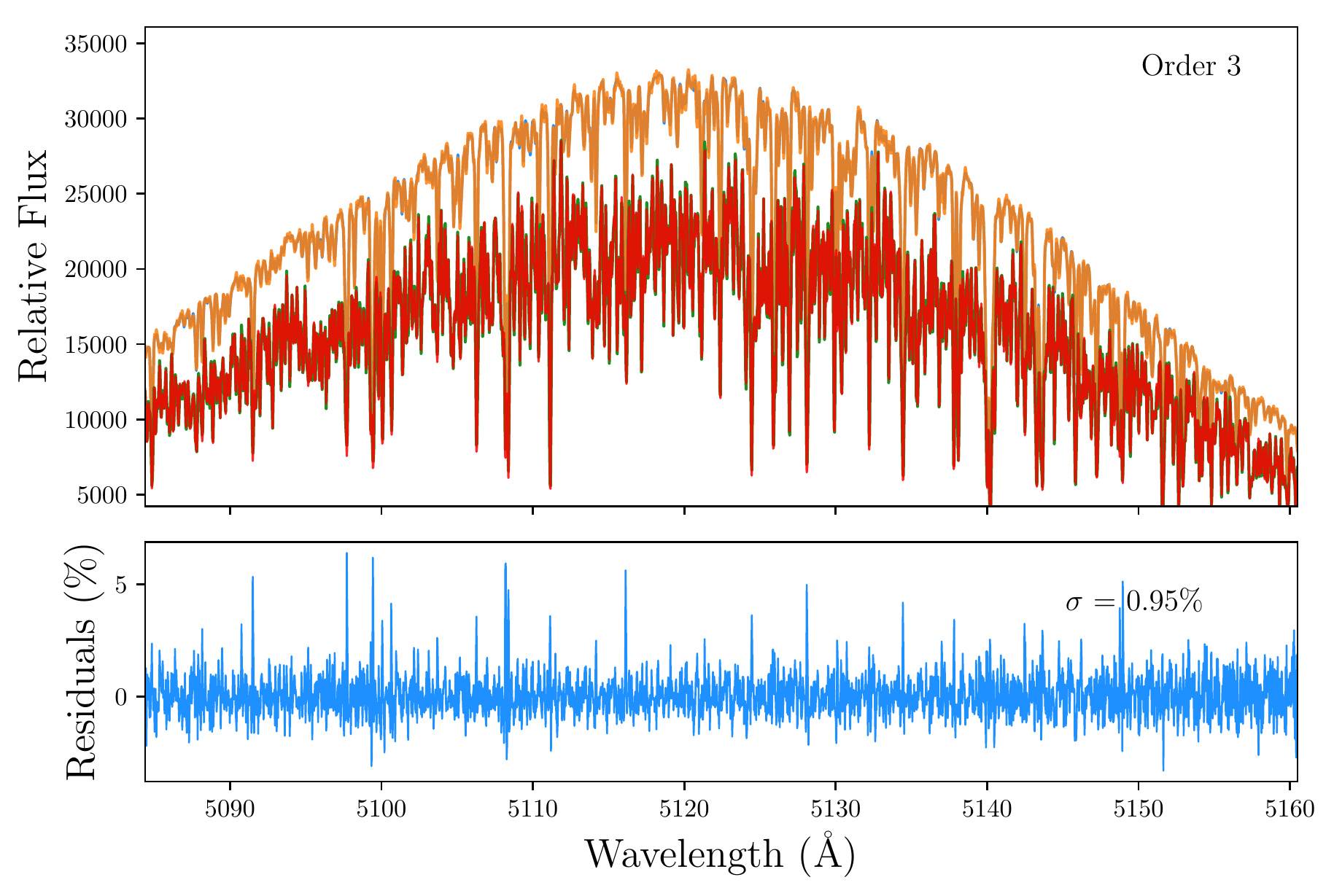}
\includegraphics[scale=0.34]{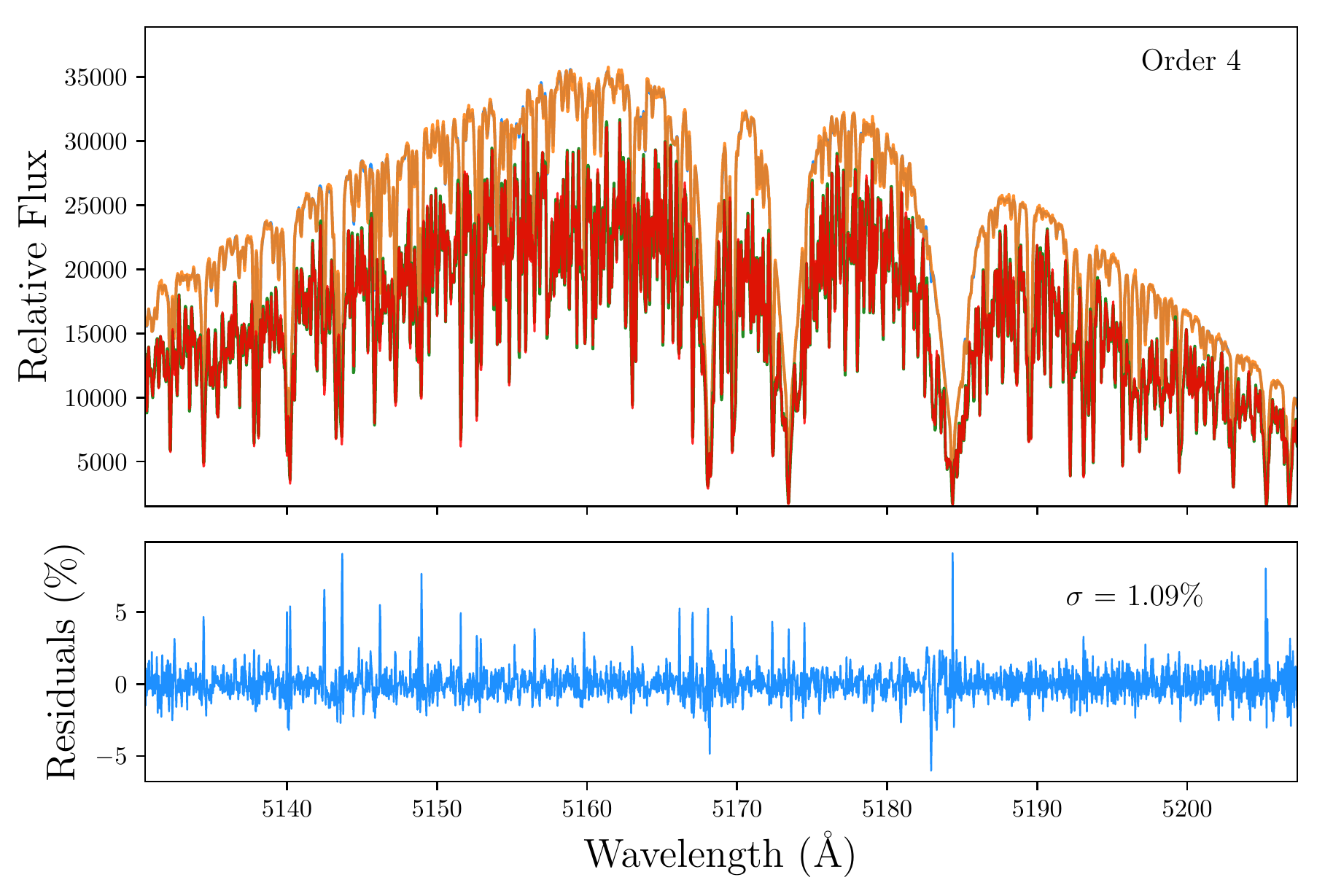}
\includegraphics[scale=0.34]{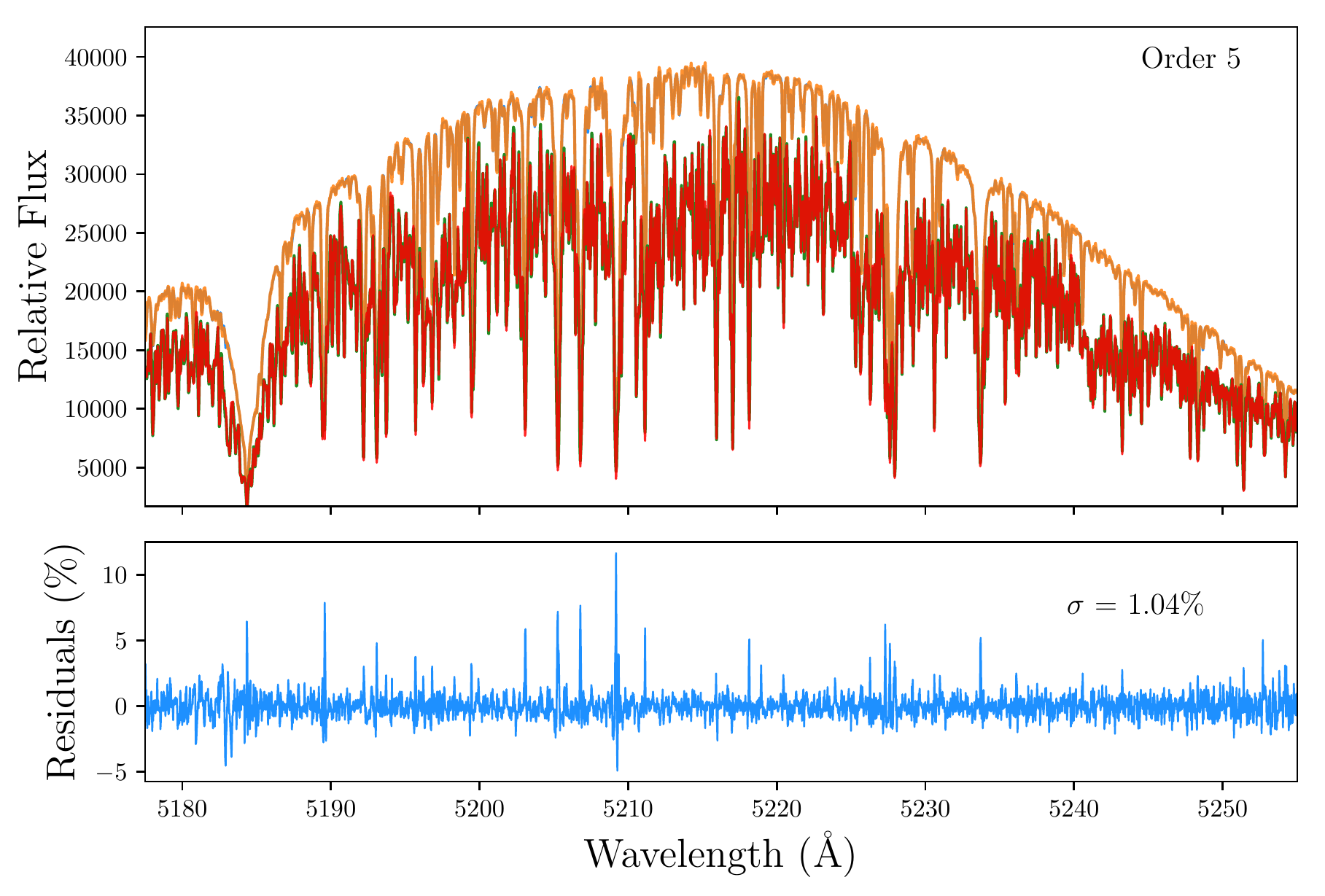}
\includegraphics[scale=0.34]{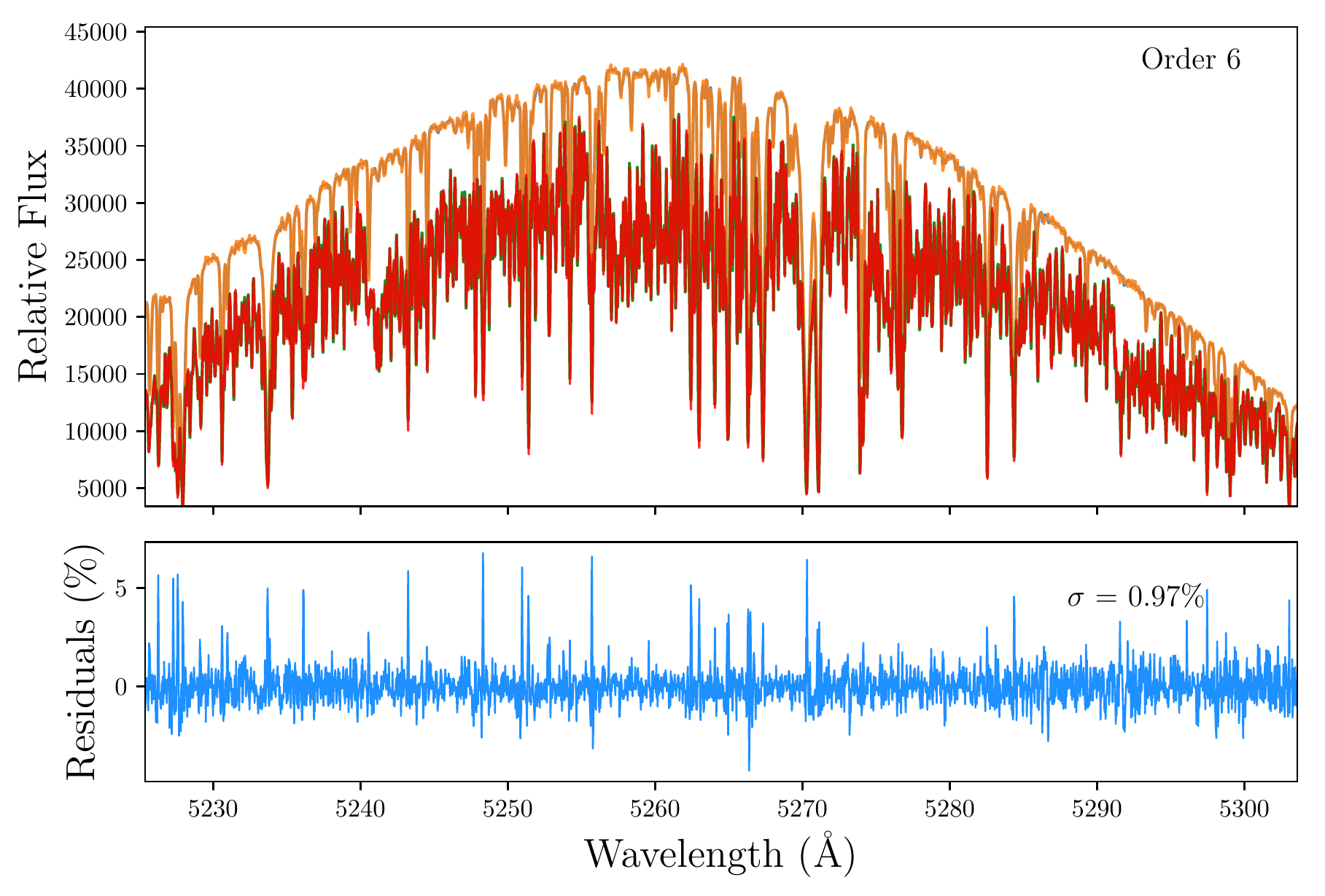}
\includegraphics[scale=0.34]{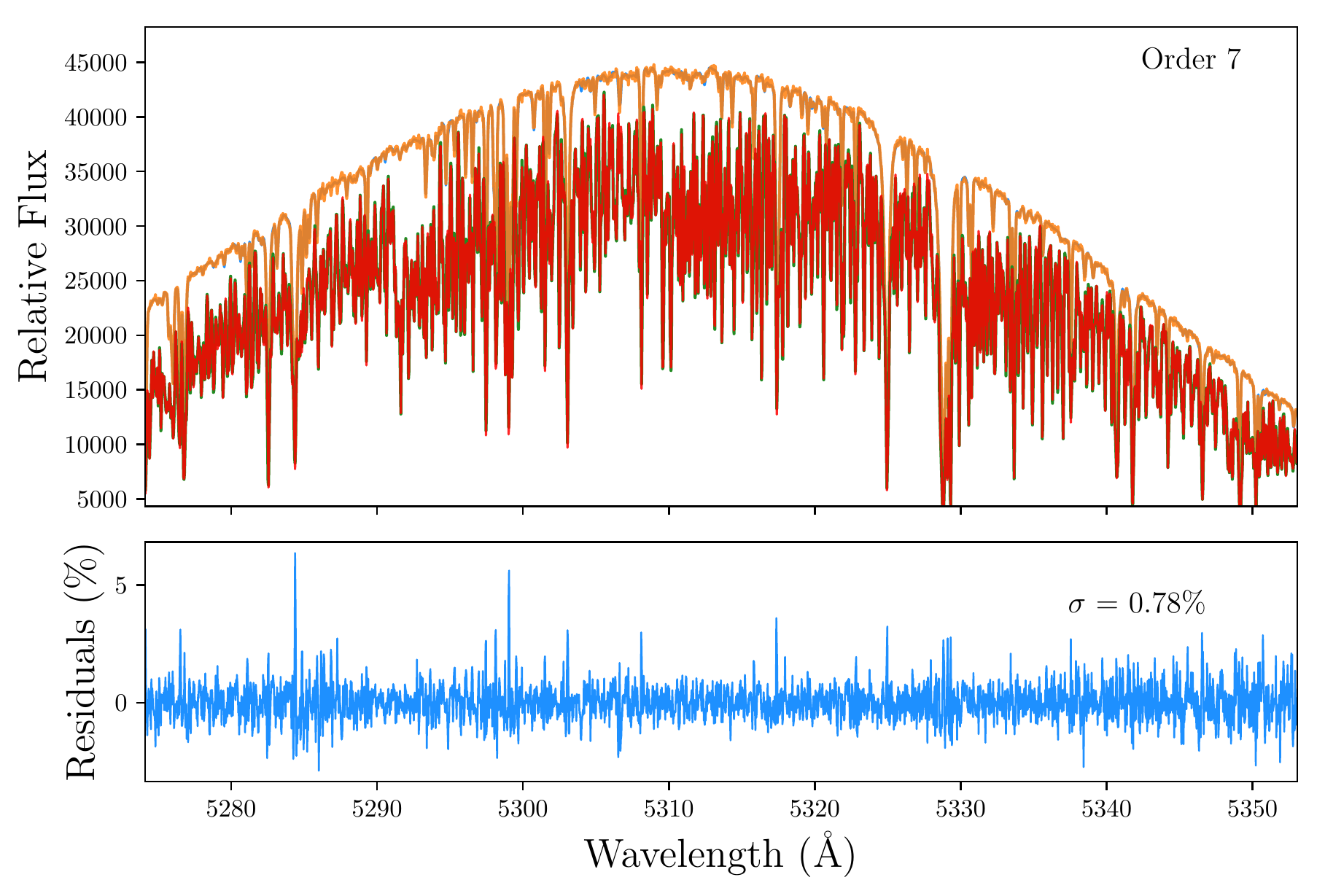}
\includegraphics[scale=0.34]{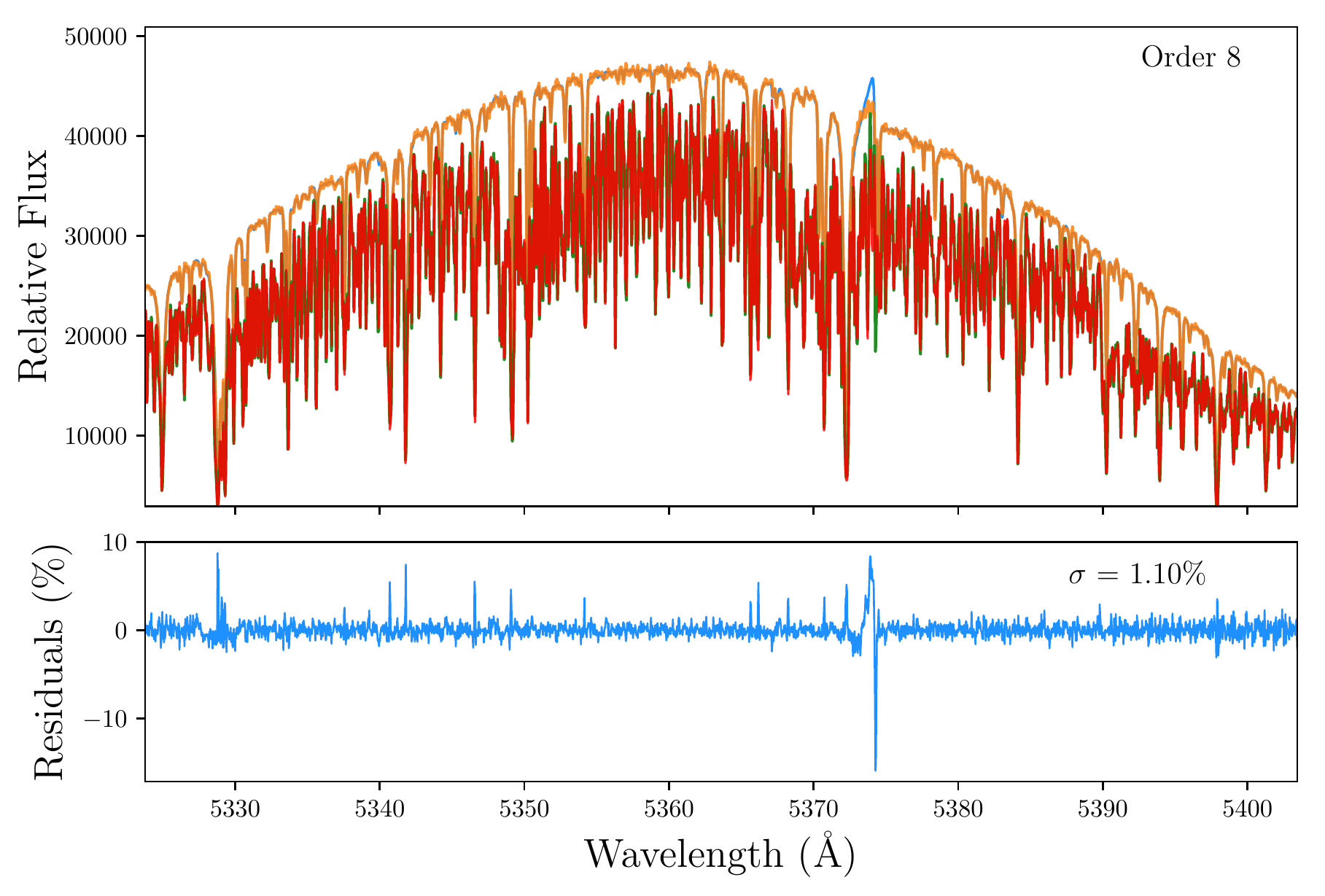}
\includegraphics[scale=0.34]{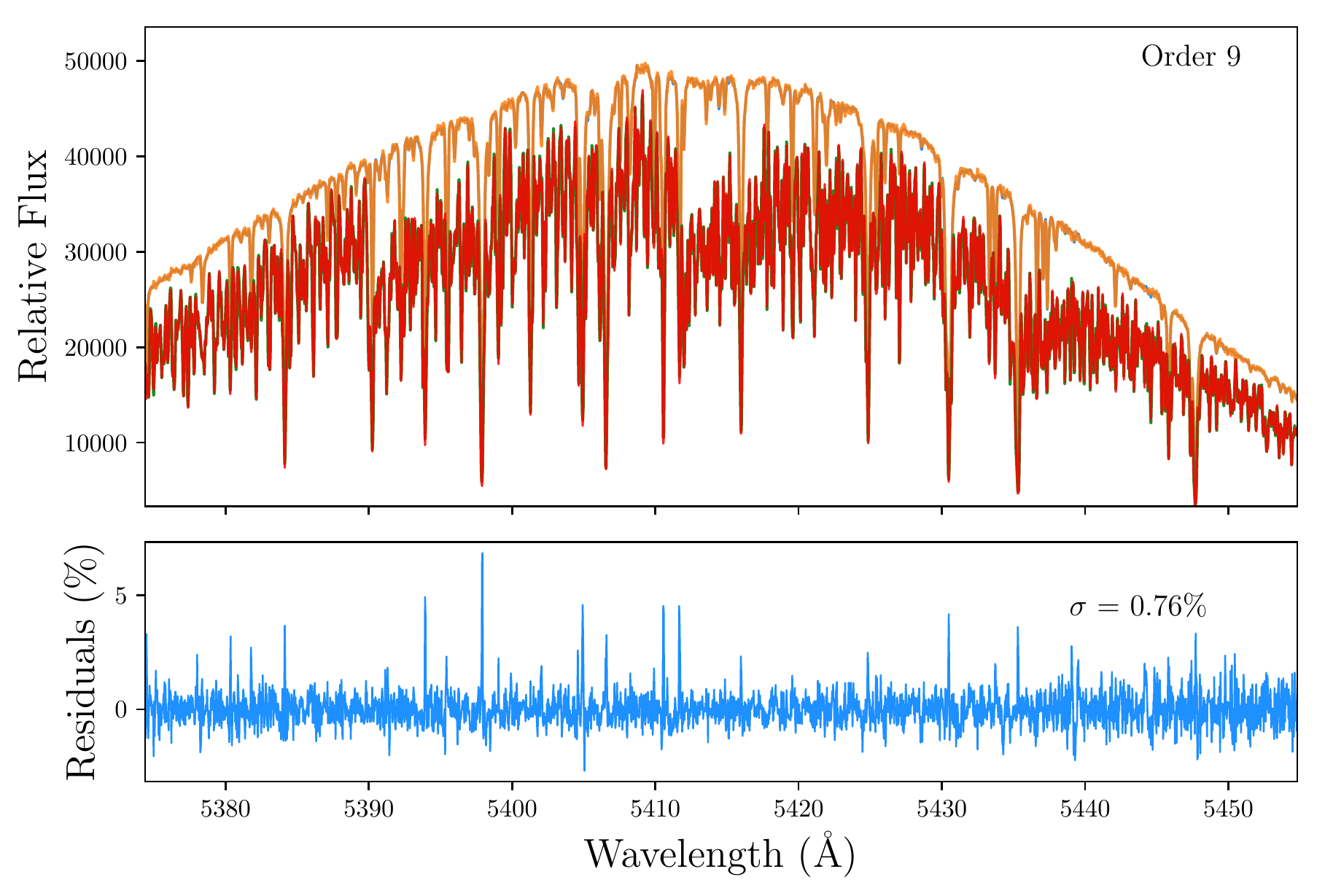}
\includegraphics[scale=0.34]{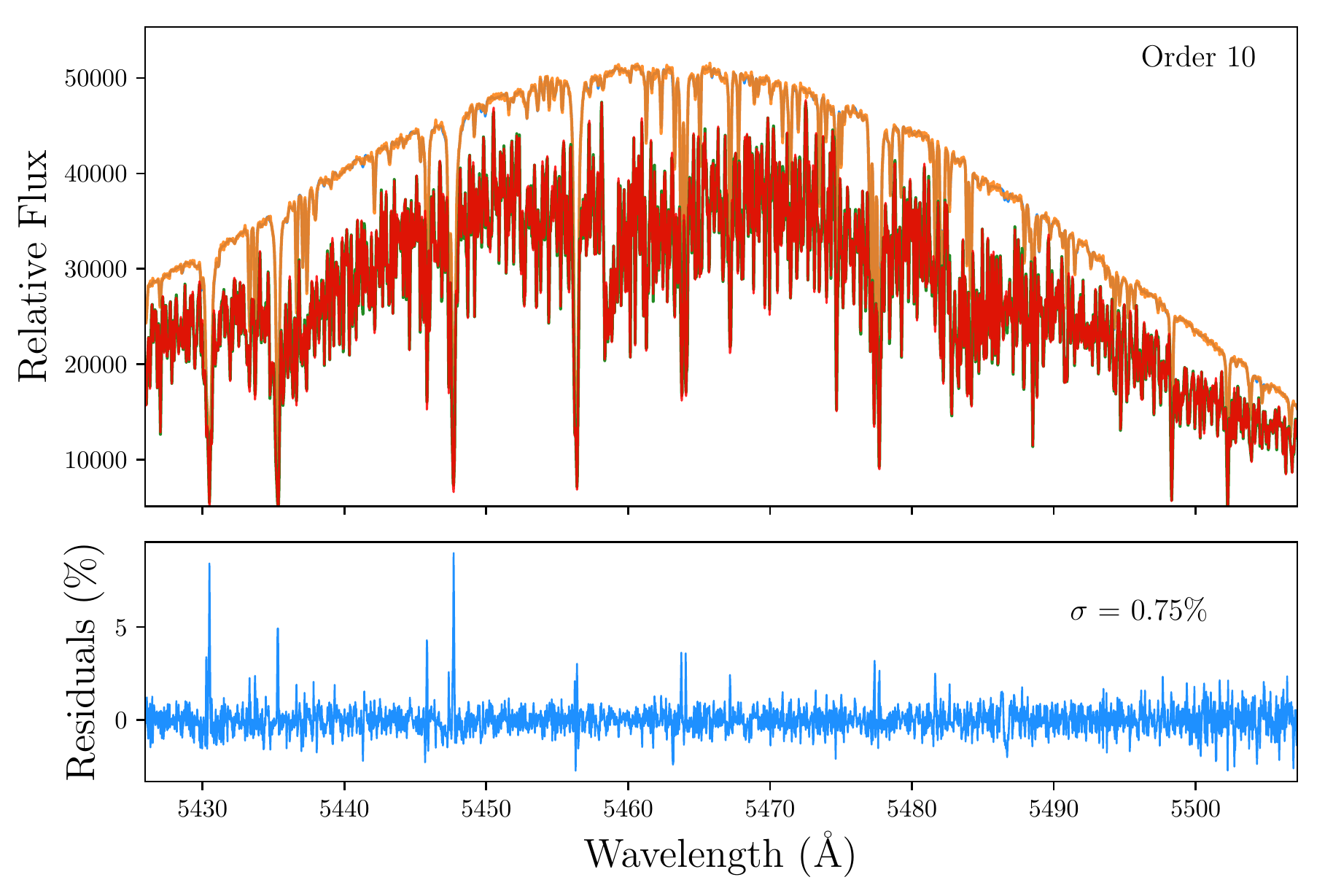}
\includegraphics[scale=0.34]{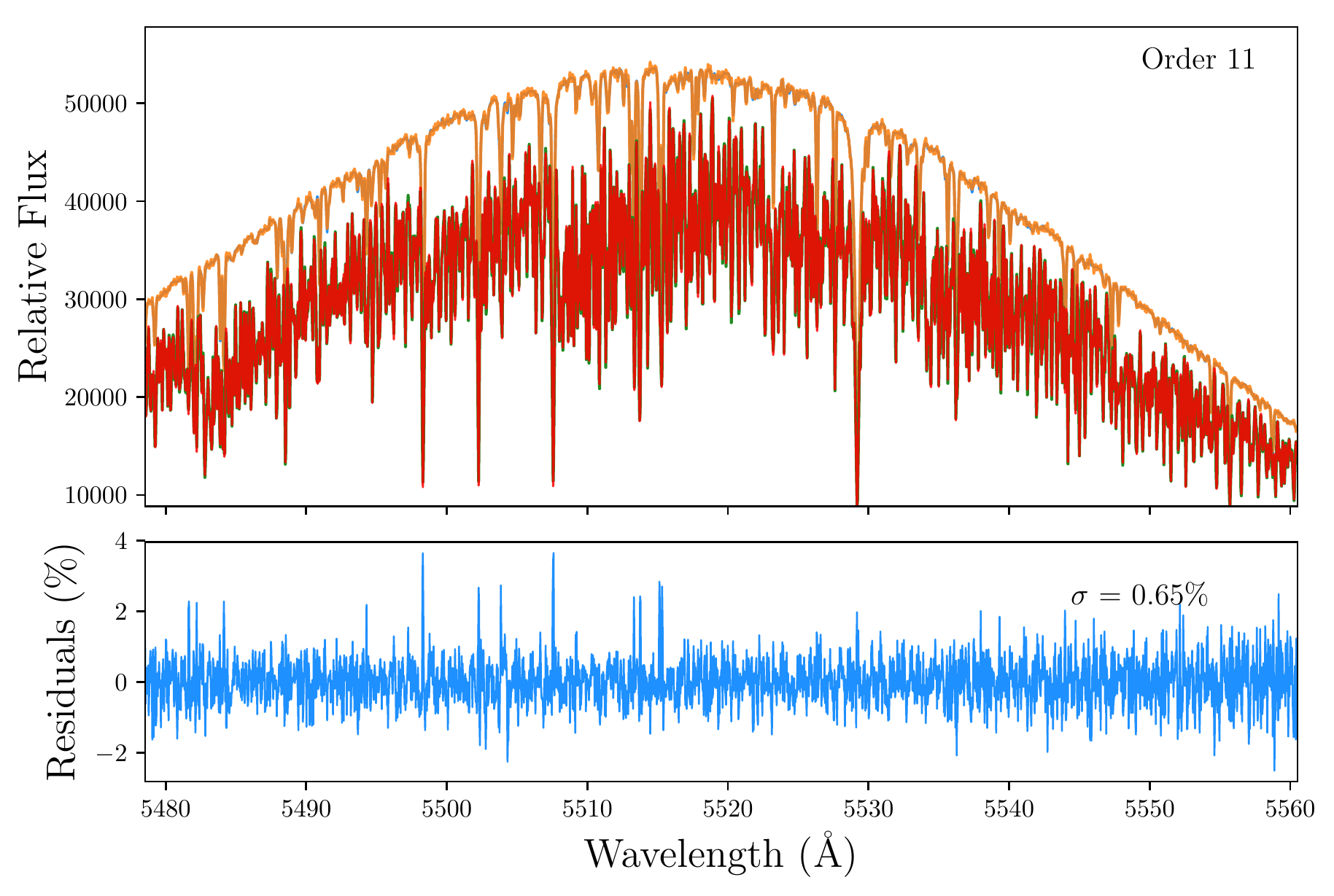}
\includegraphics[scale=0.34]{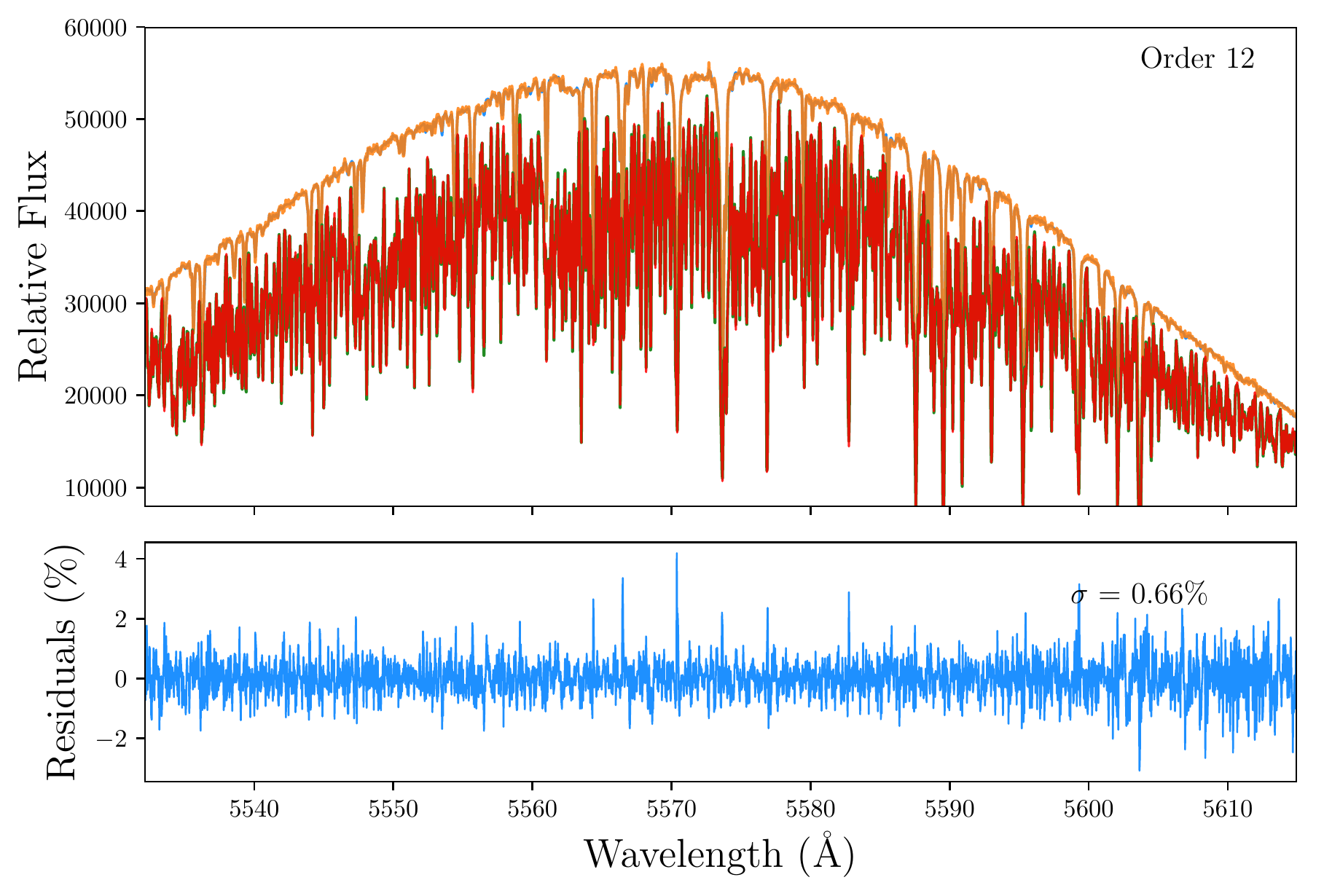}
\includegraphics[scale=0.34]{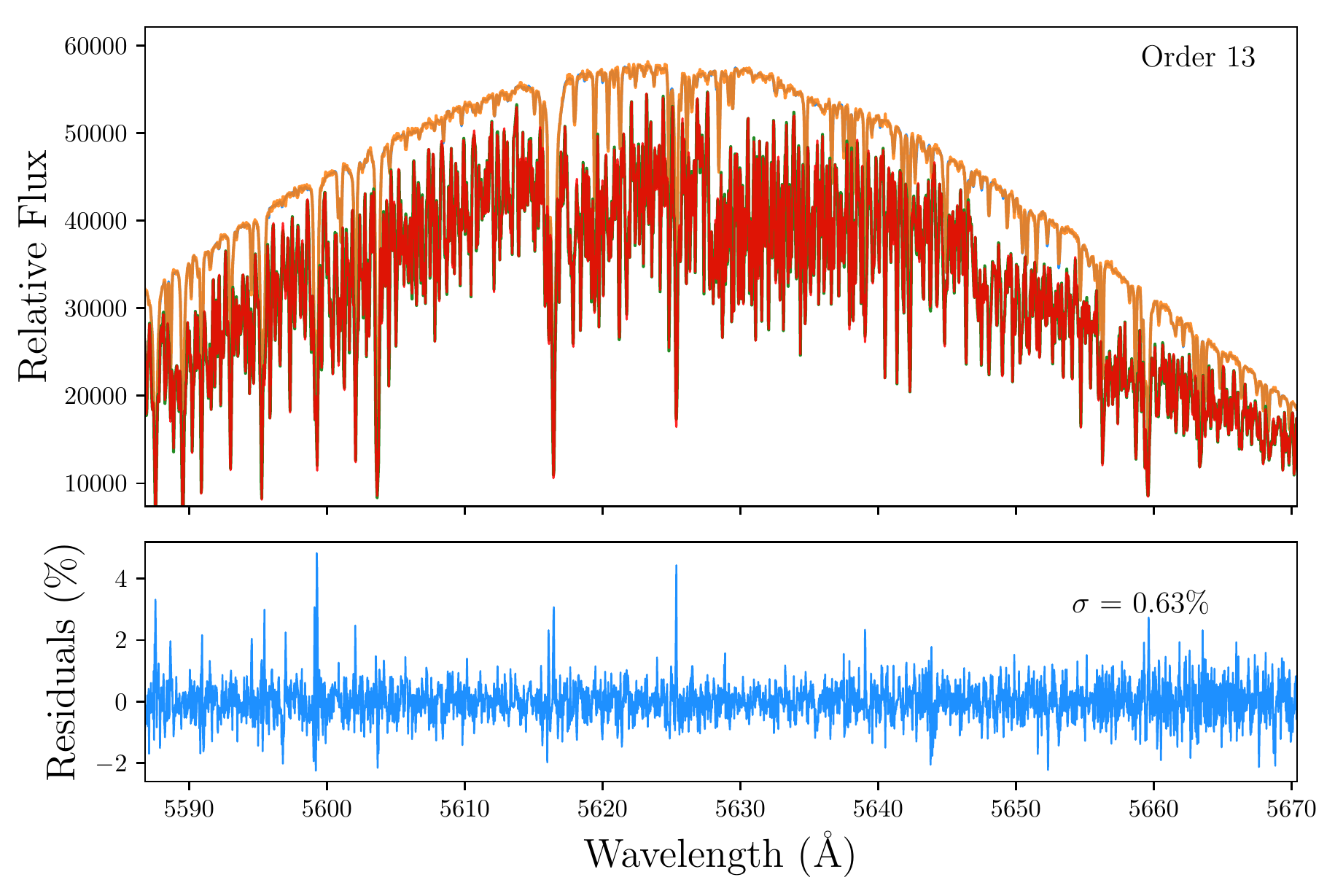}
\includegraphics[scale=0.34]{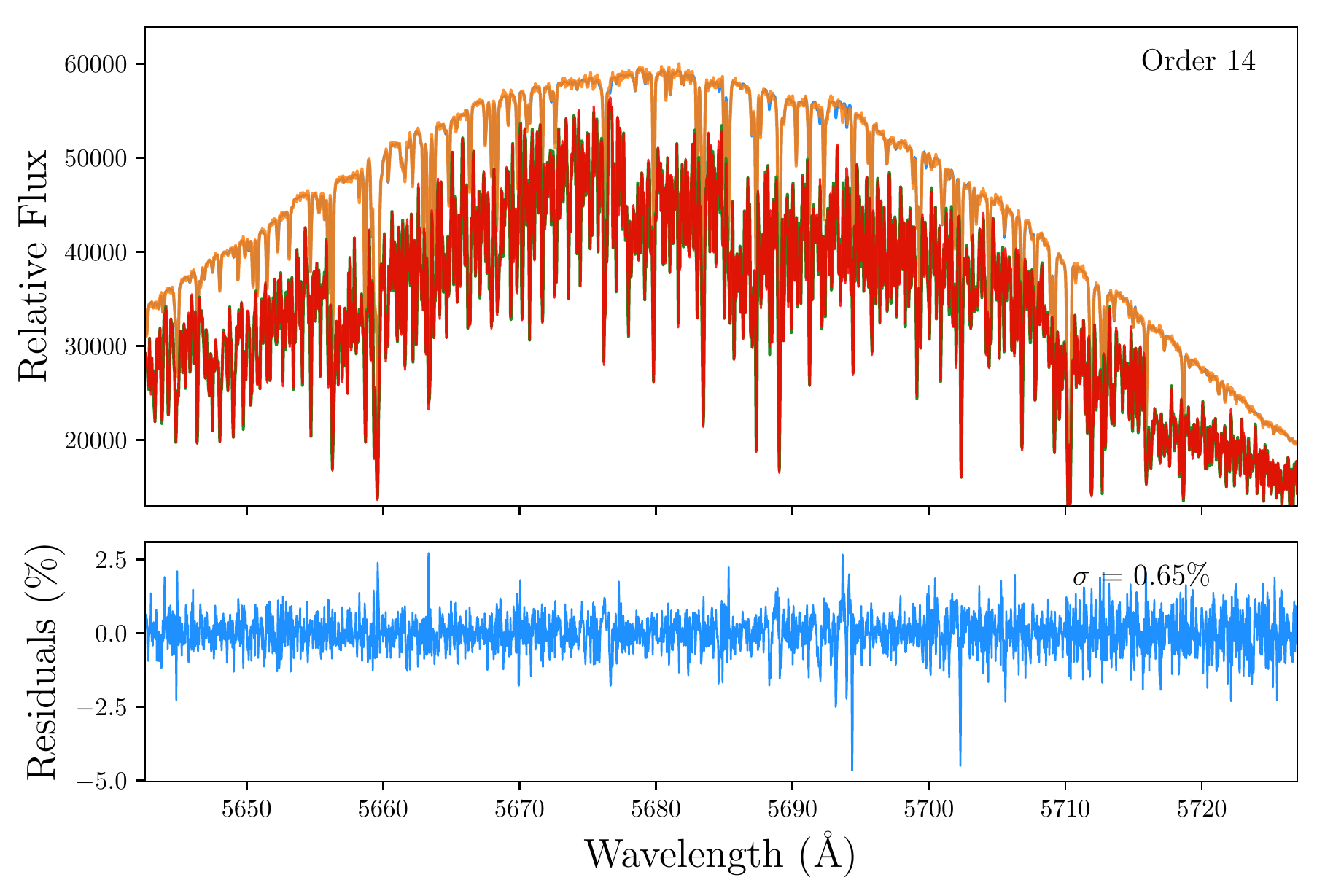}
\includegraphics[scale=0.34]{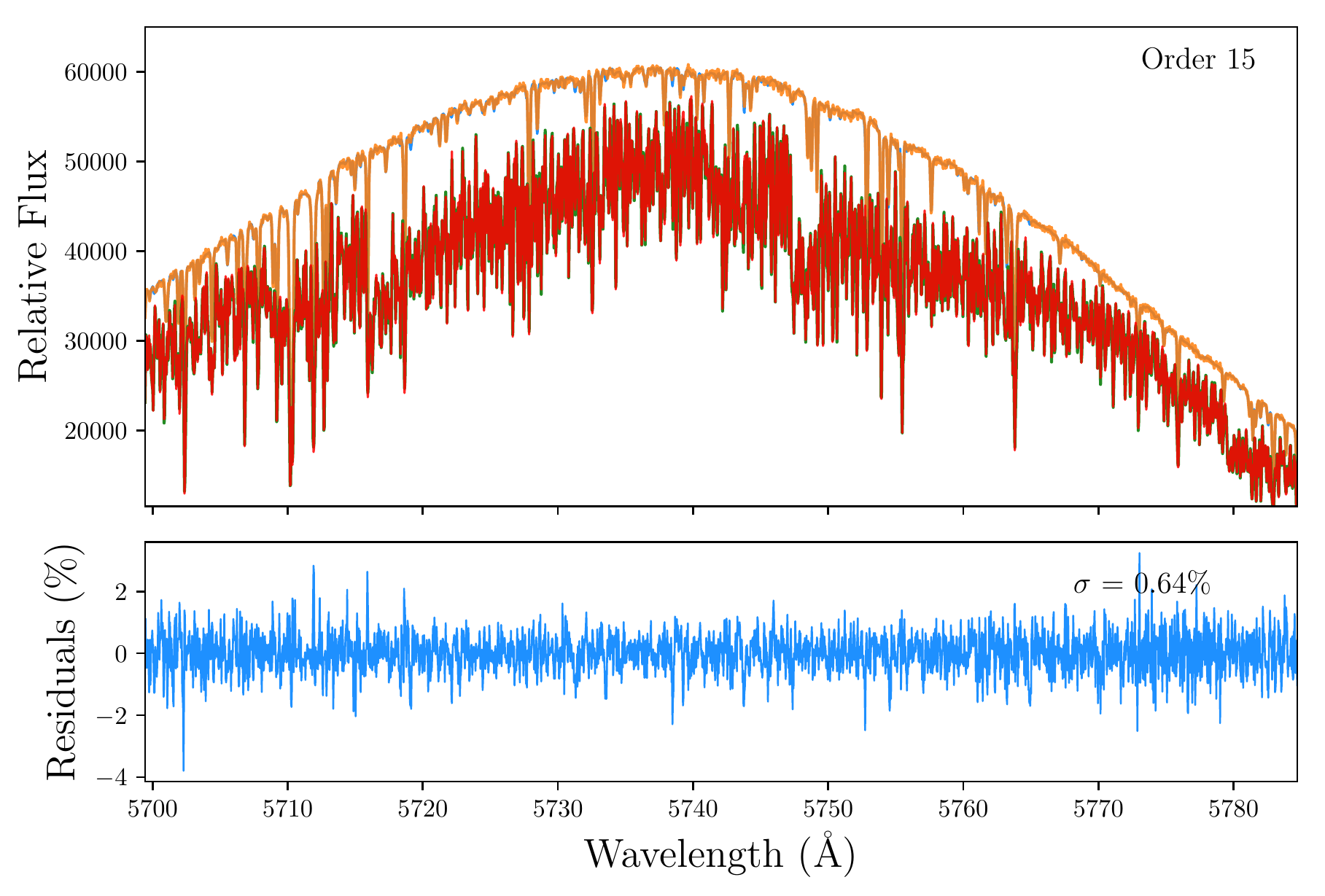}
\caption{Results of the iodine-free derivation for the star $\tau$ Ceti. We show the 20 orders in the iodine region from UCLES. On each plot, we show the observation through iodine (red), the template observation without iodine (blue), the forward model of the observations through iodine (green) and the recovered iodine-free spectrum (orange). Bottom panels on each plot show the level of precision of the recovered spectrum compared to the template (in $\%$).}
\label{fig:plot_orders_ucles1}
\end{figure*}
\renewcommand{\thefigure}{\arabic{figure} (Cont.)}
\addtocounter{figure}{-1}

\begin{figure*}
\includegraphics[scale=0.34]{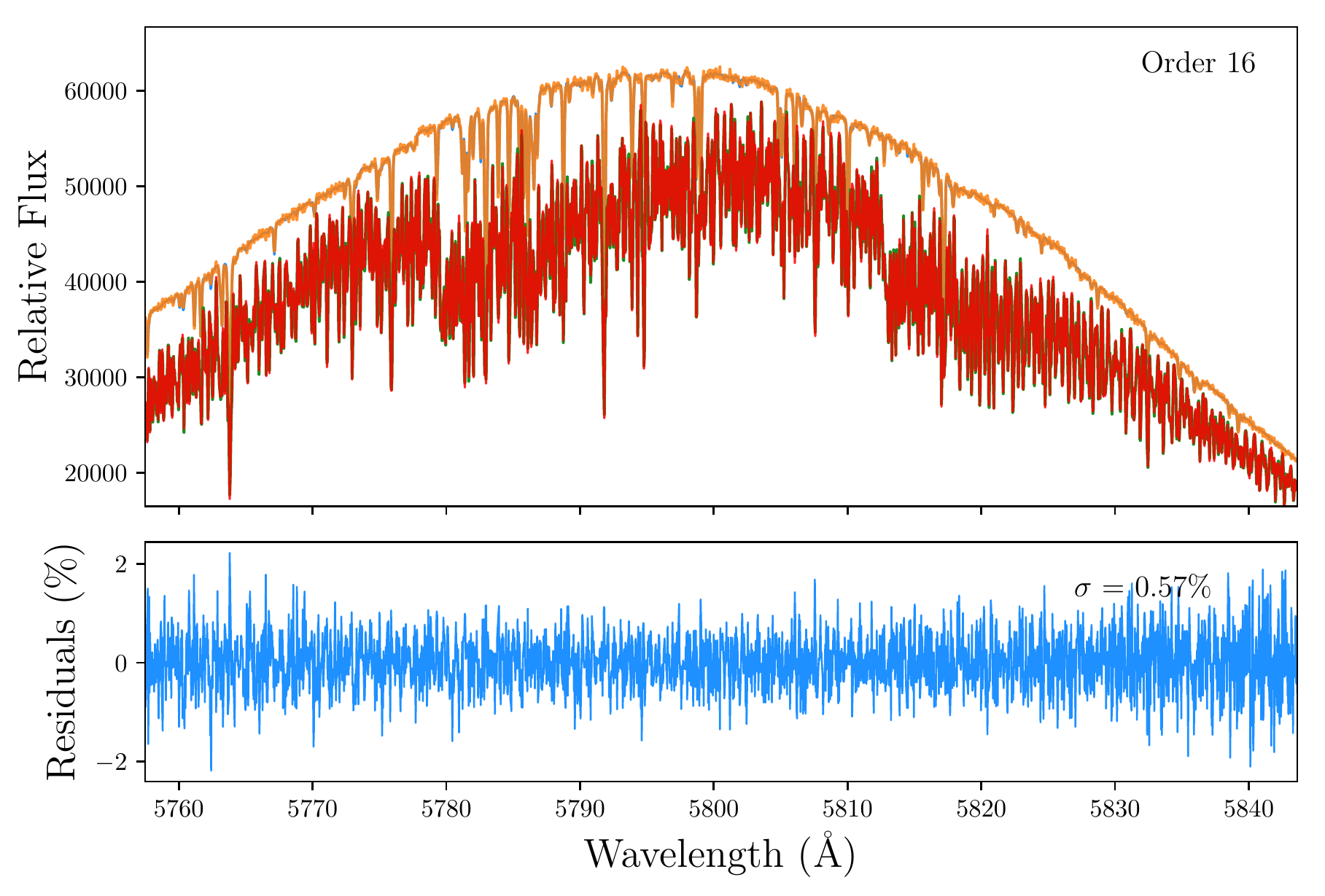}
\includegraphics[scale=0.34]{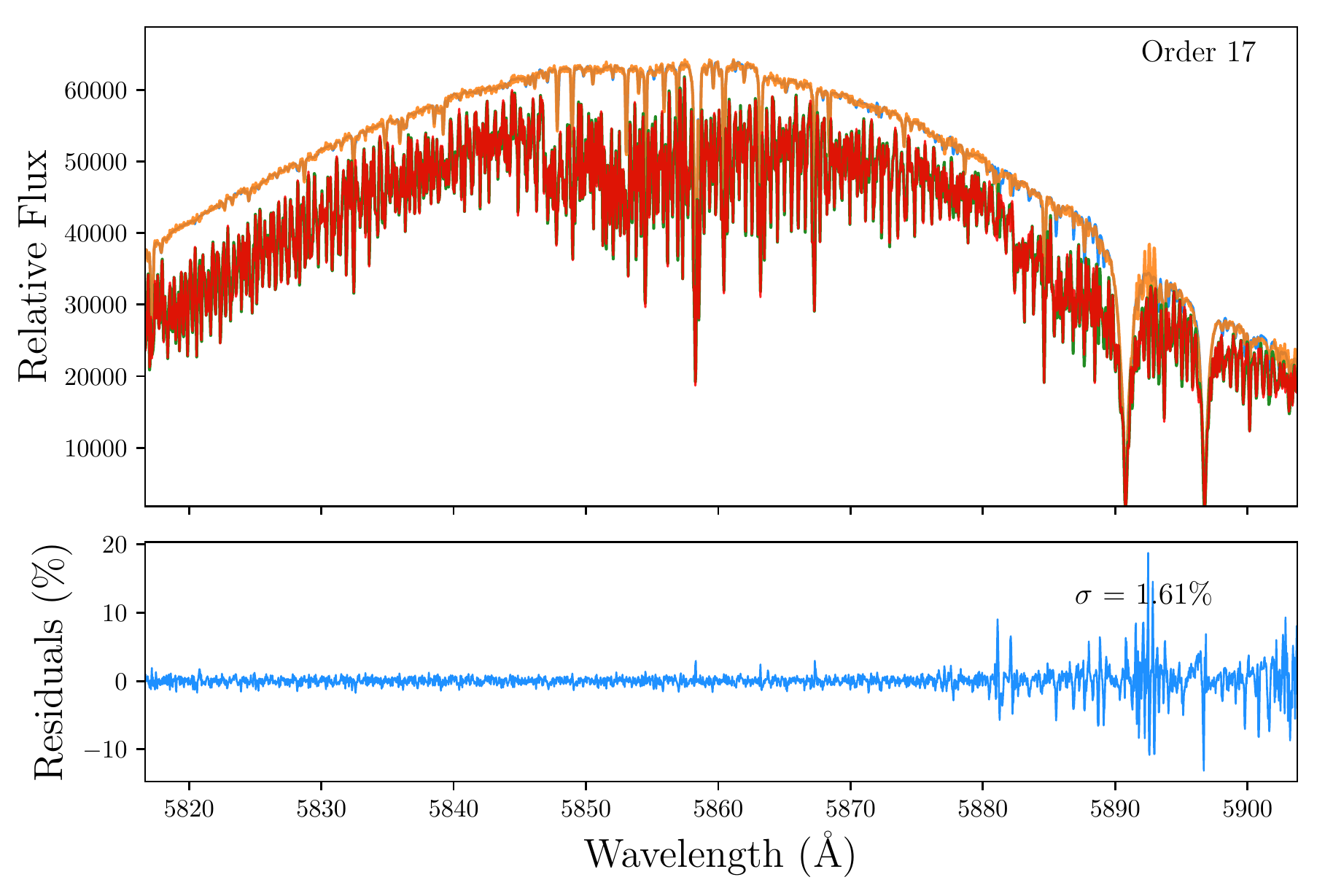}
\includegraphics[scale=0.34]{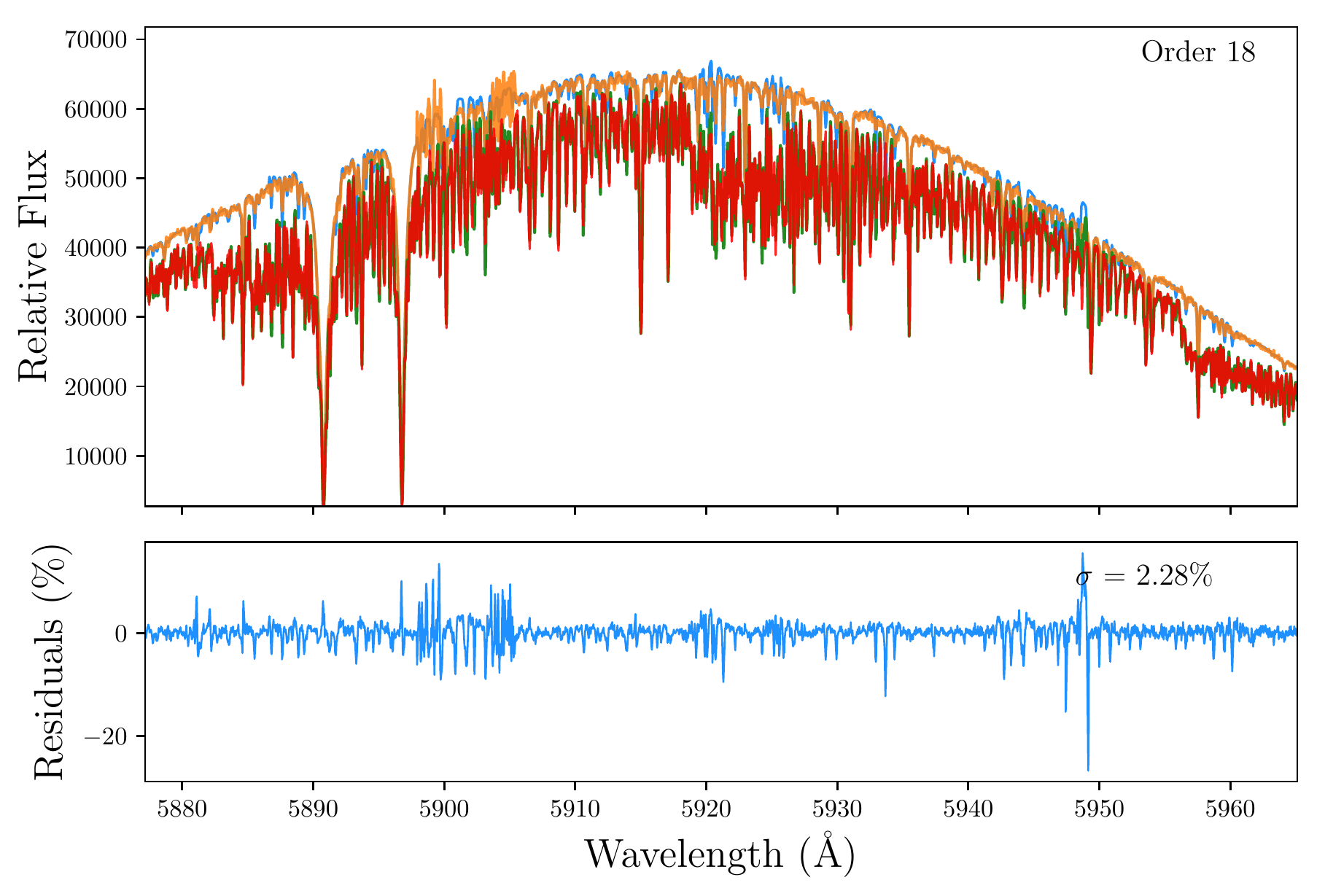}
\includegraphics[scale=0.34]{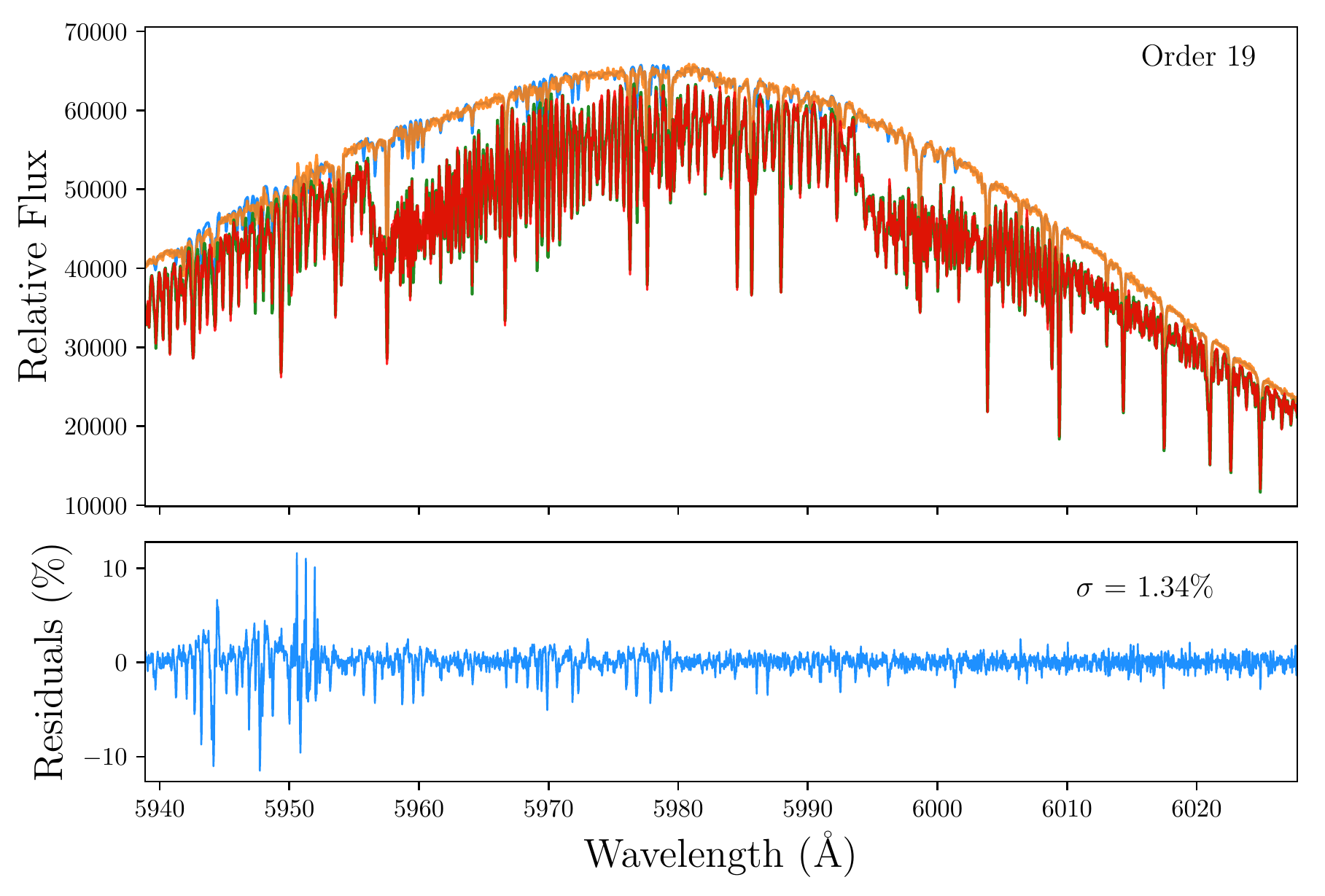}
\includegraphics[scale=0.34]{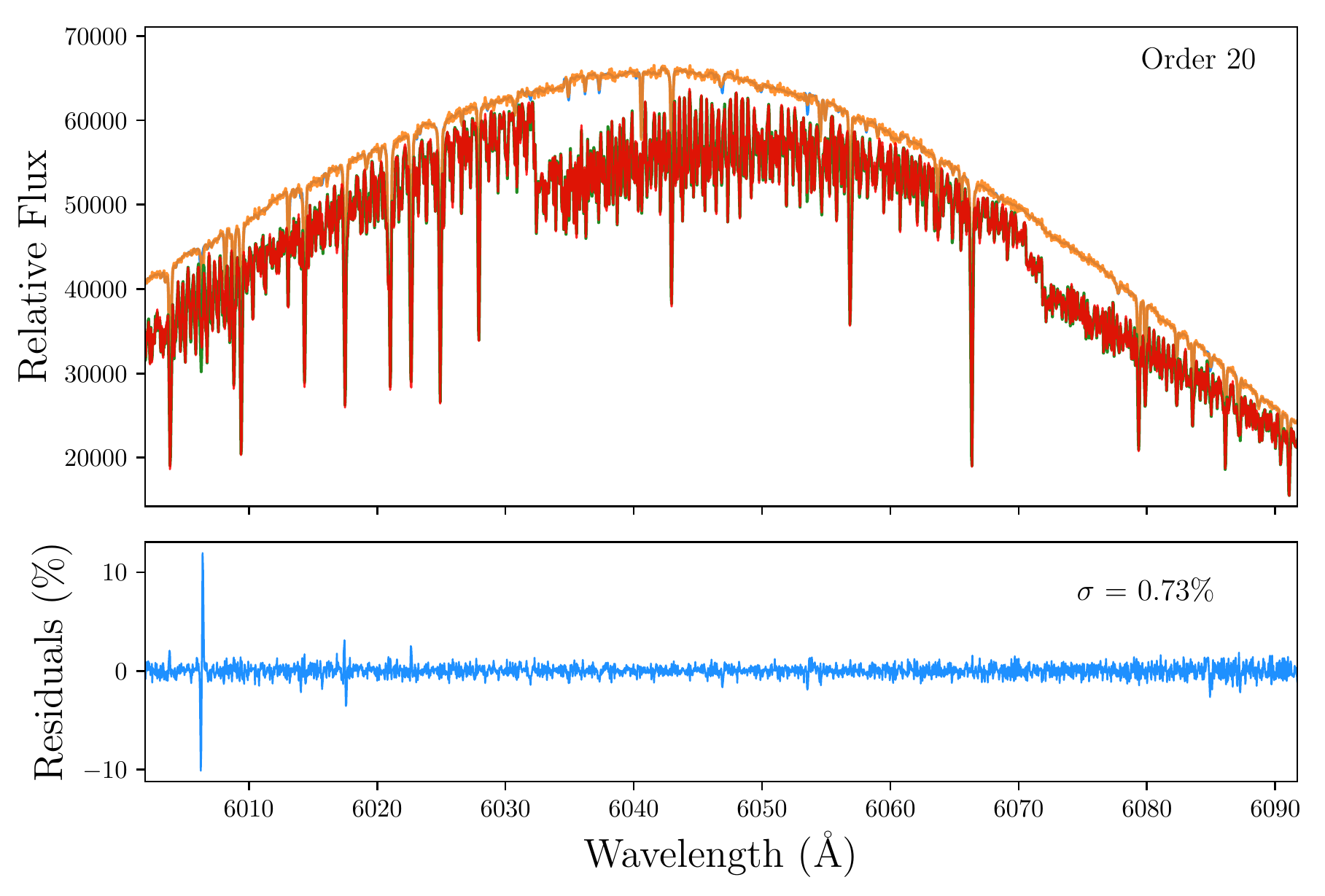}
\caption{Results of the iodine-free derivation for the star $\tau$ Ceti. We show the 20 orders in the iodine region from UCLES. On each plot, we show the observation through iodine (red), the template observation without iodine (blue), the forward model of the observations through iodine (green) and the recovered iodine-free spectrum (orange). Bottom panels on each plot show the le< vel of precision of the recovered spectrum compared to the template (in $\%$).}
\label{fig:plot_orders_ucles2}
\end{figure*}
\renewcommand{\thefigure}{\arabic{figure}}

\section{Direct Applications, Limitations and Future Work}

One immediate application for this method is to derive stellar activity indicators for absorption lines that lie within the iodine region, mostly between 5000\AA~ and 6300\AA. As future work, we plan to derive activity indicators using the iodine-free spectra, as we now can use portions of the spectra that were originally `contaminated' with the iodine spectrum. This means, generating activity indicators for lines such as the Sodium D doublet (Na D, $\lambda$=5889.95\AA, $\lambda$=5895.92\AA), or a host of other lines like Fe {\sc i} or Mg {\sc i} that fall within the iodine region of the spectrum \citep{Wise2018}. 
It should also be possible to carry out diagnostic analyses, by computing line-bisectors and so on, in an attempt to correlate the velocities for the effects of rotation, or at least to identify the observations which are questionable because of such effects. At this point, it is important to note, that the iodine-free spectrum has a LSF which is variable and depends on the optical aberration of the spectrograph and on the slit illumination. Therefore, the use of bisectors to trace line asymmetries is possible if they are larger compared with the LSF variations. Line equivalent widths and activity indices obtained by integrating flux can be useful, as they are independent of the LSF. 

The use of the LSF information derived from the iodine lines as part of the forward model is outside the scope of the present paper and will the subject of future work. 

The idea of creating so called `super templates' has also been a major motivation to build this code. A similar idea has been proposed and successfully implemented by \citet{Gao2016} for NIR observations. 
In our case, we would take out the distorting effect of the iodine cell on the spectrum for each individual spectrum, and then combine them altogether, after sorting for their mutual velocity offsets, to provide a single, high signal to noise and high resolution template observation. Since this template would be constructed using the individual spectra themselves, the signal to noise ratio is only limited by the number of observations considered (assuming the individual spectra are also considered high signal to noise). This method can potentially provide higher precision radial velocities to be calculated, given that the new `super-templates' can allow an increase in the stability of the measurements. They will also provide us with an excellent spectrum to use in the calculation of the stellar parameters like $T_\mathrm{eff}$, log\,$g$, [Fe/H], mass, radius, etc \citep[e.g.][]{Soto2018}.

Another direct application of this method consists of deriving iodine-free spectra for spectroscopic observations of transits.  In particular, many spectra of transiting planets can be taken when the planet is blocking the stellar light as it passes in front of the star, with the express aim of measuring the spin-orbit alignment of the planet through the Rossiter-McLaughlin effect \cite[RM,][]{Rossiter1924, McLaughlin1924}. When studying the RM effect, observations are acquired both outside the transit and in-transit. Hence, removing the iodine from both sets of observations enables the possibility of performing differencing, and hence tracing some elements that could be present in an planet candidate's atmosphere, which otherwise could not be possible when having the iodine spectrum superimposed on the stellar spectrum.  Hence, this method could allow possible transmission spectroscopy to be performed, or planetary reflected light studies. It should be noted however, that the precision to which this method can be considered accurate, depends on observing a high-quality template. Also, defects on the CCD can also impact the derived iodine-free spectrum, specifically they can negatively impact the deconvolved stellar spectrum.

\section{Conclusions}

We have presented a straightforward methodology to derive iodine-free spectra directly from observations taken through an iodine cell for high resolution spectrographs. This method has successfully been implemented for observations carried out with PFS, HIRES and UCLES. We are currently working on the implementation for other spectrographs that use the iodine cell method, such as APF, UVES, and CHIRON. Although the results shown in this paper consider only one well studied star, we have successfully tested the implementation of this algorithm on other stars with different spectral types, producing similar results when the spectra are of a high quality. The method works well for F, G and K spectral types, which constitute the bulk population of stars for most spectra types. Stars of spectral type M are currently not supported, but we plan to integrate a module to derive iodine-free spectra for these objects, particularly since they represent $\sim$15\% of the sample we currently monitor as part of the PFS Exoplanet Search.

One particular disadvantage of this implementation is that it currently relies upon the acquisition of a high-quality template (high signal to noise, good weather conditions) as the smallest flaw in the template is carried to the iodine-free spectrum, giving rise to large residuals for a small number of chunks. In any case, if the template observation is good, deriving 1\% or better iodine correction is possible with this method. 

Derivation of spectral activity indices is the subject of future work. In particular, the case of line bisectors that will be more affected by the asymmetries in the LSF.


\acknowledgments
M.R.D. thankfully acknowledges the support of The Observatories of the Carnegie Institution for Science in Pasadena, CA, where this work has been carried out as part of the Carnegie-Chile Graduate Program 2015-2017, funded by The Observatories. M.R.D. also acknowledges the support of CONICYT-PFCHA/Doctorado Nacional-21140646, Chile and partial funding of Proyecto Basal AFB-170002. R.P.B. gratefully acknowledges support from NASA OSS Grant NNX07AR40G, the NASA Keck PI program, and from the Carnegie Institution of Washington. This material is based upon work supported by the National Aeronautics and Space Administration (NASA) through the NASA Astrobiology Institute under Cooperative Agreement Notice NNH13ZDA017C issued through the Science Mission Directorate. J.S.J. acknowledges support by FONDECYT grant 1161218 and partial support by CATA-Basal (PB06, CONICYT). The authors wish to recognize and acknowledge the very significant cultural role and reverence that the summit of Mauna Kea has always had within the indigenous Hawaiian community. We are most fortunate to have the opportunity to conduct observations from this mountain. 


\bibliographystyle{aa}
\bibliography{references}
\listofchanges

\end{document}